\documentclass[a4paper,11pt]{article}
\usepackage{a4wide}
\usepackage{ascii}
\usepackage[T1]{fontenc}
\usepackage{amsmath,amssymb,amsthm,mathtools,array,booktabs}
\usepackage{stmaryrd}
\usepackage{graphicx,hyperref}
\usepackage{xcolor}
\usepackage{cancel}
\usepackage{tikz}
\usetikzlibrary{cd}
\usepackage{amscd}
\usepackage{latexsym,cite}
\usepackage{subcaption}
\usepackage[boxsize=0.8em]{ytableau}
\usepackage{pictexwd,dcpic}
\usepackage{youngtab}

\Yboxdim{5pt}

\input xy
\xyoption{all}

\hypersetup{
  unicode,
  bookmarksnumbered,
  linktoc = all,
  pdfborderstyle = {/S/U/W 0.5}
}

\def\mC{\mathbb{C}}

\def\cL{\mathcal{L}}

\def\cH{\mathcal{H}}

\def\cF{\mathcal{F}}
\def\cB{\mathcal{B}}
\def\cK{\mathcal{K}}
\def\cL{\mathcal{L}}

\def\cT{\mathcal{T}}
\def\cO{\mathcal{O}}
\def\cW{\mathcal{W}}
\def\cA{\mathcal{A}}
\def\cE{\mathcal{E}}
\def\cH{\mathcal{H}}
\def\cB{\mathcal{B}}

\def\CP{{{\mathbb{P}}}}
\def\C{{{\mathbb{C}}}}
\def\Z{{{\mathbb{Z}}}}

\let\Re\relax
\let\Im\relax
\DeclareMathOperator{\Re}{Re}
\DeclareMathOperator{\Im}{Im}

\DeclareMathOperator{\Res}{Res}
\DeclareMathOperator{\Sym}{Sym}

\DeclareMathOperator{\Hom}{Hom}
\DeclareMathOperator{\Cone}{Cone}
\DeclareMathOperator{\rank}{rank}
\DeclareMathOperator{\Ch}{ch}
\DeclareMathOperator{\Td}{Td}

\DeclareMathOperator{\Tr}{Tr}
\DeclareMathOperator{\Id}{Id}

\numberwithin{equation}{section}

\begin{document}
\thispagestyle{empty}
\begin{flushright}
preprint
\end{flushright}
\vspace{1cm}
\begin{center}
{\LARGE\bf B-brane Transport and Grade Restriction Rule for Determinantal 
Varieties} 
\end{center}
\vspace{8mm}
\begin{center}
Ban Lin\footnote{{\tt lin-b19@mails.tsinghua.edu.cn}}
\\
Mauricio Romo\footnote{{\tt mromoj@simis.cn}}
\end{center}
\vspace{6mm}
\begin{center}
${}^{1,2}$Yau Mathematical Sciences Center, Tsinghua University, Beijing, 100084, China
\\
${}^{1,2}$Department of Mathematical Sciences, Tsinghua University, Beijing 100084, China
\\
${}^{2}$Center for Mathematics and Interdisciplinary Sciences, Fudan University, Shanghai, 200433, China
\\
${}^{2}$Shanghai Institute for Mathematics and Interdisciplinary Sciences (SIMIS), Shanghai, 200433, China
\end{center}
\vspace{15mm}
\begin{abstract}
\noindent
We study autoequivalences of $D^{b}Coh(X)$ associated to B-brane transport 
around loops in the stringy K\"ahler moduli of $X$. We consider the case of 
$X$ being certain resolutions of determinantal varieties embedded in 
$\mathbb{P}^{d}\times G(k,n)$. Such resolutions have been modeled, in general, 
by nonabelian gauged linear sigma models (GLSM). We use the GLSM construction 
to determine the window categories associated with B-brane transport between 
different geometric phases using the machinery of grade restriction rule and 
the hemisphere partition function. In the family of examples analyzed the monodromy around phase boundaries enjoy the interpretation as loop inside 
link complements. We exploit this interpretation to find a decomposition of 
autoequivalences into simpler spherical functors and we illustrate this in two examples of Calabi-Yau 3-folds $X$, modeled by an abelian and nonabelian GLSM respectively. In addition we also determine explicitly the action of the autoequivalences on the Grothendieck group $K(X)$ (or equivalently, B-brane charges).
\end{abstract}
\newpage
\setcounter{tocdepth}{3}
\tableofcontents
\setcounter{footnote}{0}

\section{Introduction}

The study of derived equivalences or equivalences between 
triangulated categories using gauged linear sigma models 
\cite{Witten1993} (GLSM) is already a 
well established subject in physics and mathematics 
\cite{herbst2008b,halpern2015derived,ballard2019variation,hori2013exact}. In 
the present work we 
proposed the analysis of autoequivalences of derived categories $D^{b}Coh(X)$ 
of coherent sheaves of Calabi-Yau (CY) varieties $X$ that are given by 
resolutions of determinantal varieties. The GLSMs having a geometric phase 
corresponding to a NLSM on $X$ where constructed and their deformation moduli 
space thoroughly analyzed in \cite{2012determinantal}. In particular, their 
stringy K\"ahler moduli $\mathcal{M}_{K}(X)$ was completely determined for an extensive family of examples. We will focus on the families termed linear PAX models in \cite{2012determinantal}, which we review in section 
\ref{sec:section2}. Let us mention the main characteristics of these models, 
some of them
that encompasses the main novelties of our analysis:
\begin{enumerate}
 \item The CY variety $X$ is presented as a non-complete intersection inside 
the space
$\mathbb{P}^{n(n-k)-1}\times G(n-k,n)$. Therefore, standard techniques, for 
example, to determine $\mathcal{M}_{K}(X)$, such as the 
ones coming from mirror symmetry do not apply generically in these examples.
\item The GLSM realizing $X$ has gauge group $G=U(1)\times U(n-k)$. Therefore 
$\mathrm{dim}\mathcal{M}_{K}(X)=2$ and it has shown to have three geometric 
phases \cite{2012determinantal} denoted by $X_{A}$, $X_{A^{T}}$ and $Y_{A}$ 
\footnote{In general, we can have four geometric phases, however we will focus 
on these three.}
\end{enumerate}
The B-branes of the GLSM are, in part, specified by a representation 
$\rho_{M}:G\rightarrow GL(M,\mathbb{C})$, whose weights are constrained by the 
grade restriction rule at the different phase boundaries\footnote{The grade 
restriction rules and their corresponding window categories wehere originally 
defined for abelian non-anomalous GLSMs in \cite{herbst2008phases} and 
mathematically in 
\cite{Segal:2010cz,halpern2015derived,ballard2019variation} for 
general gauge group and including anomalous GLSMs, for a physics 
perspective of this latter generalizations see 
\cite{hori2013exact,Clingempeel:2018iub}.} \cite{herbst2008phases}. We will 
first 
focus on determining the window categories defined by the grade restriction 
rules at the $(X_{A},X_{A^{T}})$ phase boundary and at the $(X_{A},Y_{A})$ 
phase boundary which we call $X$ and $Y$ phase boundary, respectively. Denoting 
the weights of $\rho_{M}$ as $q^{0}$ for the $U(1)$ factor and $q^{\alpha}$, 
$\alpha=1,\ldots,n-k$ for the $U(n-k)$ factor, the window category 
$\mathbb{W}_{Y}(\theta_{0})$ at the $Y$ boundary is determined (see section 
\ref{sec:section4}) to be given by
\begin{equation}
-\frac{n(n-k)}{2}< \frac{\theta_0}{2\pi}+q^0 < 
\frac{n(n-k)}{2},
\end{equation}
and the window category 
$\mathbb{W}_{X}(\theta_1)$ at the $X$ boundary is determined by the hypercube
\begin{equation}
-\frac{k+1}{2}< \frac{\theta_1}{2\pi}+q^\alpha < \frac{k+1}{2}, \qquad 
\text{For all \ }\alpha
\end{equation}
As an application/check of these results, we compute the monodromy around the 
phase boundaries in two specific examples, one is an abelian model, the 
determinantal quintic $X_{5}\subset \mathbb{P}^{4}\times  \mathbb{P}^{4}$ and 
the Gulliksen-Negard (GN) determinantal variety $X_{GN}\subset  
\mathbb{P}^{7}\times G(2,4)$, both CY 3-folds. In the case of $X_{5}$, the 
computation is standard and can be performed using the already known techniques 
of \cite{herbst2008phases}, so, it serves as a good check for our general 
results. We find the that the autoequivalence 
$T_{X}\in\mathrm{Aut}(D^{b}Coh(X_{5}))$ associated to the monodromy around the 
$X$ phase 
boundary is given by:
\begin{equation}
T_{X}=\mathcal{D}_{x}^{-5}(\mathcal{D}_{x}
T_{\mathcal{O}_{X_{5}}})^{5}\qquad \mathcal{D}_{x}:=-\otimes 
\mathcal{O}_{X_{5}}(1,0)
\end{equation}
and the autoequivalence $T_{Y}$ corresponding to the monodromy around the $Y$ 
phase boundary:
\begin{equation}
T_{Y}=\mathcal{D}_{y}^{-5}(\mathcal{D}_{y}
T_{\mathcal{O}_{X_{5}}})^{5}\qquad \mathcal{D}_{y}:=-\otimes 
\mathcal{O}_{X_{5}}(0,-1)
\end{equation}
where $T_{E}$ 
denotes 
the spherical twist with respect to the object $E$. Likewise for $X_{GN}$ we 
obtain
\begin{equation}
T_{X}=-\otimes \mathcal{O}_{X_{GN}}(-4,0)
\end{equation}
and
\begin{equation}
T_{Y}=\mathcal{D}^{-4}
(T_{S_{X}}T_{\mathcal{O}_{X_{GN}}}\mathcal{D})^{2}
(T_{\mathcal{O}_{X_{GN}}}\mathcal{D})^{2}
\end{equation}
where in this case $\mathcal{D}:=-\otimes \mathcal{O}_{X_{GN}}(0,-1)$ and 
$\mathcal{O}_{X_{GN}}(0,-1)$ stands for $\mathcal{O}_{X_{GN}}$ twisted by 
$\mathrm{det}^{-1}S_{X}$ where $S_{X}$ stands for the tautological sub-bundle of 
$G(2,4)$ pulled back to $X_{GN}$. The expressions for $T_{X}$ in both examples 
are directly found from functors mapping between the windows. On the 
other hand, the expression for $T_{Y}$ is found in an analogous way, however, its decomposition in terms of simpler spherical functors is found after a careful analysis of the fundamental group $\pi_{1}(\mathcal{M}_{K}(X))$ around the $Y$ phase boundary, as it was done in \cite{aspinwall2001some} for some specific abelian models (see section \ref{sec:section5}). A similar decomposition in principle exists for $T_{X}$ as well but we can only determine it in the case of $X_{5}$, being $\pi_{1}(\mathcal{M}_{K}(X_{GN}))$ near the $X$ boundary too technically challenging to analyze.\\

The paper is organized as follows. In section \ref{sec:section2} we review the 
general aspects about GLSMs for determinantal varieties from 
\cite{2012determinantal} and in particular the ones for linear PAX models as 
well as the two working examples $X_{5}$ and $X_{GN}$. In section 
\ref{sec:section3} after reviewing the construction of B-branes and the 
hemisphere partition function $Z_\cB(t)$ (\ref{ZD2}) for GLSMs, 
following mostly 
\cite{herbst2008phases,hori2013exact,hori2019notes} we compute explicitly the 
B-branes corresponding to generators of $D^{b}Coh(X_{A})$ and check against 
their central charges (via the hemisphere partition function) that they indeed 
give the correct geometric content. We present explicit expressions for 
the central charges in the $X_{A}$ phase. In \ref{sec:section4} we compute the 
grade restriction rule determining the window categories at the 
$X$ and $Y$ phase boundaries and  the empty branes at the different phases that will be used to compute monodromy later on. In \ref{sec:section5} we use the 
window categories and empty branes to compute the monodromy as 
autoequivalences acting on $D^{b}Coh(X_{A})$ and using the B-brane central charges from \ref{sec:section3}, we compute the matrices corresponding to $T_{X}$ and $T_{Y}$ as acting on the basis of the K-theory/brane charges.   

\section*{Acknowledgements}

We thank J. A. Cruz-Morales, W. Donovan, R. Eager, K. Hori, S. Hosono, J. Knapp, E. Scheidegger and
L.Smith  
for helpful and enlightening discussions. We thank L. Santilli for carefully reading the draft. MR acknowledges support from the
National Key Research and Development Program of China, grant No. 
2020YFA0713000, the Research Fund for International Young Scientists, NSFC grant No. 1195041050. MR also acknowledges IHES, Higher School of Economics,  Simons Center for Geometry 
and Physics, St. Petersburg State University and Fudan University for hospitality at the final stages of this work. BL acknowledges the hospitality of C.Closset and University of Birmingham for hosting a short term visit of BL in 2023 when this work was being carried on.
\section{\label{sec:section2} GLSMs for Determinantal Varieties}

In this section we review the construction of GLSMs for certain determinantal 
CYs. We will focus on the construction of a class of $U(1)\times 
U(k)$ gauge theories, termed 'PAX models' in \cite{2012determinantal}. In 
particular we will be interested in linear PAX models (defined below). These 
models includes our main working examples, the determinantal quintic 
\cite{abpax1,abpax2} in $\CP^4$ and the Gullinksen-Negard (GN) CY 
\cite{GN1971} (or more precisely, its resolution) in $\CP^7\times G(2,4)$. 

\subsection{Determinantal Varieties and PAX Model}

Let $B$ be a compact algebraic variety of dimension $d$, and two holomorphic vector bundles $\cE$, $\cF$ over $B$ of rank $n$ and $m$ respectively. Without loss of generality, assume $m\leq n$. In addition, consider a holomorphic section $A$ of the bundle $\Hom(\cE,\cF)$. Then, the determinantal variety $Z(A,k)$, for $0\leq k<m$ is defined by:
\begin{equation}
Z(A,k)=\lbrace \phi\in B\ \vert\ \rank A(\phi)\leq k \rbrace,
\end{equation}
$Z(A,k)$ is an algebraic variety, with $I(Z(A,k))$ the ideal generated by all 
$(k+1)\times(k+1)$ minors of $A$. Since, in general, the number of 
$(k+1)\times(k+1)$ minors generally exceeds the codimension of $Z(A,k)$ in $B$, 
which is $(m-k)(n-k)$, determinantal varieties are generically non-complete 
intersections. When $\Hom(\cE,\cF)$ is generated by global sections, all 
singularities on $Z(A,k)$ arise only at $Z(A,k-1)\subset Z(A,k)$, or at 
singularities induced from $B$.

Denote $B_{\cE,k}$ the fibration of the (relative) Grassmannian of $k$-planes 
plane in the fibers of $\cE$ i.e. the fiber of $B_{\cE,k}$ over $\phi\in B$ is 
given by $G(k,\cE_{\phi})$: 
\begin{equation}
\begin{tikzcd}
G(k,\cE) \arrow{r}
    & B_{\cE,k}\arrow{r}{\pi}
        & B
\end{tikzcd}
\end{equation}
We can use $B_{\cE,n-k}$ (with fibers $G(n-k,\cE)$) to construct a 
desingularization $\tilde{Z}(A,k)$ of $Z(A,k)$ through an incidence 
correspondence \cite{harris1992}:
\begin{equation}
X_A:=\tilde{Z}(A,k)=\lbrace p\in B_{\cE,n-k}\ \vert\ 
A(\pi(p))\circ\pi^{-1}(\phi)=0 \rbrace,
\end{equation}
then, the incidence correspondence $X_{A}$  
resolves the singularities located at $Z(A,k-1)\subset Z(A,k)$ and $X_A$ is
birational to $Z(A,k)$.

Consider now $B_{\cE^\vee,k}$ (with fibers $Gr(k,\cE^\vee)$) where $\cE^\vee$ 
denotes the dual bundle to $\cE$ and let $\mathcal U$ be the rank $k$ 
sub-bundle and $\mathcal Q$ the rank $n-k$ quotient sub-bundle on 
$B_{\cE^\vee,k}$, which restrict to the corresponding bundles on 
$Gr(k,\cE^\vee)$. Define then the bundle $\mathcal X$ by\footnote{We abuse 
notation by denoting the projection $\pi:B_{\cE^\vee,k}\rightarrow B$ also in 
this case.}
\begin{equation}
\mathcal X:=(\pi^*\cE^\vee/\mathcal U)\otimes\pi^*\cF\cong\Hom(\mathcal 
Q^\vee,\pi^*\cF),
\end{equation}
which has a global holomorphic section $\tilde{A}$ induced from 
$A$ \footnote{Here $f$ denotes the map appearing on the short exact sequence 
$0\rightarrow\mathcal{U}\otimes 
\pi^{*}\mathcal{F}\rightarrow
\pi^{*}\mathcal{E}^{\vee}\otimes\pi^{*}\mathcal{F}\stackrel{f}{\longrightarrow
} \pi^{*}\mathcal{Q} \otimes\pi^{*}\mathcal{F}\rightarrow 0$}:
\begin{equation}
\tilde{A}:B_{\cE^\vee,k}\rightarrow\mathcal{Q}\otimes\pi^*\cF,\quad 
\tilde{A}=f\circ\pi^*A.
\end{equation} 
It is then straightforward to show that $X_{A}$ is isomorphic to the variety 
$X_{\tilde{A}}$ given by the zeroes of $\tilde{A}$ in $B_{\cE^\vee,k}$ 
\cite{2012determinantal}. That is
\begin{equation}
X_{\tilde{A}}=\{ \tilde{A}=0\}\cap B_{\cE^\vee,k}\cong X_{A}
\end{equation}


The advantage of having $X_{\tilde{A}}$ is that, it is a complete 
intersection, hence, topological invariants can be computed 
straightforwardly in $X_{\tilde{A}}$ and then related to the topological invariants in $X_{A}$ using the isomorphism. For instance, the total Chern class of $X_{\tilde{A}}$ is given by
\begin{equation}
c(X_{\tilde{A}})=\frac{c(B_{\cE^\vee,k})
} { c(\mathcal X)}.
\end{equation}
In particular,
\begin{equation}
\begin{aligned}
c_1(X_{\tilde{A}})=&c_1(B_{\cE^\vee,k})-c_1(\mathcal X)
\\
&=\pi^*c_1(B)-(m-k)\pi^*c_1(\cE^\vee)-(n-k)\pi^*c_1(\cF)-(n-m)c_1(\mathcal U).
\end{aligned}
\end{equation}
The Euler characteristic $\chi(X_{\tilde{A}})$ can be evaluated by
\begin{equation}
\chi(X_{\tilde{A}})=\int_{X_{\tilde{A}}}c_{top}(X_{\tilde{A}})=\int_{B_{\cE^\vee
,
k } } c_{ top} (\mathcal X)\wedge c_{top}(X_{\tilde{A}}).
\end{equation}
Furthermore, the intersection numbers $I(\gamma_{k_1},\cdots,\gamma_{k_s})$ for 
cohomology classes $\gamma_{k}\in H^{ev}(X_{\tilde{A}},\Z)$ induced from $B_{\cE^\vee,k}$ are computed by
\begin{equation}
I(\gamma_{k_1},\cdots,\gamma_{k_s})=\int_{X_{\tilde{A}}}\gamma_{k_1}
\wedge\cdots\wedge\gamma_{k_s}=\int_{B_{\mathcal{E}^{\vee},k}}c_{top}(\mathcal 
X)\wedge\gamma_{k_1}\wedge\cdots\wedge\gamma_{k_s}.\label{intfor}
\end{equation}

In the following we will work with particular families of smooth determinantal 
varieties $X_{A}$, that were termed 'PAX models' in \cite{2012determinantal}. 
These families are also called linear determinantal varieties and arise whenever 
$\cE$ takes the form ($\cF$ remains arbitrary)
\begin{equation}
\cE=\cL\otimes\cO_B^{\oplus n}
\end{equation}
where $\cL$ is a line bundle. In such case, the fibration becomes trivial:
\begin{equation}
B_{\cE^\vee,k}\cong B\times G(k,n)
\end{equation}
and the variety $X_{A}$ can be written as
\begin{equation}
\begin{gathered}
X_A=\lbrace (\phi,x)\in B\times G(n-k,n)\ \vert\ A(\phi)x=0 \rbrace
\end{gathered}\label{Zresol}
\end{equation}
For our interest on determinantal Calabi-Yau 3-folds, we require
\begin{equation}
\dim X_A=d-(n-k)(m-k)=3,\quad c_1(X_{A})=0.
\end{equation}
Then the dimension of singular loci is $2k-(n+m-2)$. Hence a 3-fold has
non-empty singular locus if $n-1=m-1=k$. Otherwise, there is no locus of reduced 
rank with generic $A$. Furthermore, we will restrict to the case 
\begin{equation}
m=n,
\end{equation}
i.e. $A$ 
will be a square matrix. Several possible generalizations of these conditions 
are discussed in \cite{2012determinantal}.

\subsection{GLSM Construction}
In this subsection we introduce a construction for a GLSM that implements the 
linear determinantal varieties discussed above. More precisely, here we review 
the construction of \cite{2012determinantal} for a linear determinantal variety 
i.e. a GLSM that has a geometric phase corresponding to a nonlinear sigma model 
with target space $X_{A}$. In this construction we will restrict to the case 
that the variety $B$ is a projective space. In general we can consider $B$ to 
be any toric variety, in a straightforward way. For this purpose, consider the 
gauge group  
\begin{equation}
G=U(1)^{s}\times U(n-k)
\end{equation}
and matter superfields (chiral superfields) in the following representation of 
$G$: $\Phi_a$, $a=1,\cdots,d+s$ in the trivial representation of $U(n-k)$ and 
in with weights $Q_{a}$ under $U(1)^{s}$. $P_{i}$, $i=1,\ldots,n$ in 
the fundamental of $U(n-k)$ and weights $-\mathfrak{q}_i$ under 
$U(1)^{s}$. $X_{i}$, $i=1,\ldots,n$ in 
the anti-fundamental of $U(n-k)$ and weights $q_i$ under $U(1)^{s}$. Choosing a basis for $U(1)^{s}=U(1)_{1}\times\cdots\times U(1)_{s}$ we can summarize the 
matter representation in the following table ($l=1,\ldots, s$)
\begin{equation}
\begin{array}{c|ccc} 
  & \Phi_a & P_i & X_i \\
  \hline
U(1)_l & Q_a^l & -\mathfrak{q}_i^l & -q_i^l \\
U(n-k) & 0 & \mathbf{n-k} & \overline{\mathbf{n-k}}
\end{array}
\end{equation}
The matter interact via the $G$-invariant superpotential
\begin{equation}\label{PAXpot}
W=\sum_{i,j=1}^n \mathrm{Tr}(P_{ 
i}A(\Phi)^{ij}X_{j}),
\end{equation}
there the trace is taken over $U(n-k)$ and
$A(\Phi)^{ij}$ is homogeneous in $\Phi$, however its degree depends on the 
particular model. Then, (\ref{PAXpot}) is $G$-invariant, provided that
\begin{equation}
\deg_l\left(A(\phi)^{ij} \right)= \mathfrak{q}_i^l +q_j^l,\quad \forall 
i,j=1,\cdots,n.
\end{equation}
where $\deg_l(f)$ denotes the weight of $f$ under $U(1)_{l}$. In the following 
we will denote $\phi,x$ and $p$ the lowest components of the superfields 
$\Phi,X$ and $P$ respectively.

This model was called PAX \textit{model} in \cite{2012determinantal} and we 
will refer in general to this family of models, as such. 
The anomaly-free $U(1)_A$ condition is given by
\begin{equation}\label{anomalyfree}
\sum_{a=1}^{d+s}Q_a^l=(n-k)\deg_l\left( \det A(\phi) \right),\quad \forall 
l=1,\cdots,s.
\end{equation}
and the central charge of PAX model is given by
\begin{equation}\label{centralch}
\hat{c}:=\frac{c}{3}=d-(n-k)^2.
\end{equation}
In \cite{2012determinantal} these models are shown to correspond to geometric 
GLSMs for $X_{A}$ and the $c_{1}(X_{A})$ condition to be equivalent to 
\ref{anomalyfree}. In the following subsection we will review a subfamily of 
these models termed linear PAX models.

\subsection{Linear PAX model for Calabi-Yau 3-folds}

\begin{figure}[h]
\centering
\includegraphics[scale=0.3]{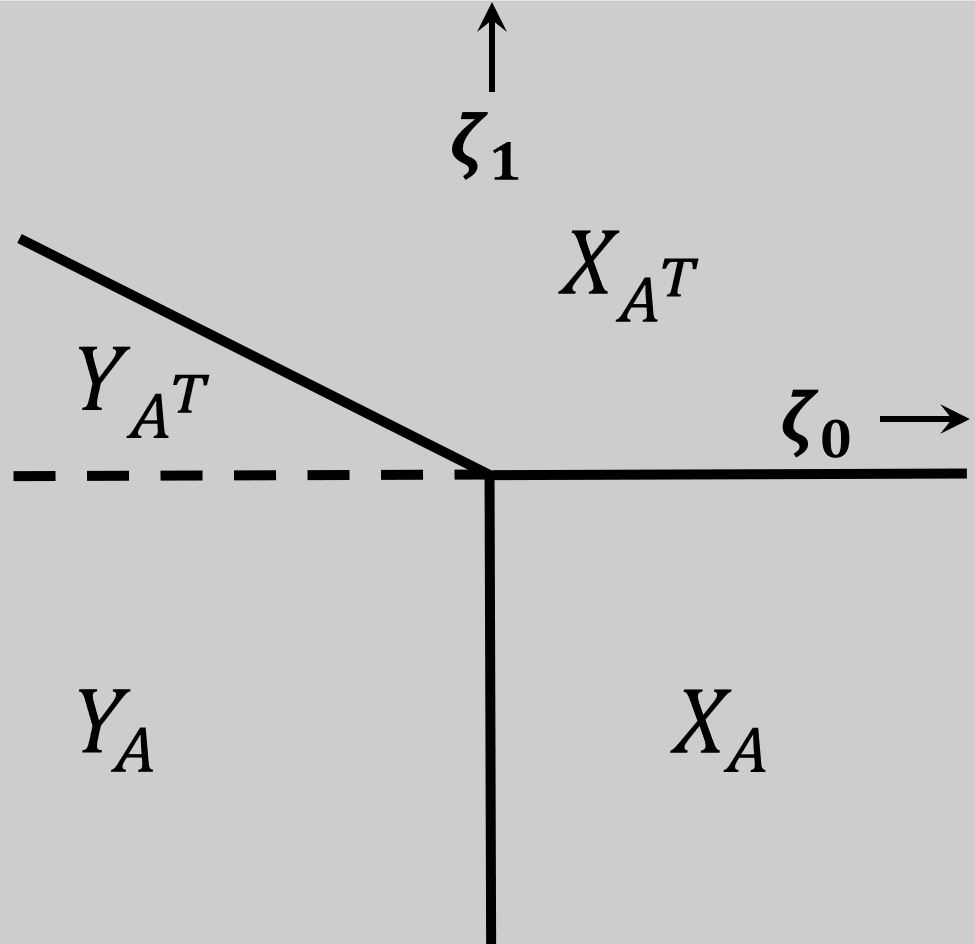}
\caption{Classical moduli space for linear PAX model. The phase boundary between 
$Y_A$ and $Y_{A^T}$ only exists for $\lfloor\sqrt{nk-1}\rfloor\geq k+1$ and 
$n-k\geq2$, thus is absent in our explicit examples.}\label{fig:phasesPAX}
\end{figure}

A linear PAX model is defined by the condition that $A(\Phi)$ is 
linear in $\Phi$. Therefore we can write
\begin{equation}
A(\Phi)=\sum_{a=1}^{d+s}A^a\Phi_a,\qquad A^{a}\in\mathrm{Mat}_{n\times 
n}(\mathbb{C}).
\end{equation}
In the following we will set
\begin{equation}
s=1\qquad Q_{a}=1,
\end{equation}
then this model corresponds to a two parameter GLSM with a geometric 
phase corresponding to $X_{A}\subset \mathbb{P}^{d}\times G(n-k,n)$. The 
condition 
\ref{anomalyfree} can then be fulfilled with the following choice of matter 
representations:
\begin{equation}
\begin{array}{c|ccc} 
  & \Phi_a & P_i & X_i \\
  \hline
U(1) & 1 & -1 & 0\\
U(n-k) & 0 & \mathbf{n-k} & \overline{\mathbf{n-k}}
\end{array}
\end{equation}
Recall the FI-theta parameters can be written as
$t=\zeta-i\theta$, where \cite{hori2019notes}
  \begin{eqnarray}
  t\in \left(\frac{\mathfrak{t}^{\vee}_{\mathbb{C}}}{2\pi i 
\mathrm{P}}\right)^{W_{G}}\cong\frac{\mathfrak{z}^{\vee}_{\mathbb{C}}}{2\pi i 
\mathrm{P}^{W_{G}}},
  \end{eqnarray}
  here $\mathrm{P}$ denotes the weight lattice, $W_{G}$ the Weyl subgroup of 
$G$, 
$\mathfrak{t}$ the Cartan sub-algebra of $\mathfrak{g}=\mathrm{Lie}(G)$ and 
$\mathfrak{z}=\mathrm{Lie}(Z(G))$. In the present case we can choose a basis 
where the FI-theta parameters for $U(1)$ and $U(n-k)$ can be denoted $t_0$ and 
$t_1$ respectively, $t_\alpha=\zeta_\alpha-i\theta_\alpha$, $\alpha=0,1$. The 
phases of these 
models are all weakly coupled and were determined in \cite{2012determinantal} 
as we proceed to review. All the phases correspond to NLSMs whose target space is given by the classical geometry of the Higgs branch, that is
\begin{equation}
X_{\zeta}:=\mu^{-1}(0)/G\cap \{dW^{-1}(0)\}
\end{equation}
where $\mu:V\rightarrow \mathfrak{g}^{\vee}$ denotes the moment map on the 
vector space $V$ (whose coordinates are $(\phi,x,p)$). Then, the equations 
$\mu^{-1}(0)=0$ corresponds to the D-term 
equations, for each factor of $G$:
\begin{equation}
\begin{gathered}
U(1):\quad 
\sum_{a=1}^{d+1}|\phi_a|^2-\sum_{i=1}^{n}\mathrm{Tr}(p_{i}p^{\dagger,i}
)-\zeta_0=0
\\
U(n-k):\qquad \sum_{i=1}^{n}\left(p_{i}p^{\dagger,i}-x^{\dagger,i} 
x_{i}\right)-\zeta_1\mathbf{1}_{n-k}=0
\end{gathered}\label{Dterm}
\end{equation}
where $\mathbf{1}_{n-k}$ denotes the identity in $\mathrm{Mat}_{n-k\times 
n-k}(\mathbb{C})$. The F-term 
equations $dW^{-1}(0)$ from the superpotental (\ref{PAXpot}) become:
\begin{equation}
\sum_{j=1}^{n}(A^{a,ij}\phi_a)x_{j}=0,\quad 
\sum_{i=1}^{n}p_{i}(A^{a,ij}\phi_a)=0,\quad 
\sum_{i,j=1}^{n}\Tr(p_{i}A^{a,ij}x_{j})=0.
\end{equation}
The phase space of this result, generically, in three 
phases\cite{2012determinantal}, see fig. \ref{fig:phasesPAX}, each corresponding 
to NLSMs whose target spaces are the smooth determinantal varieties $X_{A}$, 
$X_{A^{T}}$ and $Y_{A}$ respectively. Explicitly, they are given by 
($Y_{A}\cong Y_{A^{T}}$):
\begin{eqnarray}
\zeta_0>0,\zeta_1<0 &:& X_A =\lbrace (\phi,x)\in\CP^d\times Gr(n-k,n)\ \vert\ 
A(\phi)^{ij}x_{j}=0 \rbrace,\nonumber\\
\zeta_0+(n-k)\zeta_1>0,\zeta_1>0 &:& X_{A^T}= \lbrace (\phi,p)\in\CP^d\times 
Gr(n-k,n)\ \vert\ A(\phi)^{ij}p_{i}=0 \rbrace,
 \nonumber\\
\zeta_0+(n-k)\zeta_1<0, \zeta_0<0 &:& Y_{A}=\lbrace (p,x)\in 
\CP(S^{\oplus n})\rightarrow Gr(n-k,n)\ \vert\ \Tr (p_{i}A^{a,ij}x_{j})=0 
\rbrace.\nonumber
\end{eqnarray}

The classical phase boundaries get quantum corrected because of the existence 
of mixed Coulomb-Higgs mixed branches where the theory breaks down. The locus 
where 
these branches occur is a hypersurface (which can have multiple components) in 
the FI-theta space \cite{Witten1993}, usually called simply the discriminant 
of the GLSM. When projected to the $\zeta$-plane (i.e. the FI space), it 
asymptotically coincides with the classical phase boundaries (up to possibly a constant shift). These discriminant where computed for abelian models in \cite{Witten1993,morrison1995summing} and for nonabelian models in \cite{hori_tong2007}. The nonabelian case is much less 
straightforward but there is strong evidence \cite{Jockers:2012dk} supporting
the discriminant of PAX models, computed in \cite{2012determinantal}. Set
$q_\alpha:=\exp(-t_\alpha)$ then the 
discriminant is described by the (reducible) parameterized 
curve\cite{2012determinantal}
\begin{equation}
(q_0(\xi),q_1(\xi))=\left(
(-1)^{n(n-k)}\prod_{\alpha=1}^{n-k}\frac{(1+\omega^{n_\alpha}\xi)^n}{\xi^n},
\xi^n \right),
\end{equation}
where $\omega:=e^{2\pi i/n}$, $n_\alpha\neq n_\beta$ for 
$\alpha\neq\beta=1,\cdots,n-k$.

Imposing $X_A$ to be a CY threefold restricts $(k,n)$ to be
\begin{equation}
(k,n)=(4,5),\ (2,4),\ (1,5)
\end{equation}
embedded in $\CP^4\times \CP^4$, $\CP^7\times G(2,4)$ and $\CP^{19}\times\CP^4$ 
respectively. The last one is isomorphic to the first one by 
Seiberg-like duality \cite{Hori:2011pd}. Thus $(4,5)$ and $(2,4)$ will serve as 
the main examples for 
this paper, which we proceed to analyze their GLSMs in detail in the following 
subsection.

\subsubsection{Determinantal quintic in $\CP^4$}
This model corresponds to $(k,n)=(4,5)$ and is an abelian GLSM, whose matter content and vector R-charges $U(1)_R$ are summarized in the following table: 
\begin{equation}
\begin{array}{c|ccc} 
  & \Phi_a & P_i & X_i \\
  \hline
U(1)_0 & 1 & -1 & 0\\
U(1)_1 & 0 & 1 & -1\\
U(1)_R & 2(1-\epsilon-\delta) & 2\epsilon & 2\delta
\end{array}
\end{equation}
for $a,i,j=1,\cdots,5$ and $2\epsilon,2\delta\in[0,2)$. Its 
classical Higgs branch is given by
\begin{equation}
\begin{gathered}
X_A=\quad \lbrace (\phi,x)\in \CP^4\times\CP^4\ \vert\ A(\phi)_i\cdot x=0 
\rbrace\\
X_{A^T}=\quad \lbrace (\phi,p)\in \CP^4\times\CP^4\ \vert\ p\cdot A(\phi)^j=0 
\rbrace \\
Y_A=\quad \lbrace (p,x)\in \CP(\cO(-1)^{\oplus 5})\rightarrow\CP^4\ \vert\ 
p\cdot A^a\cdot x=0 \rbrace,
\end{gathered}
\end{equation}
the variety $X_{A}$ corresponds to the resolved determinantal quintic studied in \cite{abpax1,abpax2}. Passing through each phase boundary corresponds to a flop transition between geometries of both sides.

The determinantal variety $Z(A,4)\subset\CP^4$ generically has isolated singular nodal points. This can be verified by the Thom-Porteus formula. In this case $\cE=\cO^{\oplus5}$, $\cF=\cO(1)^{\oplus5}$, then the cohomology class of the singular loci is given by 
\begin{equation}
[Z(A,3)]=\left\vert\begin{array}{cc} c_2(\cF)&c_3(\cF)\\c_1(\cF) & c_2(\cF) \end{array}\right\vert=100H^4-50H^4=50H^4,
\end{equation}
where $H$ denotes the hyperplane class of $\CP^4$. Therefore, $Z(A,3)$ is 
generically composed of $50$ singular nodal points. Using 
the intersection formulae (\ref{intfor}), one can determine the topological 
data 
of $X_A$ to be
\begin{equation}
\begin{gathered}
\chi(X_A)=-100 \quad c_2(X_A)\cdot H=50 \quad 
c_2(X_A)\cdot\varsigma_1=50  \\
H^3=5 \quad H^2\cdot\varsigma_1=10 \quad H\cdot\varsigma_1^2=10 \quad 
\varsigma_1^3=5
\end{gathered}\label{topdata1}
\end{equation}
where $\varsigma_1$ denotes the hyperplane class of the 
Grassmannian $G(1,5)\cong \CP^4$. Then we can compute the Hodge numbers of 
$X_A$ to be $(h^{1,1},h^{2,1})=(2,52)$. In particular, positive K\"ahler cone 
generators $H$ and $\varsigma_1$ are asymptotically identified with 
$\zeta_0+\zeta_1$ and $\zeta_1$ in $X_{A^T}$, and $\zeta_0$ and $-\zeta_1$ in 
$X_A$.

\begin{figure}[h]
\centering
\includegraphics[scale=0.4]{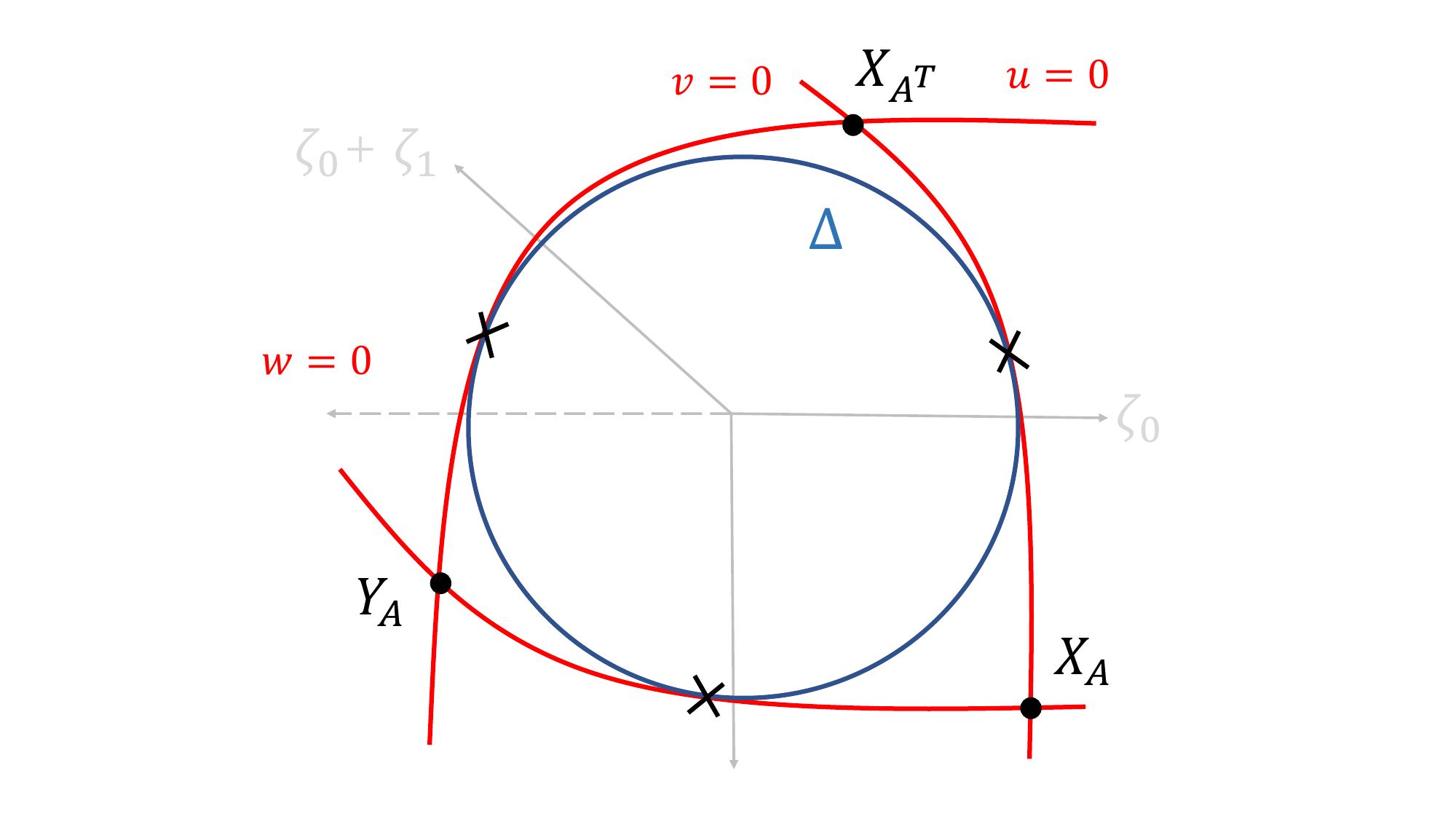}
\caption{Moduli space for determinantal quintic}
\label{moduliabpax}
\end{figure}

The discriminant in FI parameter space is given by 
\begin{equation}
\Delta:\quad (q_0,q_1)=\left(-\frac{(1+\xi)^5}{\xi^5},\xi^5\right),
\end{equation}
which exactly match the result from mirror symmetry \cite{abpax3}:
\begin{equation}
\Delta:\quad (u+v+w)^5-5^4uvw(u+v+w)^2+5^5uvw(uv+vw+wu)=0\subset\CP^2
\end{equation}
where we identify $[u,v,w]=[q_1,q_0q_1,1]\in\CP^2$ in the $X_{A^{T}}$ phase 
and $[u,v,w]=[1,q_0,q_{1}^{-1}]$ in the $X_{A}$ phase. This identification also 
matches the FI 
parameters for generators on each phase. Furthermore, three large volume points corresponding to classical phases are given by
\begin{equation}
u=v=0,\quad v=w=0,\quad w=u=0.
\end{equation}
The full moduli space is illustrated in fig. \ref{moduliabpax}. $\Delta$ intersects with the divisors $\{w=0\}$, $\{u=0\}$ and $\{v=0\}$ at the points $[1:-1:0]$, $[0:1:-1]$ and $[1:0:-1]$, respectively. All intersections are tangential and of degree 5.

\subsubsection{GN Calabi-Yau in $\CP^7$}
This model is a two parameter non-abelian GLSM whose gauge group is 
$G=U(1)\times U(2)$ and with matter content
\begin{equation}
\begin{array}{c|ccc} 
  & \Phi_a & P_i & X_{i} \\
  \hline
U(1) & 1 & -1 & 0\\
U(2) & 0 & \mathbf{2} & \overline{\mathbf{2}}\\
U(1)_R & 2(1-\epsilon-\delta) & 2\epsilon & 2\delta
\end{array}
\end{equation}
for $a=1,\cdots,8$, $i=1,\cdots,4$. The Higgs branch geometries are given by
\begin{equation}
\begin{gathered}
X_A:\quad \lbrace (\phi,x)\in \CP^7\times Gr(2,4)\ \vert\ A(\phi)_i\cdot x^\alpha=0 \rbrace\\
X_{A^T}:\quad \lbrace (\phi,p)\in \CP^7\times Gr(2,4)\ \vert\ p_\alpha\cdot A(\phi)^j=0 \rbrace  \\
Y_A:\quad \lbrace (p,x)\in \CP(S^{\oplus 4})\rightarrow Gr(2,4)\ \vert\ p\cdot A^a\cdot x=0 \rbrace
\end{gathered}
\end{equation}
The geometry of $X_A$ is an incidence correspondence for $Z(A,2)$ as
\begin{equation}
Z(A,2)=\lbrace \phi\in\CP^7\ \vert\ \rank(A(\phi))\leq2 \rbrace,
\end{equation}
with topological data 
\begin{equation}
\begin{gathered}
\chi(X_A)=-64 \quad c_2(X_A)\cdot H=56 \quad c_2(X_A)\cdot \varsigma_1=56 \\
H^3=20 \quad H^2\cdot\varsigma_1=20 \quad H\cdot\varsigma_1^2=16 \quad 
\varsigma_1^3=8
\end{gathered}\label{topdata2}
\end{equation}
where hyperplane classes are $H$ of $\CP^7$ and Schubert class 
$\varsigma_1=c_1(Q)$ of $Gr(2,4)$, asymptotically correspond to 
$\zeta_0+2\zeta_1$ and $\zeta_1$ in $X_{A^T}$, and $\zeta_0$ and $-\zeta_1$ in 
$X_A$. The resulting number $(h^{1,1},h^{2,1})=(2,34)$ matches the result in 
\cite{nabpax1}.  Note that Gulliksen and Neg{\aa}rd first implicitly described 
this 
particular CY 3-fold in terms of a resolution of the corresponding ideal 
sheaf \cite{GN1971}. The variety was analyzed as a determinantal variety in 
\cite{nabpax1},\cite{nabpax2}. 

Similar analysis shows that $X_A$ is the resolution of desingularized variety $Z(\cA,7)$, for
\begin{equation}
Z(\mathcal A,7)=\lbrace x\in Gr(2,4)\ \vert\ \rank\mathcal A\leq7 \rbrace
\end{equation}
in term of $8\times8$ matrix $\mathcal A=A^a_i\cdot x_\alpha$. It has 
generically 56 nodal points, and the K\"ahler parameter $2\varsigma_1-H$ in 
$X_A$ 
measures the volume of the 56 blown-up $\CP^1$s.

\begin{figure}[h]
\centering
\includegraphics[scale=0.4]{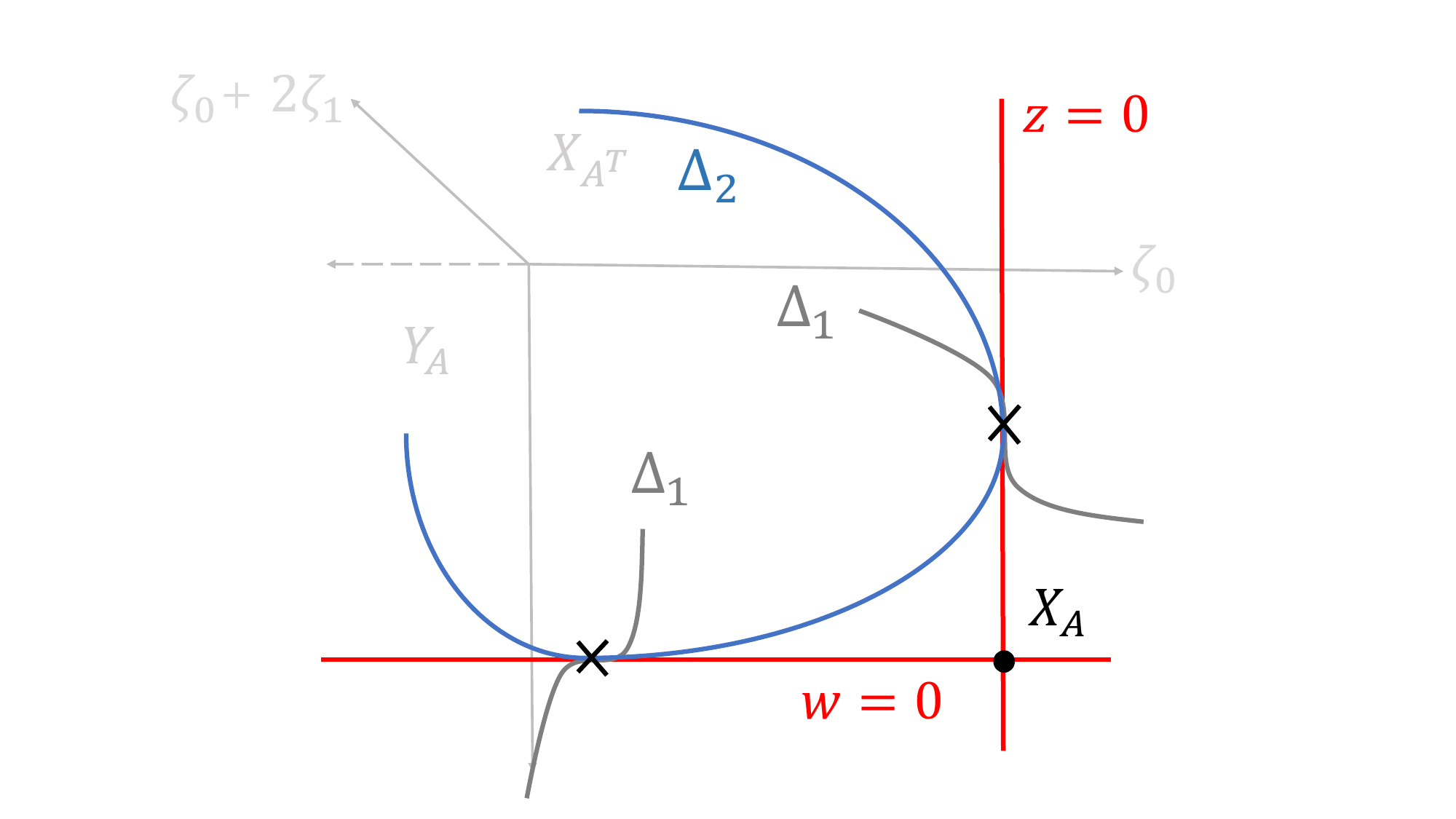}
\caption{Moduli space for GN Calabi-Yau around $X_A$. The $X_{A^T}$ side is identical.}
\label{modulinabpax}
\end{figure}

The discriminant from GLSM is given by curves\footnote{Note that equations 
(\ref{desc}) differ slightly from the discriminant found in 
\cite{2012determinantal,Jockers:2012dk}, specifically the difference is given by the change of sign $q_{1}\rightarrow -q_{1}$. This change of sign can be traced back to the correction to the twisted potential for the $\sigma$ fields by W-bosons (proportional to $\pi\sum_{\alpha>0}\alpha(\sigma)$). This 
correction was overlooked in \cite{hori_tong2007,2012determinantal} and also in the computation of the $S^{2}$ partition function 
\cite{Benini:2012ui,Doroud:2012xw}. The correct factor due to W-bosons was found in the computation of the $D^{2}$ partition function\cite{hori2013exact,Honda:2013uca,Sugishita:2013jca}.} 
\begin{equation}
\begin{gathered}
\Delta_1:\quad (1+w)^4-2z(1-6w+w^2)+z^2=0 
\\
\Delta_2:\quad -(1+w)^8+4z(1+34w+w^2)(1+w)^4-2z^2(3-372w+1298w^2-372w^3+3w^4)
\\
+4z^3(1+34w+w^2)-z^4=0
\end{gathered}\label{desc}
\end{equation}
in where $(w,z)$ should be identified to $(q_1,q_0q_1^2)$ for $X_{A^T}$ and to $(q_1^{-1},q_0)$ for $X_A$ according to K\"ahler parameters. Then the moduli structure is symmetric under $q_1\mapsto q_1^{-1}$. However, the full structure of moduli is not known. The moduli structure for $X_A$ is illustrated in fig. \ref{modulinabpax}. Note that $\Delta_1,\Delta_2$ intersects divisors at $(z,w)=(0,-1)$ tangentially of degree $4,8$, and at $(1,0)$ of degree $2,4$ 
respectively.

The mirror family for this CY is not known, however, its GLSM predicts that the mirror family contains three large complex structure points characterized by maximal unipotent monodromy, and it has a discriminant locus consisting of the two rational curves in (\ref{desc}), and boundary divisors associated to the 
FI parameters tending to infinity.

\section{\label{sec:section3} B-branes and their central charges for determinantal varieties}

In this section we will review the construction of B-branes on GLSMs from 
\cite{herbst2008phases,hori2013exact} and their central charges as defined in 
\cite{hori2013exact}. Afterwards we will apply these results to the specific 
case of the determinantal quintic and the GN CY defined in section 
\ref{sec:section2}.

 \subsection{B-branes in GLSMs}
B-branes on GLSM are defined as boundary conditions
that preserve the B-type subset of the $\mathcal{N}=(2,2)$ supersymmetry. 
Namely if the left and right moving supercharges are denoted 
$\mathbf{Q}_{-},\overline{\mathbf{Q}}_{-}$ and 
$\mathbf{Q}_{+},\overline{\mathbf{Q}}_{+}$ respectively, the B-type 
supersymmetry is spanned by the supercharges 
$\mathbf{Q}_{B}:=\overline{\mathbf{Q}}_{+}+\overline{\mathbf{Q}}_{-}$ and its 
charge conjugate $\mathbf{Q}^{\dag}_{B}:=\mathbf{Q}_{+}+\mathbf{Q}_{-}$. In 
general a GLSM involves chiral and twisted chiral fields, denote them 
collectively as $\phi$, (taking values in $V\cong \mathbb{C}^{N}$) and 
$\sigma$ (taking values in $\mathfrak{t}_{\mathbb{C}}$), 
respectively\footnote{Strictly speaking the fields $\phi$ and $\sigma$ are the 
lowest component of the chiral multiplet and the vector multiplet 
respectively.}. One has to define 
boundary conditions for both types of fields, when dealing with B-branes \cite{herbst2008phases,hori2013exact}. The boundary conditions for the chirals $\phi$, transforming on a representation $\rho_{m}:G\rightarrow GL(V)$ and $R:U(1)_{R}\rightarrow GL(V)$ of the gauge group and (vector) R-charge\footnote{The weights of $R$ are allowed to be real, as $U(1)_{R}$ is not a gauge group.} 
respectively, will involve the superpotential 
$W\in(\mathrm{Sym}V^{\vee})^{G}$. These boundary conditions are algebraic. We will label them by $\mathcal{B}:=(M,\rho_{M},R_{M},\mathbf{T})$, where 
the elements in this 4-tuple are defined as:
\begin{itemize}
\item \textbf{Chan-Paton vector space}: a $\mathbb{Z}_{2}$-graded, finite dimensional free $\mathrm{Sym}(V^{\vee})$ module denoted by $M=M_{0}\oplus 
M_{1}$.
\item \textbf{Boundary gauge and (vector) R-charge representation}: $\rho_{M}: 
G\rightarrow GL(M)$, and $R_{M}:U(1)_{R}\rightarrow GL(M)$ are a pair 
of commuting and $\mathbb{Z}_{2}$-even 
representations, where the weights of $R_{M}$ are allowed to be real.
\item \textbf{Matrix factorization of $W$}: Also known as the tachyon profile, 
a 
$\mathbb{Z}_2$-odd endomorphism $\mathbf{T} \in 
\mathrm{End}^{1}_{\mathrm{Sym}(V^{\vee})}(M)$ satisfying 
$\mathbf{T}^{2}=W\cdot\mathrm{id}_{M}$.
\end{itemize}
The group actions $\rho_{M}$ and $R_{M}$ must be compatible with $\rho_{m}$ and 
$R$, i.e.
  for all $\lambda\in U(1)_{R}$ and $g\in G$ the identities
    \begin{equation}\label{rhodef}
   \begin{aligned}
      R_{M}(\lambda)\mathbf{T}(R(\lambda)\phi)R_{M}(\lambda)^{-1} & = \lambda 
\mathbf{T}(\phi) , \\
\rho_{M}(g)^{-1}\mathbf{T}(\rho_{m}(g)\cdot \phi)\rho_{M}(g) & = 
\mathbf{T}(\phi)
    \end{aligned}
  \end{equation}
must hold. Denote the weights of $\rho_{m}$ as 
$Q_{j}:\mathfrak{t}^{\vee}\rightarrow\mathbb{R}$ and the weights of $R$ as 
$R_{j}\in\mathbb{R}$ for $j=1\ldots,N=\mathrm{dim}_{\mathbb{C}}V$. Define
the collection of hyperplanes $\mathcal{H}\subset\mathfrak{t}_{\mathbb{C}}$ by
\begin{eqnarray}
  \mathcal{H}=\bigcup_{j=1}^{N}\bigcup_{n\in\mathbb{Z}_{\geq 
0}}\left\{\sigma\in 
\mathfrak{t}_{\mathbb{C}}\Big|Q_{j}(\sigma)-i\frac{R_{j}}{2}-in=0\right\}
\end{eqnarray}
The other piece of data that we need to fully define the B-brane are the 
boundary conditions for the twisted chirals. That is, the specification of a 
profile $L_{t}$ (denoted to emphasize, in general, it may depend on $t$) for 
the zero modes of $\sigma\in\mathfrak{t}_{\mathbb{C}}$. $L_{t}$ corresponds to a gauge-invariant middle-dimensional sub-variety 
$L_{t}\subset \mathfrak{g}_{\mathbb{C}}\setminus \mathcal{H}$ of the 
complexified Lie algebra of $G$ or equivalently its intersection 
$L_{t}\subset\mathfrak{t}_{\mathbb{C}}\setminus \mathcal{H}$ invariant under 
the action of the Weyl group $W_{G}\subset G$ \cite{hori2013exact}. 
In \cite{hori2013exact}, an 
\textbf{admissible contour} is defined as a profile
$L_{t}$ that is a continuous deformation of the real contour 
$L_{\mathbb{R}}:=\{\Im\sigma=0|\sigma\in 
\mathfrak{t}_{\mathbb{C}}\}$, such that 
the 
imaginary part of the boundary effective twisted superpotential 
$\widetilde{W}_{\text{eff},q}:\mathfrak{t}_{\mathbb{C}}\rightarrow \mathbb{C}$
\begin{equation}\label{twistedbdry}
  \widetilde{W}_{\text{eff},q}(\sigma):= \left(\sum_{\alpha>0}\pm 
i\pi\,\alpha\cdot\sigma\right)-\left(\sum_j 
(Q_j(\sigma))\left(\log\left(\frac{iQ_j(\sigma)}{\Lambda}
\right)-1\right)\right)-t(\sigma)+2\pi i q(\sigma)
\end{equation}
approaches $+\infty$ in all asymptotic directions of $L_{t}$ and for all the 
weights $q\in \mathfrak{t}^{\vee}$ of $\rho_{M}$.
Signs in the sum over positive roots $\alpha$ of $G$ depend on the Weyl chamber 
in which $\Re(\sigma)$ lies.

The GLSM B-branes, are then given by pairs $(\mathcal{B},L_{t})$. These are known to form a category which we denote as $MF_{G}(W)$ which was defined purely in terms of the data $\mathcal{B}$ in 
\cite{ballard2019variation}. However the category 
$MF_{G}(W)$ with objects given by pairs $(\mathcal{B},L_{t})$ has its origins 
on the dynamics of B-branes on GLSMs \cite{hori2013exact} and it can be reduced to the data $\mathcal{B}$ when we are in a specific phase (and so, the data $L_{t}$ can be dropped), in the IR SCFT. However, in our approach, $L_{t}$ is crucial for deriving the relation between the categories of B-branes at different phases as was used for the derivation of the grade restriction rule in \cite{herbst2008phases,hori2013exact} and we will review it in section \ref{sec:section4}.

\subsection{Hemisphere Partition Function and Central Charge of B-branes}

The central charge of B-branes (A-branes), for $\mathcal{N}=(2,2)$ theories is 
defined \cite{Cecotti:1991me,hori2000d} by the partition function the A-twisted 
(B-twisted) theory on a disk with an infinitely flat cylinder attached to it 
and boundary conditions corresponding to a B-brane (A-brane). This coupling of 
A/B-twist in the bulk of $\mathcal{N}=(2,2)$ theories and B/A-brane boundary 
conditions is a very 
natural object to study, for instance in 
SCFTs \cite{Ooguri:1996ck}. Using supersymmetric localization techniques 
\cite{hori2013exact,Honda:2013uca,Sugishita:2013jca} the 
central charge for a B-brane $(\mathcal{B},L_{t})\in MF_{G}(W)$ was computed in 
the context of GLSMs and found to be given by
\begin{equation}
\label{ZD2}
Z_\mathcal{B}(t):=\int_{L_{t}\subset 
\mathfrak{t}_{\mC}} d^{l_{G}}\sigma 
\prod_{\alpha>0}\alpha(\sigma)\sinh(\pi\alpha(\sigma))\prod_{j=1}
^{N}\Gamma\left(iQ_
{j}(\sigma)+\frac{R_{j}}{2}\right)e^{it(\sigma)}f_{\mathcal{B}}(\sigma).
\end{equation}
where
\begin{equation}
f_{\mathcal{B}}(\sigma):=\mathrm{tr}_{M}\left(R_{M}(e^{i
\pi})\rho_{M}(e^{2\pi\sigma
} )\right)
\end{equation}
the symbol $\prod_{\alpha>0}$ denotes the product over the positive roots of 
$G$  and $l_{G}:=\mathrm{dim}(\mathfrak{t})$.
The function (\ref{ZD2}) is conjectured to coincide with the IR central charge 
as defined by \cite{Cecotti:1991me}, for SCFTs i.e. non-anomalous GLSMs and to be related in a certain limit in the anomalous case (see 
\cite{hori2013exact,hori2019notes}). Is important to remark that in (\ref{ZD2}) 
the integration variable is dimensionless since it corresponds to $r\sigma$ 
(where $r$ is the radius of the hemisphere) but we denoted it $\sigma$ in order 
not to introduce more notation. In addition, (\ref{ZD2}) has a normalization of 
the form $C(r\Lambda)^{\frac{\hat{c}}{2}}$ where $\Lambda$ corresponds to the 
UV cut-off and $C$ is just a numerical constant. We will simply set this factor 
to a convenient numerical constant (specified in each example) in the 
following, but in general it should be considered for 
applications involving anomalous models \cite{hori2019notes}, for instance. 
We will be concerned in this work only with geometric phases. The central 
charge for B-branes on NLSMs with a CY target space $X$ takes the form 
\cite{Cheung:1997az,Green:1996dd,minasian1997k}
\begin{equation}
\begin{gathered}
Z_{\mathcal{B}}(t)=\int_X 
e^{J}\hat\Gamma_X\mathrm{ch}(\cB)+\text{instantons}=:Z^0_{\cB}(t)+\text{
instantons } ,
\end{gathered}\label{LVZ}
\end{equation}
where $J:=B+i\frac{\omega}{2\pi}\in H^2(X,\mathbb{C})$, $B\in 
H^{2}(X,\mathbb{Z})$ is the $B$-field and $\omega\in 
\mathcal{K}_{X}\subset H^{2}(X,\mathbb{R})$. $\mathcal{K}_{X}$ denotes the K\"ahler cone of $X$ and the instantons are weighted by
\begin{equation}\label{instantonexp}
\exp\left(2\pi i\int_{\beta}J\right)\qquad \beta\in H_{2}(X,\mathbb{Z})
\end{equation}
where $\beta$ is an effective curve class.
The Gamma class 
$\hat\Gamma_X$ is a multiplicative characteristic class, given by 
\begin{equation}
\hat\Gamma_X:=\prod_{j}\Gamma\left(1-\frac{\lambda_{j}}{2\pi i}\right)
\end{equation}
where $\lambda_{j}$ are the Chern roots of the holomorphic tangent bundle $TX$ 
of 
$X$. It satisfies the important property
\begin{equation}\label{gammaroot}
\hat\Gamma_X\hat{\Gamma}^{*}_{X}=\hat{A}_{X}
\end{equation}
with $\hat{A}_{X}$ the $\hat{A}$-genus of $X$. For $X$ CY, the $\hat{A}$-genus 
equals the Todd class $\mathrm{Td}_{X}$. Because of (\ref{gammaroot}), we can 
regard $\hat\Gamma_X$ as a root of $\hat{A}_{X}$ moreover, this root is not 
unique and $\hat\Gamma_X$ is just a particular choice \cite{Halverson:2013qca}, 
however this choice is different from the one corresponding to the 
Ramond-Ramond (RR) charge computed in 
\cite{Cheung:1997az,Green:1996dd,minasian1997k}. The choice $\hat\Gamma_X$ 
encodes the perturbative corrections to the central charge.
Fix $X$ to be a CY 3-fold, then $\hat\Gamma_{X}$ is explicitly given 
by\footnote{The Gamma class appears implicitly in the works 
\cite{libgober1999chern,Hosono:2000eb} and then it was further defined in a 
mathematical context in \cite{iritani2009integral,katzarkov2008hodge}.}
\begin{equation}\label{gammaCY3}
\hat\Gamma_X=1+\frac{1}{24}c_2(X)+\frac{\zeta(3)}{(2\pi i)^3}c_3(X).
\end{equation}
Fix a basis $\{J_{\alpha}\}$ of $H^2(X,\C)$ and so we can write 
$J=\kappa_\alpha J_\alpha$, $\kappa_\alpha\in\mathbb{C}$, then we can write 
explicitly the central charge at the zero-instanton sector for chosen basis of 
generators of $D^{b}Coh(X)$ given by sheaves, namely
\begin{equation}
\begin{gathered}
\begin{aligned}
Z^0_{\cO_X}(t)=\frac{1}{3!}c_{\alpha\beta\gamma}\kappa_\alpha \kappa_\beta 
\kappa_\gamma +c_\alpha \kappa_\alpha+\frac{\zeta(3)}{(2\pi i)^3}\chi(X),
\end{aligned}
\\
c_{\alpha\beta\gamma}:=\int_X J_\alpha J_\beta J_\gamma,\quad 
c_\alpha:=\frac{1}{24}\int_X c_2(X)J_\alpha,\label{period}
\end{gathered}
\end{equation}
Similarly, for each generator $J_\alpha$ we have a D4-brane $\cO_{D_\alpha}$ 
corresponding to the structure sheaf of the divisor $D_{\alpha}$, dual to 
$J_{\alpha}$, 
having brane charge
\begin{equation}
Z^0_{\cO_{D_\alpha}}(t)=\frac{1}{2}c_{\alpha\beta\gamma}\kappa_\beta 
\kappa_\gamma-\frac{1}{2}c_{\alpha\alpha\beta}\kappa_\beta+\frac{1}{6}c_{
\alpha\alpha\alpha}+c_\alpha.\label{period2}
\end{equation}
The D2-branes $\cO_{C^\alpha}$ supported on the curve classes 
$C^{\alpha}$ have\footnote{In the following it would be convenient to consider 
a twist $\mathcal{E}_{\alpha}$ of $\cO_{C^\alpha}$ so that 
$Z^0(\mathcal{E}_{\alpha})=\kappa_\alpha$.}
\begin{equation}
Z^0_{\cO_{C^\alpha}}(t)=\kappa_\alpha+1
\end{equation}
while the D0-brane $\cO_p$ supported at a point $p\in X$, i.e. the skyscraper 
sheaf at $p$, is normalized to
\begin{equation}
Z^0_{\cO_p}(t)=1.
\end{equation}
The lattice of B-brane charges is isomorphic to 
 the Grothendieck group $K_{0}(X)$ of the CY $X$ 
which is spanned by homological classes of holomorphic vector bundles over 
$X$\cite{witten1999d,Hosono:2000eb}. 
Indeed the central charges $Z:D(X)\rightarrow\mathbb{C}$ factor through 
$K_{0}(X)$, hence we can always uniquely write the central charge of a B-brane 
$\cB$ as:
\begin{equation}\label{chargeofBB}
Z_\cB=a^0Z_{\cO_X}+a^\alpha Z_{\cO_{D_\alpha}}+a_\alpha 
Z_{\cE_\alpha}+a_0 Z_{\cO_P}=:\left(a^0,a^\alpha,a_\alpha,a_0\right),
\end{equation}
where $a^0,a^\alpha,a_\alpha,a_0\in\mathbb{Z}$.

\subsection{B-branes and their central charges in linear PAX models}

In this subsection we will perform the explicit evaluation of the (quantum 
corrected) central charge for the B-branes generating the derived category 
$D^{b}Coh(X_{A})$ for our choice of examples. Since all the phases are 
geometric (and weakly coupled) in the linear PAX models, the computations 
presented in this sections can be carried on in analogous way in other 
phases. In addition, the linear PAX models we are considering are CY 3-folds 
and have $h^{1,1}=2$, hence a canonical choice of generators for   
$D^{b}Coh(X_{A})$ will be chosen to be  
$\{\cO_X,\cO_{D_\alpha},\cE_{\alpha}\}$, $\alpha=0,1$ where $\cE_{\alpha}$ 
is a twist of $\cO_{C^{\alpha}}$. $\cO_{C^{\alpha}}$ and
$\cO_{D_{\alpha}}$ are the structure sheaves of curves and divisors on $X_{A}$, 
respectively. 

The convergence condition can be solved by requiring $\Im\sigma_0>0$ and 
$\Im\sigma_{1,\mathsf{k}}<0$ in the $X_{A}$ phase. Then, an admissible contour 
$\gamma$ can be taken as
\begin{equation}
\gamma:\quad 
\left(\sigma_0,
\sigma_{1}\right)=\left(\Re\sigma_0+i\left(\Re\sigma_0\right)^2,\Re\sigma_{
1,\mathsf{k}}-i\left(\Re\sigma_{
1,\mathsf{k}}\right)^2\right), \quad \mathsf{k}=1,\ldots,k.\label{contour}
\end{equation}
Then, in the limit of the R-charge of $\phi$ and $x$ to be zero, and 
that of $p$ to be $2$ (the exact R-charges at the IR fixed point in the $X_{A}$ 
phase) the integral (\ref{ZD2}) can be be evaluated by 
multidimensional residues in a neighbourhood of the poles.

\subsubsection{Determinantal quintic in $\CP^4$}
\begin{figure}[h]
\centering
\includegraphics[scale=0.35]{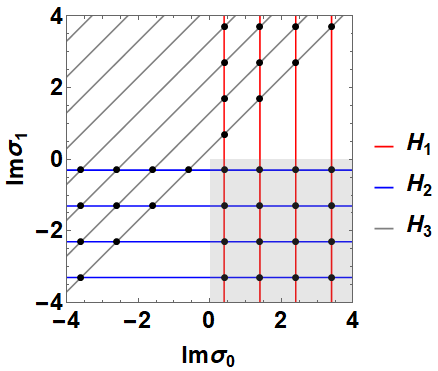}
\caption{Gamma poles of the determinantal quintic. The admissible contour 
envelops poles in the lower-right quadrant.}
\label{GDivisor1}
\end{figure}

The hemisphere partition function (\ref{ZD2}) of a B-brane $\cB$ is given 
by\footnote{In general, we will choose a basis for the generator of 
$\mathfrak{g}$ so that $t(\sigma)=t_{0}\sigma_{0}+t_{1}(\sigma_{1})$ and 
$t_{1}(\sigma_{1})=\sum_{\mathsf{k}=1}^{k}t_{1}^{\mathsf{k}}\sigma_{1,\mathsf{k}
} $. }
\begin{equation}\label{ZforDet}
Z_{\cB}(t)=\frac{1}{(2\pi i)^{8}}\int_\gamma d^{2}\sigma e^{it(\sigma)} 
\Gamma(i\sigma_0+1-\epsilon-\delta)^5\Gamma(i(\sigma_1-\sigma_0)+\epsilon)^5
\Gamma(-i\sigma_1+\delta)^5 f_\cB(\sigma),
\end{equation}
where we have taken a normalization factor $(2\pi i)^{-8}$. The contour 
$\gamma$ is chosen so that is admissible in the $X_{A}$ phase. $Z_{\cB}(t)$ 
is convergent, for any $\cB$,provided $\Im\sigma_0>0$ and $\Im\sigma_1<0$, then, 
is straightforward to show that the following integration contour satisfy all 
the admissibility conditions:
\begin{equation}
\gamma:\quad 
\left(\sigma_0,
\sigma_1\right)=\left(\Re\sigma_0+i\left(\Re\sigma_0\right)^2,
\Re\sigma_1-i\left(\Re\sigma_1\right)^2\right).\label{contourP4}
\end{equation}
The poles of Gamma functions lies in the families of hyperplanes
\begin{equation}
\cH_1=\left\lbrace \sigma_0=i(k+1-\epsilon-\delta) 
\right\rbrace_{k=0}^{\infty},\ \cH_2=\left\lbrace \sigma_1=-i(k+\delta) 
\right\rbrace_{k=0}^{\infty},\ \cH_3=\left\lbrace 
\sigma_1-\sigma_0=i(k+\epsilon) \right\rbrace_{k=0}^{\infty}
\end{equation}
where $k\in\Z_{\geq0}$. The projection of the hyperplanes to the 
$\mathrm{Im}\sigma$ plane is illustrated in fig. \ref{GDivisor1}. Given 
$\gamma$, using the prescription of 
\cite{passare1994multidimensional,zhdanov1998computation}, the integral on 
$Z_{D^{2}}(\cB)$ can be reduced to an infinite sum of residues. When the FI 
parameters $\zeta$ are chosen on the chamber of the phase $X_A$, this 
sum is over the the residues of the poles located at $\cH_1\cap \cH_2$.

Then the partition function (\ref{ZforDet}) has a
series expression as a sum over residues of admissible poles:
\begin{equation}
\begin{aligned}
Z_\cB(t)=&\frac{1}{(2\pi i)^8}\sum_{n,m=0}^\infty \Res_{(z_0,z_1)=(n,m)}\left(\frac{\pi}{\sin\pi z_0}\right)^5\left(\frac{\pi}{\sin\pi z_1}\right)^5
\\
\times&\frac{\Gamma(z_0+z_1+1)^5}{\Gamma(z_0+1)^5\Gamma(z_1+1)^5}e^{
-t_0z_0+t_1z_1}f_\cB(iz_0,-iz_1).
\end{aligned}\label{ZXA1}
\end{equation}
where we have taken the limit $(\delta,\epsilon)\rightarrow (0,1)$ to set the 
R-charges to their exact value in the $X_{A}$ phase.
Since $\gamma$ is admissible for any element $\cB\in MF_{G}(W)$, in the $X_{A}$ 
phase, in the following we will present a particular choice of representatives 
$\cB$ that flows to our desired set of generators in $D^{b}Coh(X_{A})$. 
The tachyon profile
\begin{equation}
Q_{\cO_X(p,q)}:= p_i\bar{\eta}^i+ A(\phi)^{ij}x_j\eta_i,
\end{equation}
which is conveniently expressed in terms of free fermions $\eta_{i}$, 
$\bar{\eta}^{i}$ satisfying the Cifford algebra
\begin{equation}
\{\eta_{i},\bar{\eta}^{j}\}=\delta^{j}_{i},\quad\{\bar{\eta}^{i},\bar{\eta}^{j}
\} =\{ \eta_{i},\eta^{j}\}=0\qquad\text{ \ for all \ }i,j=1,\ldots, 5
\end{equation}
the representations $\rho_{M}$ and $R_{M}$ are uniquely determined by the 
choice of Clifford vacuum $|p,q\rangle$, on an irreducible representation 
of the gauge group $U(1)\times U(1)$, satisfying
\begin{equation}
\bar{\eta}^{i}|p,q\rangle=0\text{ \ \ for all \ }i,\qquad 
\rho_{M}(\lambda_{1},\lambda_{2})|p,q\rangle=\lambda_{1}^{p}\lambda_{2}^{q}|p,
q\rangle\qquad 
R_{M}(\lambda)|p,q\rangle=|p,
q\rangle
\end{equation}
for all $(\lambda_{1},\lambda_{2})\in U(1)\times U(1)$ and $\lambda\in 
U(1)_{R}$. Then its brane factor $f_{\cO_X(p,q)}(\sigma)$ can be 
straightforwardly 
computed:
\begin{equation}
f_{\cO_X(p,q)}(\sigma)=\left( 1-e^{2\pi(\sigma_1-\sigma_0)} 
\right)^5e^{2\pi(p\sigma_0+q\sigma_1)}.
\end{equation}
These B-branes RG-flow to the objects $\cO_{X_{A}}(p,q)\in D^{b}Coh(X_{A})$. It 
is possible to choose some subset of the $(p,q)$ pairs such that 
$\cO_{X_{A}}(p,q)$ generate $D^{b}Coh(X_{A})$ but we are not going to use this 
basis here. The zero instanton sector of the B-brane $Q_{\cO_X(p,q)}$ is given 
by
\begin{equation}
\begin{aligned}
&Z^0_{\cO_X(p,q)}
\\
=&-\frac{100\zeta(3)}{(2\pi i)^3}+\frac{5}{12}(p-q)(5+2(p^2-5pq+q^2))
\\
 &+\left(50+60(p^2-4pq+2q^2)\right)\frac{\kappa_0}{24}
\\
 &+\left(50+60(2p^2-4pq+q^2)\right)\frac{\kappa_1}{24}
\\
 &+\frac{5}{2}\left( (p-2q) \kappa_0^2+4(p-q)\kappa_0\kappa_1+(2p-q)\kappa_1^2 \right)
\\
 &+\frac{5}{6}\kappa_0^3+5\kappa_0^2\kappa_1+5\kappa_0\kappa_1^2+\frac{5}{6}\kappa_1^3,
\end{aligned}\label{ZOXquin}
\end{equation}
where K\"ahler parameters are identified with
\begin{equation}\label{identdet}
\kappa_0=-\frac{t_{0}}{2\pi i}-5/2,\qquad 
\kappa_1=\frac{t_{1}}{2\pi i}-5/2.
\end{equation}
We remark that the identification (\ref{identdet}) is only valid 
asymptotically i.e. it must be understood up to instanton corrections, in 
general $\kappa_0=-\frac{t_{0}}{2\pi i}-5/2+\mathcal{O}(\exp(-t_{0}))$ and 
likewise for $\kappa_{1}$. When $p=q=0$, (\ref{ZOXquin}) exactly matches 
equation (\ref{period}) for the topological data of the determinantal quintic.

We expect two divisor classes on $X_{A}$. We can choose these divisors for 
example to be given by $D_{\phi}:=\{\phi_{1}=0\}$ and $D_{x}:=\{x_{1}=0\}$. The 
B-branes RG-flowing to the structure sheaves $\mathcal{O}_{D_{\phi}}$ and  
$\mathcal{O}_{D_{x}}$ are given by the tachyon profiles
\begin{equation}
Q_{D_{\phi}}=Q_{\cO_{X}}+\phi_{1}\chi\qquad 
Q_{D_{x}}=Q_{\cO_{X}}+x_{1}\chi
\end{equation}
respectively, where $\chi$ is an additional free fermion. By choosing the 
Clifford vacuum in the trivial representation of $G$ and $U(1)_{R}$, is 
straightforward to compute their brane factors:
\begin{equation}
\begin{gathered}
f_{\cO_{D_\phi}}(\sigma)=f_{\cO_X}(\sigma)(1-e^{-2\pi\sigma_0})=f_{\cO_X}
(\sigma)-f_{ \cO_X(-1, 0) } (\sigma),
\\
f_{\cO_{D_x}}(\sigma)=f_{\cO_X}(\sigma)(1-e^{2\pi\sigma_1})=f_{\cO_X}(\sigma)-f_
{\cO_X(0 ,1)}(\sigma).
\end{gathered}
\end{equation}
Their zero instanton partition functions are given by
\begin{equation}
\begin{gathered}
Z^0_{\cO_{D_\phi}}=\frac{5}{2}\kappa_0^2+10\kappa_0\kappa_1+5\kappa_1^2
-\frac{5}{2}\kappa_0-5\kappa_1+\frac{35}{12},
\\
Z^0_{\cO_{D_x}}=\frac{5}{2}\kappa_1^2+10\kappa_0\kappa_1+5\kappa_0^2
-\frac{5}{2}\kappa_1-5\kappa_0+\frac{35}{12},
\end{gathered}
\end{equation}
which matches the expected geometric formula (\ref{period2}). 

To obtain the curve and point class, is convenient to work at a generic but 
fixed 
smooth point in complex structure. 
Consider a generic $A$ such that the superpotential is given by
\begin{equation}
W=\sum_{i=1}^5p_iF_i(\phi,x), \label{quinticW}
\end{equation}
where 
\begin{equation}
F_i:=\phi_1x_{i}+\phi_2x_{i+1}+\cdots+\phi_5x_{i+4}.
\end{equation}
All the sub-indices are understood mod 5. Let us show that this choice is 
indeed smooth. The Jacobian of $X_{A}$ is given by
\begin{equation}
\mathrm{Jac}(X_{A})=\left(\begin{array}{ccc|ccc}
x_1 & \cdots & x_5 & \phi_1& \cdots & \phi_5
\\
x_2 & \cdots & x_4 & \phi_5 & \cdots & \phi_4
\\
 \vdots &  & \vdots & \vdots & &\vdots\\
x_{5} & \cdots & x_{4} & \phi_{4} &\cdots  &\phi_{1}
\end{array}\right)=:\left( \vec{x}_i | \vec{\phi}_i \right)_{i=1}^{5}.
\end{equation}
where we denoted the rows of $\mathrm{Jac}(X_{A})$ by $( \vec{x}_i | 
\vec{\phi}_i )$. Define the rank $5$ matrices:
\begin{equation}
\mathcal{P}:=\left(\begin{array}{ccccc}
0 & 1 & 0 & 0 & 0 \\
0 & 0 & 1 & 0 & 0 \\
0 & 0 & 0 & 1 & 0 \\
0 & 0 & 0 & 0 & 1 \\
1 & 0 & 0 & 0 & 0\\
\end{array}\right),\qquad 
\mathcal{C}:=\mathrm{diag}(1,\omega_{5},\omega_{5}^{2},\omega_{5}^{3},\omega_{5}
^{4} )
\end{equation}
where $\omega_{5}$ is a fifth root of unity, $\omega_{5}^{5}=1$. Then, 
$\vec{x}_{i}=P^{i-1}\vec{x}_{1}$ and $\vec{\phi}_{i}=P^{1-i}\vec{\phi}_{1}$ for 
$i=1,\ldots 5$. WLOG, the equation defining a point in $X_{A}$, where 
$\mathrm{rank}(\mathrm{Jac}(X_{A}))<5$ can be written as
\begin{equation}\label{LDeqs}
\vec{x}_{1}=\lambda\mathcal{P}^{r}\vec{x}_{1},\qquad 
\vec{\phi}_{1}=\lambda\mathcal{P}^{-r}\vec{\phi}_{1}
\end{equation}
for $r=1,2,3$ or $4$ and $\lambda\in\mathbb{C}^{*}$. It is easy to show that a 
necessary condition for eqs. (\ref{LDeqs}) to have a nontrivial solution is 
that $\lambda=\omega_{5}$. Then a general solution of (\ref{LDeqs}) can be 
written as
\begin{equation}
\vec{x}_{1}=x\mathcal{C}^{a}\vec{q},\qquad 
\vec{\phi}_{1}=\phi\mathcal{C}^{a'}\vec{q},\qquad \vec{q}:=(\omega_{5}
^{4},\omega_{5}
^{3},\omega_{5}
^{2},\omega_{5}
,1   )^{t},\quad x,\phi\in\mathbb{C}
\end{equation}
where $a,a'$ are the unique integers mod 5 satisfying $1-r+ra=0$ 
mod 5 and $1+r-ra'=0$ mod 5. It is straightforward to show that $a+a'=2$, for any 
$r$. Then, for example, the equation $F_{1}(\phi,x)=0$ implies $x\phi=0$ which 
implies either $\vec{x}_{1}=0$ or $\vec{\phi}_{1}=0$ which 
are points excluded in $\CP^4\times\CP^4$. Hence $\mathrm{Jac}(X_{A})$ is full 
rank everywhere in $X_{A}$.

Next, in order to define the B-brane defining a skyscraper sheaf in $X_{A}$, we 
use the fact that the point $P\in\CP^4\times\CP^4$, defined by the linear 
equations: 
\begin{eqnarray}\label{pointeqs}
&& (\phi_1+\phi_2) =\phi_3=\phi_4=\phi_5=0,\nonumber\\
&& x_1-x_{2} =x_{2}-x_{3}=x_{3}-x_{4}=x_4-x_5=0
\end{eqnarray}
belong to $X_{A}$. At (\ref{pointeqs}), the superpotential $W$ vanishes, then, 
by Hilbert's Nullstellensatz there exists homogeneous functions 
$g_I(p,\phi,x)$, $I=1,\ldots,8$, such that $W$ can be written as
\begin{equation}
W=\sum_{i=1}^5p_iF_i(\phi,x)=\sum_{I=1}^8 g_I(p,\phi,x) l_I(\phi,x)
\end{equation}
with $l_{I}(\phi,x)$ the linear equations (\ref{pointeqs}) defining the 
point. Therefore, we can write a tachyon profile, defining a matrix 
factorization of $W$ by
\begin{equation}
Q_{\cO_P}=\sum_{I=1}^8\left( g_I\bar{\eta}_I+ l_I\eta_I\right)
\end{equation}
choosing the trivial representation for the Clifford vacuum, we obtain the 
brane factor
\begin{equation}
f_{\cO_P}(\sigma)=\left( 1-e^{2\pi\sigma_1} \right)^4\left( 1-e^{-2\pi\sigma_0} 
\right)^4
\end{equation}
that gives $Z^{0}_{\mathcal{O}_{P}}=1$. This B-brane flows to the skyscraper 
sheaf $\mathcal{O}_{P}$ on $X_{A}$. 

The B-branes corresponding to structure sheaves of curves in $X_{A}$, can be 
obtained likewise. Consider the $7$ hyperplanes
\begin{eqnarray}
&& (\phi_1+\phi_2)=(\phi_3+\phi_4)=\phi_5=0\nonumber\\
&& x_1-x_{2} =x_{2}-x_{3}=x_{3}-x_{4}=x_4-x_5=0
\end{eqnarray}
This defines the curve $C^{0}$: 
$([\phi_1:-\phi_1:\phi_3:-\phi_3:0]\times[1:1:\cdots:1])\in\CP^4\times\CP^4$. 
Since this curve lies in $X_{A}$ using the Nullstellensatz we can write the 
tachyon profile:
\begin{equation}
Q_{\cO_{C^0}}=\sum_{I=1}^7\left( \tilde{g}_I\bar{\eta}_I+ 
\tilde{l}_I\eta_I\right)
\end{equation}
where $\tilde{l}_I$ are the linear equations defining $C^{0}$ and 
$\tilde{g}_I$ the homogeneous polynomials that guarantees 
$Q_{\cO_{C^0}}^{2}=W\mathrm{id}$. 
The same construction can be applied to obtain a matrix factorization 
corresponding to the structure sheaf of a curve $C^1$ in $X_{A}$, just by 
exchanging the roles of $x$ and $\phi$. Thus, we obtain the following brane 
factors:
\begin{equation}\label{fbofcurves}
\begin{gathered}
f_{\cO_{C^0}}(\sigma)=\left( 1-e^{-2\pi\sigma_0} \right)^3\left( 
1-e^{2\pi\sigma_1} 
\right)^4,
\\
f_{\cO_{C^1}}(\sigma)=\left( 1-e^{2\pi\sigma_1} \right)^3\left( 
1-e^{-2\pi\sigma_0} 
\right)^4
\end{gathered}
\end{equation}
It is convenient to use the twisted B-branes whose zero-instanton central 
charge is given by $\kappa_{\alpha}$, $\alpha=0,1$. These are given by changing 
the charge of the vacuum in (\ref{fbofcurves}), in order to obtain the sheaves 
$\cO_{C^0}(1,0)$ and $\cO_{C^1}(0,1)$. The corresponding brane factors are
\begin{equation}
\begin{gathered}
f_{0}(\sigma):=f_{\cO_{C^0}(1,0)}(\sigma)=e^{-2\pi \sigma_{0}}\left( 
1-e^{-2\pi\sigma_0} 
\right)^3\left( 1-e^{2\pi\sigma_1} 
\right)^4,
\\
f_{1}(\sigma):=f_{\cO_{C^1}(0,1)}(\sigma)=e^{2\pi \sigma_{1}}\left( 
1-e^{2\pi\sigma_1} 
\right)^3\left( 1-e^{-2\pi\sigma_0} 
\right)^4
\end{gathered}
\end{equation}
We are ready to compute the exact central charges of the B-branes generating 
$D^{b}Coh(X_{A})$. We compute them on the $X_{A}$ phase by using the residue 
expansion (\ref{ZXA1}):
\begin{eqnarray}\label{AbPAXresexp}
Z_{\cO_P}(t)&=&\sum_{n,m=0}^{\infty}F_{nm}(t),\nonumber
\\
Z_{0}(t) &=& \kappa_0 Z_{\cO_P}(t) +\frac{5}{2\pi 
i}\sum_{n,m=0}^{\infty}\big( \psi_{nm}-\psi_n \big)F_{nm}(t)\nonumber
\\
Z_{1}(t)&=& \kappa_1 Z_{\cO_P}(t) +\frac{5}{2\pi 
i}\sum_{n,m=0}^{\infty}\big( \psi_{nm}-\psi_m\big)F_{nm}(t)\nonumber
\\
Z_{\cO_{D_\phi}}(t)&=&-\frac{5}{2}\left( 
\kappa_0^2+4\kappa_0\kappa_1+2\kappa_1^2-2 \right)Z_{\cO_P}(t)
+\frac{5}{2}(2\kappa_0+4\kappa_1-1)Z_{0}(t)+5(2\kappa_0+2\kappa_1-1)Z_{1}(t)\nonumber
\\
&+&\frac{25}{2(2\pi 
i)^2}\sum_{n,m=0}^{\infty}\bigg\lbrace5\left( 
\psi_n^2+2\psi_m^2+7\psi_{nm}^2+4\psi_n\psi_m-(6\psi_n-8\psi_m)\psi_{nm} 
\right)\nonumber
\\
&+& 7\psi'_{nm}-\psi'_n-2\psi'_m\bigg\rbrace F_{nm}(t)\nonumber
\\
Z_{\cO_{D_x}}(t)&=&Z_{\cO_{D_\phi}}(t)|_{(\kappa_0\leftrightarrow\kappa_1,
n\leftrightarrow m)}\nonumber
\\
Z_{\cO_{X_{A}}}(t)&=&\left(\frac{5}{6}
\kappa_0^3+5\kappa_0^2\kappa_1+5\kappa_0\kappa_1^2+\frac{5}{6}
\kappa_1^3-5(\kappa_0+\kappa_1)\right)Z_{\cO_P}(t)\nonumber
\\
&+&\left( 
-\frac{5}{2}\kappa_0^2-10\kappa_0\kappa_1-5\kappa_1^2+\frac{5}{2}
(\kappa_0+2\kappa_1)+\frac{25}{6} \right)Z_{0}(t)
\nonumber
\\
&+&\left( 
-\frac{5}{2}\kappa_1^2-10\kappa_0\kappa_1-5\kappa_0^2+\frac{5}{2}
(\kappa_1+2\kappa_0)+\frac{25}{6} \right)Z_{1}(t)\nonumber\\
&+&\kappa_0Z_{\cO_{D_\phi}}(t)+\kappa_1 Z_{\cO_{D_x}}(t)-\frac{25}{6(2\pi 
i)^3}\sum_{n,m=0}^{\infty}\bigg\lbrace-15(\psi_m+2\psi_n-3\psi_{nm}
)\psi'_m\nonumber
\\
&-&15(\psi_n+2\psi_m-3\psi_{nm})\psi'_n\nonumber
\\
&+& 25(\psi_n+\psi_m-2\psi_{nm})(\psi_n^2+\psi_m^2+7\psi_{nm}
^2+5\psi_n\psi_m-7(\psi_n+\psi_m)\psi_{nm})
\nonumber
\\
&+& 105(\psi_n+\psi_m-2\psi_{nm})\psi'_{nm}+\psi''_n+\psi''_m-14\psi''_{nm}
\bigg\rbrace F_{nm}(t)
\end{eqnarray}
\\
where $\psi(z):=\frac{\Gamma'(z)}{\Gamma(z)}$ is the polygamma function, 
$\psi_n:=\psi(n+1)$ and $\psi_{nm}:=\psi(n+m+1)$. The function 
$F_{nm}(t)=F_{mn}(t)$ is given by
\begin{equation}
F_{nm}(t)=\left(\frac{(n+m)!}{n!\ m!}\right)^5e^{2\pi i(n\kappa_0+m\kappa_1)}.
\end{equation}
Finally, we relate the residue expansions (\ref{AbPAXresexp}) with the 
geometric formula proposed in \cite{hori2013exact,knapp2021d,iritani2009integral,Hosono:2004jp,Hosono:1993qy,
Hosono:1994ax} . For this 
purpose, we first note the relation
\begin{equation}
\int_{X_{A}}
g(H,\varsigma_{1})=\int_{\CP^4\times\CP^4}(H+\varsigma_{1})^5g(H,\varsigma_{1}
)=\oint_{(0,0)} \frac { dz_0dz_1 } {(2\pi 
i)^2}\frac{(z_0+z_1)^5}{(z_0z_1)^5}g(z_{0},z_1),
\end{equation}
the disk partition function (\ref{ZXA1}) can be given in the integral of forms as ($\kappa(H)=H\kappa_0+\varsigma_1\kappa_1,\ \kappa(n)=\kappa_0n_0+\kappa_1n_1$)
\begin{equation}
\begin{aligned}
Z_{\cB}(t)=\int_{X_A}&\Ch(\cB^{LV})e^{\kappa(H)}\sum_{n_0,n_1=0}^\infty e^{2\pi i \kappa(n)}\times
\\
&\left(\frac{\Gamma(1+n_0+n_1+(H+\varsigma_1)/2\pi i)}{\Gamma(1+n_0+H/2\pi i)\ \Gamma(1+n_1+\varsigma_1/2\pi i)}\right)^5\Td_{X_A}
\end{aligned}
\end{equation}
where 
\begin{equation}
\Td_{X_A}=\frac{(H\cdot\varsigma_1)^5\left(1-e^{-H-\varsigma_1}\right)^5}{
(H+\varsigma_1)^5\left(1-e^{-H}\right)^5\left(1-e^{-\varsigma_1}\right)^5},\quad 
\Ch(\cB^{LV})=\frac{f_\cB(H/2\pi,-\varsigma_1/2\pi)}{f_{\cO_X}(H/2\pi,
\varsigma_1/2\pi)}.
\end{equation}
where
\begin{equation}
f_{\cO_X}(H/2\pi,
\varsigma_1/2\pi)=(1-e^{-H-\varsigma_1})^{5}.
\end{equation}
Consequently, the partition function (\ref{ZXA1}) is given by the integral
\begin{equation}
\begin{aligned}
Z_{\cB}(t)=&\int_{X_A} \Ch(\cB^{LV})\hat\Gamma_XI_X(\kappa)
\end{aligned}
\end{equation}
with the corresponding Gamma class and I-function \cite{cox1999mirror} for the 
determinantal quintic $X$ given by
\begin{equation}
\hat{\Gamma}^{*}_X=\left(\frac{\Gamma(1+H/2\pi i)\ \Gamma(1+\varsigma_1/2\pi 
i)}{\Gamma(1+(H+\varsigma_1)/2\pi i)}\right)^5,
\end{equation}
\begin{equation}
I_{X}=e^{\kappa(H)}\ \hat{\Gamma}^{*}_X\sum_{n_\alpha\geq0}e^{2\pi i 
\kappa(n)}\left(\frac{\Gamma(1+n_0+n_1+(H+\varsigma_1)/2\pi 
i)}{\Gamma(1+n_0+H/2\pi i)\Gamma(1+n_1+\varsigma_1/2\pi i)}\right)^5
\end{equation}

\subsubsection{GN Calabi-Yau in $\CP^7$}

\begin{figure}[h]
\centering
\includegraphics[scale=0.45]{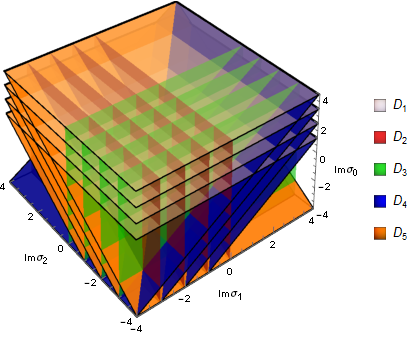}
\includegraphics[scale=0.5]{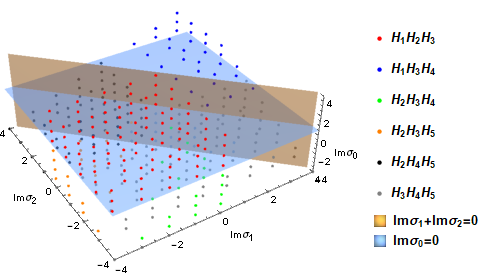}
\caption{Gamma divisors (left) and corresponding poles (right) of GN Calabi-Yau. The admissible contour envelops poles in red, in the $(-,-,+)$ octant.}
\label{GDivisors2}
\end{figure}

The hemisphere partition function (\ref{ZD2}) for the GN model is given by (to 
avoid cluttering we write $\sigma_{\mathsf{k}}:=\sigma_{1,\mathsf{k}}$)
\begin{equation}
Z_\cB(t)=\int_\gamma d\sigma_0d\sigma_1d\sigma_2\ 
Z_{U(2)}(\sigma)Z_{\Gamma}(\sigma) 
e^{it_0\sigma_0+it_1(\sigma_1+\sigma_2)}f_{\cB}(\sigma)
\end{equation}
where
\begin{equation}
\begin{gathered}
Z_{U(2)}:=(\sigma_1-\sigma_2)\sinh\pi(\sigma_1-\sigma_2)
\\
\begin{aligned}
Z_{\Gamma}:=&\Gamma\left( i\left( \sigma_1-\sigma_0 \right)+\epsilon 
\right)^4\Gamma\left( i\left( \sigma_2-\sigma_0 
\right)+\epsilon\right)^4
\\
\times&\Gamma\left( i\sigma_0+1-\epsilon-\delta\right)^8\Gamma\left( 
-i\sigma_1+\delta \right)^4\Gamma\left( -i\sigma_2+\delta 
\right)^4\nonumber 
\end{aligned}
\end{gathered}
\end{equation}
In this model B-brane $\mathcal{B}$ carries a representation of the $U(1)\times 
U(2)$ gauge group. This will be specified by a Young diagram $\lambda$, of 
height $2$ and two integers $(m,q)\in\mathbb{Z}^{2}$ specifying tensor product 
with the determinant representation $\mathrm{det}^{m}\mathbf{2}$ of $U(2)$ 
and with the one dimensional representation of $U(1)$ of weight $q$. Denote 
this representation by $(\lambda,q,m)$:
\begin{equation}\label{repnotationf}
(\lambda,q,m):=\Sigma_{\lambda}\mathbf{2}\otimes 
\mathrm{det}^{m}\mathbf{2}\otimes \mathbb{C}(q)
\end{equation}
where $\Sigma_{\lambda}$ denotes the Schur functor.
For later use, we will define the brane factor associated with the 
fundamental representation $\yng(1):=\mathbf{2}$ as
\begin{equation}
f_{\yng(1)}(\sigma):=f_{(\mathbf{2},0,0)}(\sigma)=e^{2\pi\sigma_1}+e^{
2\pi\sigma_2 } \label{rep1}
\end{equation}
and the brane factor associated with the determinant representation 
$\mathrm{det}\mathbf{2}$ as
\begin{equation}
f_{\yng(1,1)}(\sigma):=f_{(0,0,1)}(\sigma)=e^{
2\pi(\sigma_1+\sigma_2) } .\label{rep11}
\end{equation}
The Gamma poles in GN CY lie in the collection of hyperplanes 
\begin{equation}
\begin{gathered}
\cH_1=\left\lbrace \sigma_0=i(k+1-\epsilon-\delta) \right\rbrace,\ \cH_2=\left\lbrace \sigma_1=-i(k+\delta) \right\rbrace,\ \cH_3=\left\lbrace \sigma_2=-i(k+\delta) \right\rbrace,
\\
\cH_4=\left\lbrace \sigma_1-\sigma_0=i(k+\epsilon) 
\right\rbrace,\cH_5=\left\lbrace \sigma_2-\sigma_0=i(k+\epsilon) \right\rbrace,\ 
k\in\Z_{\geq0}.
\end{gathered}
\end{equation}
Is straightforward to show that the following contour is admissible in the 
$X_{A}$ phase:
\begin{equation}
\gamma:\quad 
\left(\sigma_0,
\sigma_1,\sigma_{2}\right)=\left(\Re\sigma_0+i\left(\Re\sigma_0\right)^2,
\Re\sigma_1-i\left(\Re\sigma_1\right)^2,
\Re\sigma_2-i\left(\Re\sigma_2\right)^2\right).
\end{equation}
and it encircles the poles in the intersection $\cH_1\cap \cH_2\cap \cH_3$. We 
illustrate this in fig. \ref{GDivisors2}.
Thus this integral can be similarly evaluated by taking the residue at the 
poles, giving rise to the following series (in the $X_{A}$ phase, hence 
$(\delta,\epsilon)\rightarrow (0,1)$):
\begin{equation}
\begin{aligned}
Z_\cB(t)=&\frac{-i}{(2\pi i)^{12}}\sum_{n,m,l=0}^\infty\Res_{(z_\alpha)=(l,n,m)}
\left(\frac{\pi}{\sin\pi z_0}\right)^8\left(\frac{\pi}{\sin\pi z_1}\right)^4\left(\frac{\pi}{\sin\pi z_2}\right)^4
\\
\times & \frac{\Gamma(z_0+z_1+1)^4\ \Gamma(z_0+z_2+1)^4}{\Gamma(z_0+1)^8\ \Gamma(z_1+1)^4\ \Gamma(z_2+1)^4}\left(z_1-z_2\right)\sin\pi\left(z_1-z_2\right)
\\
\times & e^{-t_0z_0+t_1(z_1+z_2)}f_\cB(iz_0,-iz_1,-iz_2).
\end{aligned}\label{ZXA2}
\end{equation}
We choose the normalization factor $-i(2\pi i)^{-12}$. for the period. The 
tachyon profile specifying $\mathcal{B}$ associated to the structure sheaf of 
$X_{A}$, $\mathcal{O}_{X_{A}}$ is straightforwadly obtained:
\begin{equation}
Q_{\cO_X}:=p_{i}^{\mathsf{k}}{\bar{\eta}^{i}_{\mathsf{k}}}+A(\phi)^{ij}x_{
j\mathsf{k} }
\eta_i^\mathsf{k}
\end{equation}
where fermions $\eta^{\mathsf{k}}_{i}$, $\bar{\eta}^{i}_{\mathsf{k}}$  
transform in the fudamental and anti-fundamental representations of $U(2)$, 
respectively. They satisfy the Clifford algebra:
\begin{equation}
\lbrace\eta^{\mathsf{k}}_{i},\bar{\eta}^{j}_{\mathsf{m}}
\rbrace= \delta_{i}^{j}\delta_{\mathsf{m}}^{\mathsf{k}}
\end{equation}
We choose $\bar{\eta}^{j}_{\mathsf{k}}$ as the creation operators. By taking 
the 
Clifford vacuum to be in the $\mathrm{det}^{m}\mathbf{2}\otimes \mathbb{C}(q)$ 
representation of $U(1)\times U(2)$, we get
\begin{equation}
f_{\cO_X(q,m)}(\sigma)=e^{2\pi q\sigma^{0}}e^{2\pi 
m(\sigma^{1}+\sigma^{2})}\left( 1- 
e^{2\pi(\sigma_1-\sigma_0)} \right)^4\left( 1- e^{2\pi(\sigma_2-\sigma_0)} 
\right)^4.
\end{equation}

The zero instanton term $Z^0_{\cO_X(q,m)}$ is given by
\begin{eqnarray}
Z^0_{\cO_X(q,m)} &=&-\frac{64\zeta(3)}{(2\pi i)^3}-\frac{1}{6}\left( 14 m 
+ 8 m^3 - 2 (7 + 24m^2 ) q + 
60mq^2 - 20 q^3 \right)
\nonumber\\
 &+& 8\left( 7  + 6 (4m^2  - 10m q + 5 q^2) 
\right)\frac{\kappa_0}{24}
+8\left( 7  + 3 (4 m^2 - 16 m q + 10 q^2) 
\right)\frac{\kappa_1}{24}
\nonumber\\
 &+&\left( 5 (2 q-2m) \kappa_0^2 + 4 (5 q-4m) \kappa_0\kappa_1 + 2 (4 
q-2m) \kappa_1^2 \right)
\nonumber\\
 &+&\left( 
\frac{10}{3}\kappa_0^3+10\kappa_0^2\kappa_1+8\kappa_0\kappa_1^2+\frac{4}{3}
\kappa_1^3 \right),
\end{eqnarray}
where $\kappa_0=-\tau_0-4$, $\kappa_1=\tau_1-2$. The case $q=m=0$ recovers the 
topological data in (\ref{topdata2}).
The divisor classes can be obtained from $\mathcal{O}_{X_{A}}$ by modifying the 
tachyon profile as
\begin{eqnarray}
Q_{D_{\phi}}&=&Q_{\mathcal{O}_{X}}+h(\phi)\chi\nonumber\\
Q_{D_{x}}&=&Q_{\mathcal{O}_{X}}+l(x)\chi
\end{eqnarray}
where $h(\phi)$ and $l(x)$ are linear functions of $\phi$ and 
$\varepsilon^{\mathsf{k}\mathsf{m}}x_{i,\mathsf{k}}x_{j,\mathsf{m}}$, 
respectively. 
The fermion $\chi$ is taken to be an annihilation operator whe acting on the 
Clifford vacuum. Then, the brane factors can be computed straightforwardly:
\begin{equation}
\begin{gathered}
f_{\cO_{D_\phi}}(\sigma)=f_{\cO_X}(\sigma)-f_{\cO_X(-1,0,0)}(\sigma),
\\
f_{\cO_{D_x}}(\sigma)=f_{\cO_X}(\sigma)-f_{\cO_X(0,0,1)}(\sigma)
\end{gathered}
\end{equation}
Their zero instanton partition functions gives the expected result:
\begin{equation}
\begin{gathered}
Z^0_{\cO_{D_\phi}}=10\kappa_0^2+20\kappa_0\kappa_1+8\kappa_1^2-10\kappa_0-10\kappa_1+\frac{17}{3},
\\
Z^0_{\cO_{D_x}}=10\kappa_0^2+16\kappa_0\kappa_1+4\kappa_1^2-8\kappa_0-4\kappa_1+\frac{11}{3}.
\end{gathered}
\end{equation}
We expect two curve classes and they can be also obtained by a simple 
reasoning. In order to obtain the tachyon profile for a curve contained in 
$\mathbb{P}^{7}$, note that we can choose a curve inside $\mathbb{P}^{7}$ that 
satisfies $A^{sj}(\phi)x_{j}=0$ for $s=1,2,3$. 
Indeed, since $x_{j,\mathsf{k}}$ can be seen as a $4\times 2$ matrix of rank 
$2$ in the $X_{A}$ phase, the equations 
$A^{sj}(\phi)x_{j}=0$ can be written as $M^{s,a}_{\mathsf{k}}\phi_{a}=0$, 
and as a $6\times 8$ matrix, $M$ is full rank, 
giving precisely a curve in the $\phi$ coordinates. WLOG, assume this curve is 
parametrized by $\phi_{1}$ and $\phi_{2}$. Then, the equations 
$A^{4j}(\phi)x_{j}=0$ can be solved by 
imposing two linear equations in $x$, namely $A^{4j,a}x_{j}=0$, 
$a=1,2$, giving a point in $G(2,4)$. Evidently this curve is contained in 
$X_{A}$. The tachyon profile then reads
\begin{eqnarray}
Q_{C^{0}}=\sum_{s=1}^{3}\left(p_{s}^{\mathsf{k}}\bar{\eta}^{s}_{\mathsf{k}}+A^{s
, j } (\phi)x_ { j , \mathsf{k} } \eta_{s}^
{ \mathsf{k} 
}\right)+A^{4j,a}x_{j,\mathsf{k}}\chi^{\mathsf{k}}_{a}+\phi_{a}p_{4}^{\mathsf{k}
} \bar { \chi}_ { \mathsf{k}}^
{ a }
\end{eqnarray}
where $\chi$ and $\bar{\chi}$ are additional free fermions, decoupled from 
$\eta$ and $\bar{\eta}$. The brane factors is then straightforwardly computed:
\begin{eqnarray}
f_0(\sigma)=e^{-2\pi\sigma_0}\left( 1- e^{2\pi(\sigma_1-\sigma_0)} 
\right)^3\left( 1- 
e^{2\pi(\sigma_2-\sigma_0)} 
\right)^3\left(1-e^{2\pi\sigma_1}\right)^2\left(1-e^{2\pi\sigma_2}\right)^2, 
\end{eqnarray}
where we have chosen the Clifford vaccum in the $\mathbb{C}(-1)$ 
representation of $U(1)$ in order to have $Z^{0}_{0}(t)=\kappa_{0}$. For a 
curve contained in $G(2,4)$, we can use the fact that generically, there exists 
points in $\mathbb{P}^{7}$ where 
$\rank A(\phi)=1$. We can 
always choose the complex structure such that one of those points is given by 
$\phi_{a}=0$ for $a\neq 1$. Then we choose $x_{j,\mathsf{k}}$ to be orthogonal 
to 
the hyperplane spanned by the image of the matrix $A^{1}$, i.e. 
$A^{ij,1}x_{j,\mathsf{k}}=0$. Then we intersect these $x$'s 
belonging to a $\mathbb{C}^{3}$ subspace of $\mathbb{C}^{4}$ with a 
hypersurface 
$b^{ij}\varepsilon^{\mathsf{k}\mathsf{m}}x_{i,\mathsf{k}}x_{j,\mathsf{m}}=0$. 
This gives the 
desired curve. The tachyon profile is given by
\begin{eqnarray}
Q_{C^{1}}=\sum_{a=2}^{8}\left(p_{i}^{\mathsf{k}}A^{ij,a}x_{j,\mathsf{k}}\bar{
\eta}_{ a}
+\phi_{a}\eta^{a}\right)+A^ { ij , 1 }\phi_{1} x_ { j , \mathsf{k}} 
\chi^{\mathsf{k}}_{i}+p_{i}^{\mathsf{k}}\bar{\chi}_{\mathsf{k}}^
{i 
}+b^{ij}\varepsilon^{\mathsf{k}\mathsf{m}}x_{i,\mathsf{k}}x_{j,\mathsf{m}}\xi.
\end{eqnarray}
where $\xi$ is an additional free fermions, decoupled from 
all the rest. We twist the Clifford vaccum by 
$\mathrm{det}\mathbf{2}$ in order to 
have $Z^{0}_{1}(t)=\kappa_{1}$, the resulting brane factor is given by
\begin{equation}
f_1(\sigma)=e^{2\pi (\sigma_1+\sigma_2)}\left( 1- e^{2\pi(\sigma_1-\sigma_0)} 
\right)\left( 1- e^{2\pi(\sigma_2-\sigma_0)} 
\right)\left(1-e^{2\pi(\sigma_1+\sigma_2)}\right)\left( 
1-e^{-2\pi\sigma_0}\right)^7.
\end{equation}
Finally, for the B-brane that RG-flows to a skyscraper sheaf 
$\mathcal{O}_{P}\in D^{b}Coh(X_{A})$, we simply intersect $C^{0}$ with a line 
$l(\phi)=l^{a}\phi_{a}$ in $\mathbb{P}^{7}$, hence the tachyon profile is given 
by
\begin{eqnarray}
Q_{\cO_{P}}=Q_{C^{0}}+l(\phi)\xi
\end{eqnarray}
but we choose the Clifford vacuum in the trivial representation, then, its 
brane factor
\begin{equation}
f_{\cO_P}(\sigma)=\left( 1- e^{2\pi(\sigma_1-\sigma_0)} \right)^3\left( 1- 
e^{2\pi(\sigma_2-\sigma_0)} 
\right)^3\left(1-e^{2\pi\sigma_1}\right)^2\left(1-e^{2\pi\sigma_2}
\right)^2\left( 1-e^{-2\pi\sigma_0} \right).
\end{equation}
evaluates to $Z^{0}_{\cO_{P}}(t)=1$. This complete our search for 
representatives for a set of generators of $D^{b}Coh(X_{A})$. The full 
expressions for the periods periods are given by ($(l,n,m)\in\Z_{\geq0}^3$)
\begin{eqnarray}
Z_{\cO_P}(t) &=& \sum_{l,n,m} \bigg\lbrace1+2(n-m)\big(\psi_{ln}-\psi_{lm}+\psi_m-\psi_n\big)  \bigg\rbrace F_{lnm}(t) \nonumber
\\
Z_{0}(t)  &=& \kappa_0Z_{\cO_P}+\frac{2}{2\pi i}\sum_{l,n,m}\nonumber
\\
& &\bigg\lbrace 2\big(\psi_{ln}-\psi_l+\psi_{lm}-\psi_l\big)\big( 1+2(n-m)\big( \psi_{ln}-\psi_{lm}+\psi_m-\psi_n \big) \big) \nonumber
\\
 &+& (n-m)\big( \psi'_{ln}-\psi'_{lm} \big) \bigg\rbrace F_{lnm}(t)
\\
Z_1(t) &=& \kappa_1 Z_{\cO_P}+\frac{1}{2\pi i}\sum_{l,n,m}\nonumber
\\
& & \bigg\lbrace 2\big( \psi_{ln}+\psi_{lm}-\psi_{n}-\psi_{m} \big)\big( 1+2(n-m)(\psi_{ln}-\psi_{lm}+\psi_m-\psi_n) \big)\nonumber
\\
&+& (n-m)(\psi'_{ln}-\psi'_{lm}-\psi'_{n}+\psi'_m) \bigg\rbrace F_{lnm}(t)
\\
Z_{\cO_{D_\phi}}(t) &=& 
\left(\frac{19}{3}-10\kappa_0^2-16\kappa_0\kappa_1-4\kappa_1^2\right)Z_{\cO_P}
\nonumber
\\
&+&(20\kappa_0+16\kappa_1-8) Z_0+(16\kappa_0+8\kappa1-4)Z_1+I^{(D_1)}
\\
Z_{\cO_{D_x}}(t) &=& 
\left(\frac{19}{3}-10\kappa_0^2-16\kappa_0\kappa_1-4\kappa_1^2\right)Z_{\cO_P}
\nonumber
\\
&+& 
\left(20\kappa_0+16\kappa_1-8\right)Z_0+\left(16\kappa_0+8\kappa1-4\right)Z_1+I^
{(D_1)}
\\
Z_{\cO_{X_A}}(t)&=&(\frac{10}{3}
\kappa_0^3+10\kappa_0^2\kappa_1+8\kappa_0\kappa_1^2+\frac{4}{3}
\kappa_1^3-9\kappa_0-\frac{19}{3}\kappa_1)Z_{\cO_P}\nonumber
\\
&+&\left(\frac{17}{3}+10\kappa_0+8\kappa_1-10\kappa_0^2+20\kappa_0\kappa_1-8\kappa_1^2\right)Z_0\nonumber
\\
&+&(5+10\kappa_0+4\kappa_1-10\kappa_0^2-16\kappa_0\kappa_1-4\kappa_1^2)Z_1\nonumber
\\
&+&\kappa_0 Z_{\cO_{D_\phi}}+\kappa_1 Z_{\cO_{D_x}}+I^{(X)}
\end{eqnarray}

where 
\begin{equation}
F_{lnm}(t)=\left(\frac{(n+l)!\ (m+l)!}{l!\ l!\ n!\ m!}\right)^4(-1)^{n+m}e^{2\pi il\kappa_0+2\pi i(n+m)\kappa_1}.
\end{equation}
we omit the lengthy closed expressions for the instanton series 
$I^{(D_{1})}$ and $I^{(X)}$ but we present some of their leading terms here:
\begin{equation}
\begin{gathered}
I^{(D_0)}=-\frac{10}{3}+\left(\frac{14}{(2\pi i)^2}-\frac{10}{3}\right)e^ {2\pi i\kappa_1}+\frac{20}{3}e^{2\pi i\kappa_2}+\cdots
\\
I^{(D_1)}=-\frac{18}{3}-\frac{8}{3}e^ {2\pi i\kappa_0}+\frac{16}{3}e^ {2\pi i\kappa_1}+\cdots
\\
I^{(X)}=-\frac{64\zeta(3)}{(2\pi i)^3}-\left(\frac{112+64\zeta(3)}{(2\pi i)^3}+\frac{32/3}{2\pi i}\right)e^ {2\pi i\kappa_0}+\left(\frac{40/3}{2\pi i}-\frac{128\zeta(3)}{(2\pi i)^3}\right)e^ {2\pi i\kappa_1}+\cdots
\end{gathered}
\end{equation}
One can always recover the exact result from the residue formula (\ref{ZXA2}).

In order to find a geometric expression for $Z_{\mathcal{B}}(t)$, we use a theorem from \cite{martin2000symplectic}, that states that given a variety $V$with a $G$-action we can write and integral over the symplectic quotient $V\sslash G$ as in integral over $V\sslash T_{G}$, where $T_{G}$ denotes the maximal torus. More precisely:
\begin{equation}\label{martin}
\int_{V\sslash G}a=\frac{1}{|W_{G}|}\int_{V\sslash T_{G}}e\cup \tilde{a}
\end{equation}
where $e=\prod_{\alpha}e(\alpha)$ is a product over all roots of $G$ and $e(\alpha)$ is the Euler class of the line bundle over $V\sslash T_{G}$ with weight $\alpha$. $a$ is a class in $H^{*}(V\sslash G)$ and $\tilde{a}\in H^{*}(V\sslash T_{G})$ its lift. In the example at hand $V$ is linear space, hence the formula (\ref{martin}) becomes
\begin{equation}
\int_{\mathbb{P}^{7}\times 
G(2,4)}f(\varsigma_1,H,\ldots)=\frac{1}{2}\int_{\mathbb{P}^{7}\times 
\mathbb{P}^{4}\times\mathbb{P}^{4}} (H_{1}-H_{2})(H_{1}+H_{2})\cup 
f(H,H_{1},H_{2})
\end{equation}
Where $f(\varsigma_1,H,\ldots)$ denotes a function in 
$H^{*}(\mathbb{P}^{7}\times G(2,4))$ and $f(H,H_{1},H_{2})$ the lift to a function in $H^{*}(\mathbb{P}^{7}\times \mathbb{P}^{4}\times\mathbb{P}^{4})$ where $H_{\alpha}$, $\alpha=1,2$ denotes 
the hyperplane classes of the $\mathbb{P}^{4}$ factors. The identification with the cohomology classes of $G(2,4)$ is $\varsigma_1=H_1+H_2$ and $H^{*}(Gr(2,4))$ is generated by $H_{1}+H_{2}$ and $H_{1}\cdot H_{2}$ as a ring:
\begin{equation}
H^{*}(Gr(2,4))\cong \frac{\mathbb{C}[H_{1},H_{2}]^{S_{2}}}{\langle
d_{3},d_{4}\rangle}
\end{equation}
where $S_{2}$ acts by $H_{1}\leftrightarrow H_{2}$ and $d_{i}(H_{1},H_{2})$ 
denotes the complete homogeneous function of degree $i$. In (\ref{ZD2}), the 
integral on the RHS of (\ref{martin}) can be immediately identified, upon taking residues on the $X_{A}$ phase. In order to write it as an integral over $X_{A}$, note that the Euler class of the normal bundle of $X_{A}$ in $\mathbb{P}^{7}\times G(2,4)$ is given by 
$(H+H_{1})^4(H+H_{2})^{4}$, then we can write:
\begin{equation}
\int_{\mathbb{P}^{7}\times 
G(2,4)}(H+H_{1})^4(H+H_{2})^{4}\cup f(\varsigma_1,H,\ldots)=\int_{X_{A}} 
f(\varsigma_1,H,\ldots).
\end{equation}
Define
\begin{equation}
\kappa(H):=\kappa_0H+\kappa_1\varsigma_1\qquad 
\kappa(n):=\kappa_0n_0+\kappa_1(n_1+n_2)
\end{equation}
Then, we can identify the power series (\ref{ZXA2}) with the following integral over $X_{A}$
\begin{equation}\label{exprcc}
\begin{aligned}
Z_{\cB}(t)=&\int_{X_A}\hat\Gamma_XI_X(\kappa)\Ch(\cB^{LV})
\end{aligned}
\end{equation}
where we defined: 
\begin{equation}
\Ch(\cB^{LV})=\frac{f_\cB(H/2\pi,-H_1/2\pi,-H_2/2\pi)}{(1-e^{-(H+H_{1})})^{4}
(1-e^{-(H+H_{2})})^{4}}
\end{equation}
\begin{equation}
\hat\Gamma_{X}=\frac{\Gamma\left(1-\frac{H}{2\pi 
i}\right)^{8}\Gamma\left(1-\frac{H_{1}}{2\pi 
i}\right)^{4}\Gamma\left(1-\frac{H_{2}}{2\pi 
i}\right)^{4}\sin\left(\frac{H_{1}-H_{2}}{2i}\right)}{\Gamma\left(1-\frac{H+H_{1
} } { 2\pi i}\right)^{4}\Gamma\left(1-\frac{H+H_{2}}{2\pi 
i}\right)^{4}\frac{H_{1}-H_{2}}{2i}},
\end{equation}
\begin{eqnarray}
I_X(\kappa)&=&e^{\kappa(H)}\frac{\Gamma\left(1+\frac{H}{2\pi 
i}\right)^{8}\Gamma\left(1+\frac{H_{1}}{2\pi 
i}\right)^{4}\Gamma\left(1+\frac{H_{2}}{2\pi 
i}\right)^{4}}{\Gamma\left(1+\frac{H+H_{1
} } { 2\pi i}\right)^{4}\Gamma\left(1+\frac{H+H_{2}}{2\pi 
i}\right)^{4}}\sum_{n_\alpha\geq0}e^{2\pi i 
\kappa(n)}\left(\frac{H_{1}-H_{2}}{2\pi i}+n_{1}-n_{2}\right)
\nonumber\\
&\times&(-1)^{n_{1}+n_{2}}\frac{\Gamma\left(1+H+H_{1}+n_{0}+n_{1}\right)^{4}
\Gamma\left(1+H+H_ { 2 } 
+n_{0}+n_{2}\right)^{4}}{\Gamma\left(1+n_{0}+H\right)^{8}\Gamma\left(1+n_{1} 
+H_{1}\right)^{4}\Gamma\left(1+n_{2}+H_{2}\right)^{4}}.
\end{eqnarray}
We remark that an expression for $I_X(\kappa)$, for linear PAX models was derived in \cite{Honma:2018fgw}. After the present work, the following preprint \cite{Priddis:2024qeb} appeared on the ArXiv, where $I_X(\kappa)$ has been computed for more general PAX models. However, neither \cite{Honma:2018fgw} and \cite{Priddis:2024qeb} present an expression for the central charge of $\cB^{LV}$ such as (\ref{exprcc}), including the factors $\Ch(\cB^{LV})$ and $\hat\Gamma_X$.

\section{\label{sec:section4} Grade Restriction Rule and B-brane transport}

Consider an arbitrary B-brane $\cB$, then, generically its brane factor takes 
the form
\begin{equation}
f_{\cB}(\sigma)=\sum_{m}R_{m}(e^{i\pi})e^{q_{m}(\sigma)}
\end{equation}
where $q_{m}$ are the weights of the representation $\rho_{M}$, corresponding 
to $\cB$. Then, denote $Z_{\cB}(t;q)$ the summand $\exp(q(\sigma))\subset f_{\cB}(\sigma)$ in (\ref{ZD2}).  $Z_{\cB}(t;q)$ has the following behavior, as $|\sigma|\gg 1$ 
\cite{hori2013exact} (as long as we keep $\sigma$ away from the singular hyperplanes $\mathcal{H}$):
\begin{equation}
Z_{\cB}(t;q)\sim \int_\gamma d^k\sigma P(\sigma)\exp \left\lbrace 
A_q(\sigma)+iB_{q}(\sigma)\right\rbrace.
\end{equation}
with $P(\sigma)$ a polynomial function and $A_{q},B_{q}$ real valued functions 
of $\sigma$. Denote 
\begin{equation}
\sigma_\alpha=\nu_\alpha+i\rho_\alpha.
\end{equation}
Absolute convergence of
$Z_{\cB}(t;q)$ is then  
determined by $A_{q}(\sigma)$, which is explicitly given by 
\begin{equation}
\begin{aligned}\label{Aq}
A_{q}(\sigma)=& \sum_{\alpha\in\Delta^+}\pi|\alpha(\nu)|+\left( 
-\zeta(\rho)+\theta(\nu)+2\pi q(\nu)\right) 
\\
+&\sum_i Q_i(\rho)\left( \log|Q_i(\sigma)|-1 \right)-|Q_i(\nu)|\left[ \frac{\pi}{2}+\arctan\frac{Q_i(\rho)}{|Q_i(\nu)|} \right].
\end{aligned}
\end{equation}
Therefore, absolute convergence requires that
$A_q(\sigma)<0$ in all asymptotic directions of $\gamma$. More 
precisely, we require $A_q(\sigma)<0$ for all the weights $q$ of $\rho_{M}$, separately. Inside a given phase, it is not hard to came up with an admissible contour $\gamma$, such as (\ref{contour}), for the $X_{A}$ phase in the linear PAX models. Indeed, the integral $Z_{\cB}(t)$ on (\ref{contour}) is absolutely convergent on the $X_{A}$ phase for any $\rho_{M}$ i.e., the asymptotic condition $A_q(\sigma)<0$ is independent of $q$.

The admissible contour (\ref{contour}) can be straightforwadly generalized to other phases. This is correlated with the fact that, in the PAX models, all the phases are weakly coupled and therefore we do not find phenomena such as the possible existence of a grade restriction rule, deep into a phase (as it is argued in \cite{eager2017beijing,EHKR}). Admissible contours on other phases can be obtained by an analysis of the absolute convergence of $Z_{\cB}(t;q)$, and they are simply given by an appropriately setting $\Im\sigma$ as an even power of a linear combination of $\Re\sigma$. For example, for the $X_{A^{T}}$ phase, characterized by $\zeta_{1}>0$ and $\zeta_{0}+2\zeta_{1}>0$ we can choose the admissible contour:
\begin{equation}
\left(\Re\sigma_{0}+i(\Re\sigma_{0})^{2},\Re\sigma_{1,\mathsf{k}}+i((\Re\sigma_{1,\mathsf{k}}-\Re\sigma_{0})^{2}+(\Re\sigma_{0})^{2})\right)
\end{equation}

Since B-branes are insensitive to deformations of the theory, by operators that are $\mathbf{Q}_{B}$-exact, they are, in particular, invariant under small deformations of the FI-theta parameters. Under a finite deformation, a GLSM B-brane $\mathcal{B}$ in general, will have a different image under RG-flow to the IR fixed point. The comparisons between different IR images of the B-brane $\mathcal{B}$ is referred to as \textit{B-brane transport}. For paths joining two values of the FI parameters, deep inside different phases, at fixed theta-angle it was studied in high detail in 
\cite{herbst2008phases,hori2013exact} and 
\cite{ballard2019variation,halpern2015derived}. 
For these paths, it was shown in \cite{herbst2008phases}, that an admissible contour $\gamma$ exists along such path in (a covering of) FI-theta space\footnote{Is important to remark, that the condition that 
$Z_{\mathcal{B}}(t)$ is absolutely convergent along a path in FI-theta space may be too strict, for nonabelian models. There are instances where absolute convergence seems to be too strong and imposing only some kind of conditional convergence is necessary \cite{GuoRomoSmith}}, if and only if we impose some restriction on the weights of $\rho_{M}$. This is called the \textit{grade restriction rule} in \cite{herbst2008phases} and the restriction on the weights, is usually called a \textit{window}, this defines a subcategory $\mathbb{W}(\theta)$ of $MF_{G}(W)$ called the window (sub)category. The notation $\mathbb{W}(\theta)$ emphasized the fact that the windows, and consequently the window categories, are not unique. They depend on the theta-angle on the covering space.

In the case at hand, of the linear PAX models, we are ultimately interested in monodromy with base point at the phase $X_{A}$. They will act as autoequivalences on $D^{b}Coh(X_{A})$ as we will see below. Therefore, we will study the window categories arising from a path joining a point $\zeta_{0}\gg 1$ and a point $\zeta_{0}\ll -1$, at a fixed $\zeta_{1}\ll-1$, we will refer to this as the $Y$ phase 
boundary. In addition we also need to study the $X$ phase boundary described analogously, but exchanging $\zeta_{0}$ with $\zeta_{1}$.

Determining the window categories of nonabelian GLSMs is not an easy 
task. However, in the family of examples at hand we can make use of 
the fact that the phases are weakly coupled and combine previously known results to get the desired windows. In the $Y$ phase boundary we can then consider first the GLSM at fixed value of the coupling $\zeta_{1}\ll-1$. At medium energies, this 
corresponds to a $U(1)$ gauged $G(n-k,n)$ sigma model where $x$ fields are the Stiefel coordinates on $G(n-k,n)$ and the $\phi$ and $p$ fields are $U(1)$-charged fields that takes values on sections of the trivial bundle and the canonical sub-bundle over $G(n-k,n)$, respectively. The fields interact via the restriction of the superpotential (\ref{PAXpot}). The $U(1)$ representations of the $p$ and $\phi$ fields are given by
\begin{equation}
\begin{array}{c|cc}
 & P^\alpha_{i} & \phi_a \\
\hline
U(1) & -1 & 1
\end{array}
\end{equation}
where $i=1,\cdots,n$, $\alpha=1,\cdots,n-k$, $a=1,\cdots,n(n-k)$. Then, we have 
locally, an abelian model fibered over $G(n-k,n)$ where $U(1)$ acts trivially 
on the base. So, the $U(1)$ gauge dynamics can be analyzed at a fixed point in 
the base. The grade restriction rule, and hence the window, for this abelian 
model can determined then using the results of  
\cite{herbst2008phases}, and is given by
\begin{equation}
-\frac{\sum_{i,Q^0_i>0} Q^0_i}{2}< \frac{\theta_0}{2\pi}+q^0 < 
\frac{\sum_{i,Q^0_i>0} Q^0_i}{2},\qquad \sum_{i,Q^0_i>0} Q^0_i=n(n-k)\label{WY}
\end{equation}
The weights of the gauge group $U(n-k)$ are unrestricted. We denote the 
window for this boundary as $\mathbb W_Y(\theta_{0})$ as it only depends on 
$\theta_{0}$.

On the other hand, at the $X$ phase boundary we can proceed in a similar way. When setting $\zeta_{0}\gg 1$, we can approximate the model locally, to a $U(n-k)$ gauge theory fibered over $\mathbb{P}^{n(n-k)-1}$, where the homogeneous coordinates of the $\mathbb{P}^{n(n-k)-1}$ are the $\phi$ fields. 
The $p$ fields are identified with fiber coordinates of 
$\mathcal{O}_{\mathbb{P}^{n(n-k)-1}}(-1)^{\oplus n}$ and $x$ fields with 
sections of the trivial bundle over 
$\mathbb{P}^{n(n-k)-1}$. Their $U(n-k)$ representations are:
\begin{equation}\label{localmodelX}
\begin{array}{c|cc}
 & P_{i} & X_{i} \\
\hline
U(n-k) & \mathbf{n-k} &  \overline{\mathbf{n-k}} 
\end{array}
\end{equation}
where $i=1,\cdots,n$. The D-term solutions of the model above describe the 
geometry $S^{\oplus n}\rightarrow G(n-k,n)$ where fiber and base is exchanged 
when the sign of $\zeta_{1}$ flips. This model is called a Grassmannian flop or 'glop' in \cite{donovan2014window,donovan2013grassmannian}. This model was also analyzed, from the perspective of GLSM and hemisphere partition function convergence in \cite{hori2013exact}, for the case of the gauge group being $U(2)$. The result is particularly simple and we can generalize the result in \cite{hori2013exact} (essentially just a re-derivation) to the $U(n-k)$ case that concern us here. The Coulomb branch singularity of the model (\ref{localmodelX}) occurs at $\zeta_{1}=0$. Then, in order to study the absolute convergence of $Z_{\cB}(t)$, near the singularity i.e. at $\zeta_{1}=0$, it suffices to use a real contour: $\sigma_{2}\equiv 0$. This reduction to the problem to a one-parameter case, justifies this choice of contour. The discriminant becomes just a point (at $\zeta_{1}=0$ in this case) and admissible contours can be straightforwadly written for $|\zeta_{1}|>0$, as we argued above, for any B-brane. Then, the grade restriction rule must arise at a contour interpolating between $\Im\sigma_{1}>0$ and $\Im \sigma_{1}<0$, therefore a natural choice is the contour satisfying $\Im\sigma_{1}=0$. Then, at this loci, (\ref{Aq}) reduces to
\begin{equation}\label{Aqreal}
\frac{1}{2\pi}A_q(\nu)=\frac{\theta_1}{2\pi}\sum_{\alpha=1}^{n-k}\nu_{\alpha} + \sum_{\alpha=1}^{n-k}q^{\alpha}\nu_{\alpha}-\frac{n}{2}
\sum_{\alpha=1}^{n-k}|\nu_{\alpha}|+\frac{1}{2} \sum_ { 
\alpha<\beta}|\nu_{\alpha}-\nu_{\beta}|.
\end{equation}
$q^\alpha$ and $\nu_{\alpha}$, $(\alpha=1,\ldots,n-k)$ denote the 
weights and the components of $\sigma_{\alpha}$, $(\alpha=1,2)$ in the chosen basis, respectively. We need to impose that (\ref{Aqreal}) remains strictly negative on all asymptotic directions in $\sigma$. Due to the symmetries of (\ref{Aqreal}) this is straightforward to analyze. In each asymptotic direction some components of $\nu$ will grow, in absolute value, faster than 
others. This amounts to set an ordering of growth between the 
$|\nu_{\alpha}|$'s. Suppose for instance that the ordering is given by $|\nu_{n-k}|<|\nu_{n-k-1}|<\cdots<|\nu_{1}|$, then 
asymptotically
\begin{equation}
\sum_ { 
\alpha<\beta}|\nu_{\alpha}-\nu_{\beta}|\sim 
\sum_{\alpha=1}^{n-k}(n-k-\alpha)|\nu_{1}|
\end{equation}
Then, we can establish the bound:
\begin{equation}\label{Aqbounded}
\begin{aligned}
\frac{1}{2\pi}A_q(\nu_{1})\leq&\frac{\theta_1}{2\pi}\sum_{\alpha=1}^{n-k}
\nu_{ \alpha} + \sum_{\alpha=1}^{n-k}q^{\alpha}\nu_{\alpha}-\frac{n}{2}
\sum_{\alpha=1}^{n-k}|\nu_{\alpha}|+\frac{n-k-1}{2} \sum_ { 
\alpha=1}^{n-k}|\nu_{\alpha}|.
\\
=& \sum_{\alpha=1}^{n-k}\ \left[\left(\frac{\theta_1}{2\pi}+ 
q^\alpha\right)\operatorname{sign}(\nu_\alpha)-\frac{k+1}{2}\right]
|\nu_\alpha|
\end{aligned}
\end{equation}
and so, asymptocically, $A_{q}(\nu)$ coincides with the bound as 
$|\nu_{\alpha}|\rightarrow\infty$, for a particular $\alpha$. Thus we 
deduce the grade restriction rule
\begin{equation}\label{grrnk}
-\frac{k+1}{2}< \frac{\theta_1}{2\pi}+q^\alpha < \frac{k+1}{2}, \qquad 
\text{For all \ }\alpha
\end{equation}
This means that the set of charges $q^{\alpha}$, $\alpha=1,\ldots,k$ lies in a 
(hyper)cube of size $k$, an shifting $\theta_{1}$ by $2\pi$ shifts the position 
of this cube along the diagonal direction $(1,\ldots,1)\in\mathbb{Z}^{n-k}$. 
Note that, for the case $(k,n)=(2,4)$ i.e. the GN model, the rule (\ref{grrnk}) 
is completely consistent with the position of the discriminant (\ref{desc}). 
Namely, if we set $z=0$ (\ref{desc}), the only solution is $w=-1$ signaling a singularity at $\theta_{1}=\pi$ mod $2\pi$, which is consistent with (\ref{grrnk}) (and also the window found in 
\cite{donovan2014window,donovan2013grassmannian}). Let us use the notation $(\lambda,q,q^{0})$ as in (\ref{repnotationf}) to denote a representation of $U(1)\times U(n-k)$ (so, in this case, $q$ denotes tensor by the representation $\mathrm{det}^{q}(\mathbf{n-k})$, $q\in\mathbb{Z}$). Then, the grade restriction rule (\ref{grrnk}) is equivalent to take the Young tableaux $\lambda$ within a $(n-k)\times k$ box $\Lambda_{(n-k)\times k}$. We will then abuse notation and denote by $\cW_{\lambda(q)}(q^{0})$ indistinctly the sheaf in $D^{b}Coh(\mathbb{P}^{n(n-k)-1}\times G(n-k,n))$ and its corresponding GLSM B-brane. The sheaf corresponds to $\Sigma_{\lambda}S$ tensored by the corresponding line bundles. Because these sheaves form a full exceptional 
collection of $D^{b}Coh(\mathbb{P}^{n(n-k)-1}\times G(n-k,n))$, they will be our building blocks for complexes in $D^{b}Coh(X_{A})$ i.e., to every object in $D^{b}Coh(X_{A})$ we can always find a quasi-isomorphic one whose factors are free sheaves of the form $\cW_\lambda(q)(q^{0})$, restricted to $X_{A}$. 
Indeed, the sheaves on the cube restriction rule (\ref{grrnk}) coincides with Kapranov's exceptional collection for $Gr(n-k,n)$ \cite{kapranov1988derived}. Denote the window category on the $X$ phase boundary by $\mathbb W_X(l)$ as:
\begin{equation}
\begin{gathered}
\mathbb W_X(l):=\langle \cW_{\lambda(l)}(q^0) \rangle,\quad q^{0}, l\in\Z,
\\
\lambda=\big(k\geq\lambda_1\geq\cdots\geq\lambda_{n-k}\geq0\big)\in\Lambda_{(n-k)\times k},
\end{gathered}
\end{equation}
which corresponding to choosing $\theta_{1}$ in the open interval
\begin{equation}
\theta_1\in ( -(k+1)\pi-2\pi l,-(k-1)\pi-2\pi l ).
\end{equation}
We may also represent the Young diagrams in the window $\mathbb 
W_X(l)$ coming from shifting all $q^\alpha$ by $l$ by adding the vector
$l\times(1,\ldots,1)$ to $\lambda=(\lambda_1,\cdots,\lambda_{n-k})$ and 
allowing the $\lambda_{\alpha}$'s to be negative. Define
\begin{equation}
\mathbb W_{X}:= \mathbb{W}_{X}(0),
\end{equation}
for simplicity. In particular, the rule (\ref{grrnk}) recovers the 
abelian case when $n-k=1$. At the $Y$ phase boundary, we denote the window 
category likewise by $\mathbb W_Y(l)$ and the abelian window according to 
(\ref{WY}) is simply given by
\begin{equation}
\mathbb W_Y(l)=\langle \cW_\lambda(l),\cdots,\cW_\lambda((n-k)n-1+l) \rangle, 
\quad l\in\mathbb Z.
\end{equation}
for
\begin{equation}
\theta_0\in ( -n(n-k)\pi-2\pi l,-(n(n-k)+2)\pi-2\pi l ).
\end{equation}
As a conclusion, the window and its shift on each phase boundary of $X_A$ in PAX model is given by:
\begin{itemize}
\item{determinantal quintic: $k=4,n=5$, $q\in\Z$}
\begin{equation}
\begin{gathered}
\mathbb W_X = \langle \cW(q,0),\cdots,\cW(q,4) \rangle
\\
\mathbb W_Y = \langle \cW(-4,q),\cdots,\cW(0,q) \rangle
\\
\\
\mathbb W_X(1) = \langle \cW(q,1),\cdots,\cW(q,5) \rangle
\\
\mathbb W_Y(-1) = \langle \cW(-5,q),\cdots,\cW(-1,q) \rangle
\end{gathered}
\end{equation}
\item{GN Calabi-Yau: $k=2,n=4$, $q\in\Z,\ \lambda=(\lambda_1\geq\lambda_2\geq0)$}
\begin{equation}
\begin{gathered}
\mathbb W_X = \left\langle \cW_{\cdot}(q),\cW_{\yng(1)}(q),\cW_{\yng(2)}(q),\cW_{\yng(1,1)}(q),\cW_{\yng(2,1)}(q),\cW_{\yng(2,2)}(q) \right\rangle
\\
\mathbb W_Y = \langle \cW_\lambda(-7),\cdots,\cW_\lambda(0) \rangle
\\
\\
\mathbb W_X(1) = \left\langle \cW_{\yng(1,1)}(q),\cW_{\yng(2,1)}(q),\cW_{\yng(3,1)}(q),\cW_{\yng(2,2)}(q),\cW_{\yng(3,2)}(q),\cW_{\yng(3,3)}(q) \right\rangle
\\
\mathbb W_Y(-1) = \langle \cW_\lambda(-8),\cdots,\cW_\lambda(-1) \rangle
\end{gathered}
\end{equation}
\end{itemize}

\section{\label{sec:section5} Application: Monodromy around singular divisors}

\subsection{\label{subsec:monwin} Monodromy from window categories and 
discriminants}

\begin{figure}[h]
\centering
\includegraphics[scale=0.3]{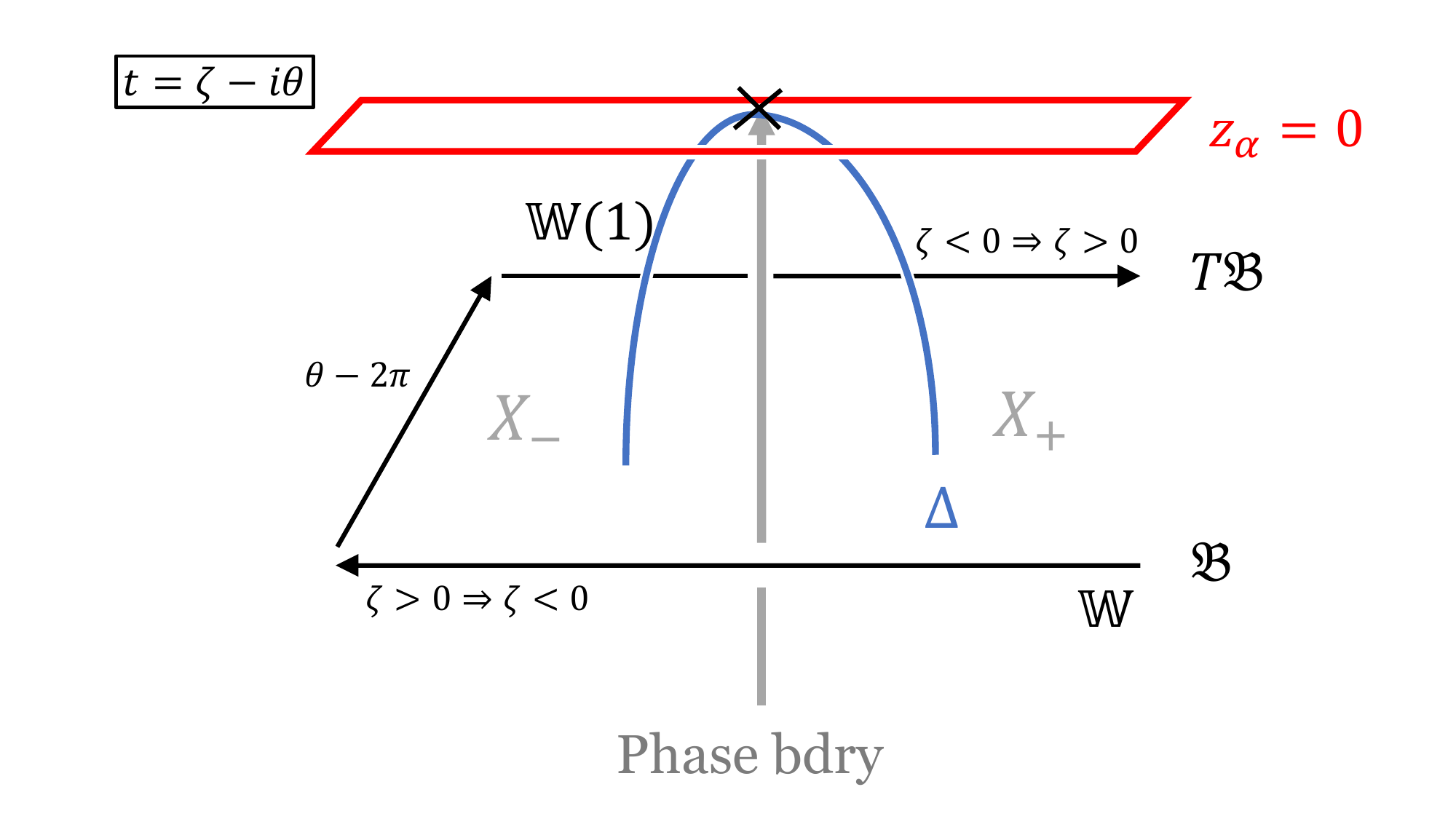}
\caption{Illustration of monodromy along K\"ahler singularity by brane 
transport 
proposal, whose path winds the tangential intersection between discriminant $\Delta$ and asymptotic plane $z_\alpha=0$}\label{monsketch}
\end{figure}

The main premise of the concept of B-brane transport has its origin on the fact 
that, given a $\mathcal{N}=(2,2)$ SCFT, marginal deformations 
constructed by $(a,c)$ operators are $\mathbf{Q}_{B}$-exact. Therefore, the 
space/category of B-branes on a SCFT will remain unaffected under such 
deformations. In the UV theory, described by the GLSM in the class of examples 
that concern us, these marginal deformations are implemented by deformations 
of the FI-theta parameters $t$. The FI-theta parameter space, when described 
by coordinates $e^{t}$, takes the form $(\mathbb{C}^{*})^{r}\setminus 
\Delta$, with the discriminant $\Delta$ given by some closed divisor in 
$(\mathbb{C}^{*})^{r}$. The discriminant $\Delta$ was analyzed thoroughly for the case of abelian GLSMs, in the mathematics literature in 
\cite{gelfanddiscriminants}, using the language of complete intersections in toric varieties. These constructions have a direct interpretation in terms of mirror symmetry for CYs given by complete intersections in toric varieties 
\cite{batyrev1994dual,batyrev1994calabi}. From a GLSM point of view, the loci $\Delta$ is the subspace in the FI-theta parameter space, where the theory is ill-defined due to the appearance of non-compact mixed Coulomb-Higgs branches. Even though $\Delta$ can be defined more or less straightforwardly in the case of abelian GLSMs, the systematic result for the general (nonabelian) case remains an 
open problem. In either case, $(\mathbb{C}^{*})^{r}\setminus 
\Delta$ gets a chamber structure, when projected to the $\mathrm{Re}(t)$-space with each chamber describing a phase of the GLSM. For the non-anomalous GLSMs that concern us, each phase has a well defined IR fixed point which can be described by a NLSM. Moreover, when $r=2$, the asymptotic direction along a phase boundary we can take WLOG to be $e^{-t_{\alpha}}=0$ for some specific $\alpha$. Then, locally, the FI-theta parameter space, when intersected with the plane $e^{-t_{\alpha}}=0$, looks like a punctured plane $\mathbb{C}^{*}$ with some points removed, corresponding to $\Delta$. We interpret the windows $\mathbb{W}(l)$ as a grade restriction rule for this local 1-parameter model. The intersection  
$\{e^{-t_{\alpha}}=0\}\cap \Delta$, in general can be given by more than one point. In the cases at hand is given by a single point, but we cannot discard the former case to occur. Then, the interpretation of $\mathbb{W}(l)$ is taken to be associated to the path going around $\Delta$ rather than within $\Delta$. More precisely, if $\Delta$ separates two chambers $I$ and $II$ with their IR B-branes categories given by $\mathcal{C}_{I}$ and $\mathcal{C}_{II}$, respectively, the equivalence corresponding to a path 
joining the a point in both chambers is given by
\begin{equation}\label{functorphases}
\mathcal{T}_{l}:\mathcal{C}_{I}\longrightarrow \mathbb{W}(l) 
\longrightarrow\mathcal{C}_{II}
\end{equation}
where each arrow in (\ref{functorphases}) is interpreted as an equivalence between categories and the composition, denoted $\mathcal{T}_{l}$ is an equivalence between triangulated categories. We get a family of equivalences label by $l$, corresponding to the integer part of $\frac{\theta_{\alpha}}{2\pi}$ The construction of $\mathcal{T}_{l}$ is very explicit using the GLSM interpretation (see \cite{herbst2008phases} for the abelian case and \cite{eager2017beijing,Guo:2023xak,GuoRomoSmith,EHKR} for 
nonabelian examples). We can also consider paths that surround $\Delta$ i.e. a closed path with base point at an specific phase, for example phase $I$, then, we have the following composition of equivalences
\begin{equation}\label{functormonodromy}
\mathcal{T}_{l,l+1}:\mathcal{C}_{I}\longrightarrow \mathbb{W}(l)\rightarrow 
\mathcal{C}_{II}\rightarrow \mathbb{W}(l+1)
\longrightarrow\mathcal{C}_{I},
\end{equation}
where again each arrow in (\ref{functormonodromy}) is interpreted as an equivalence. We remark that the functor $\mathcal{T}_{l,l+1}$ does not depend on $l$ (we can generalize it to $\mathcal{T}_{l,l'+1}$, then it depends on $l-l'$), so we can drop it from its definition. More importantly, being a composition of equivalences, $\mathcal{T}_{l,l+1}$ is an autoequivalence of 
$\mathcal{C}_{I}$: $\mathcal{T}_{l,l+1}\in \mathrm{Aut}(\mathcal{C}_{I})$. We 
illustrate this schematically in fig. \ref{monsketch}

The implementation of the equivalences in (\ref{functorphases}) and 
(\ref{functormonodromy}) is done through the objects in $MF_{G}(W)$ that are usually termed as \emph{empty branes}. These are objects 
$\mathcal{E}_{I}\in MF_{G}(W)$ that RG-flow to null-homotopic objects in a given phase, for instance $\mathcal{C}_{I}$, but not on other phases. The projection $\pi_{I}:MF_{G}(W)\rightarrow \mathcal{C}_{I}$ is not an equivalence, however, the restriction $\pi_{I}:\mathbb{W}(l)\rightarrow \mathcal{C}_{I}$ is. Therefore, the object $\mathcal{E}_{I}$ cannot satisfy the grade restriction conditions, i.e. it does not belong to any window category 
$\mathbb{W}(l)$, $l\in\mathbb{Z}$. Given an object 
$\mathcal{B}\in MF_{G}(W)$, we can use empty branes such as $\mathcal{E}_{I}$ and its shifts and twists (which will be also empty branes)to construct a bound state $\mathcal{B}'$, via the cone operation in triangulated categories, between them and $\mathcal{B}$. Due to the fact that 
$\mathcal{E}_{I} \not\in\mathbb{W}(l)$ we can build a bound state 
$\mathcal{B}'$ satisfying 
\begin{equation}
\pi_{I}(\mathcal{B}')\cong \pi_{I}(\mathcal{B})\qquad 
\mathcal{B}'\in\mathbb{W}(l)
\end{equation}
where $\cong$ denotes quasi-isomorphism in $\mathcal{C}_{I}$, even if 
$\mathcal{B}$ do not belong to $\mathbb{W}(l)$. The B-brane 
charges are described by the Grothendieck group $K_{0}(\mathcal{C})$, which in 
the case $\mathcal{C}=D^{b}Coh(X)$ is the Grothendieck group generated by 
isomorphism classes of holomorphic vector bundles over $X$, which we will 
denote simply by $K(X)$. Then an empty brane $\mathcal{E}_{I}$ satisfies
\begin{equation}
\pi_{I}(\mathcal{E}_{I})=0\in K_{0}(\mathcal{C}_{I}),\qquad 
Z_{\mathcal{E}_{I}}|_{I}\equiv 0,
\end{equation}
where $Z_{\mathcal{E}_{I}}|_{I}$ denotes the hemisphere partition function 
restricted to the phase $I$. Likewise we can 
use an empty brane to implement the equivalences $\mathbb{W}(l)\rightarrow 
\mathcal{C}_{II}\rightarrow \mathbb{W}(l+1)$ (in this notation that will be 
a B-brane $\mathcal{E}_{II}\in MF_{G}(W)$). Then, the net effect of 
(\ref{functormonodromy}) is expected to be an autoequivalence of 
$\mathcal{C}_{I}$ given by an spherical functor 
\cite{rouquier2006categorification,anno2007spherical}. When the intersection 
between the divisor $e^{-t_{\alpha}}=0$ and $\Delta$ is transversal and the 
phase $I$ is geometrical, then, in general, we expect that the autoequivalence 
$\mathcal{T}$ is something of the form of an EZ-spherical twist, originally 
studied by Horja in \cite{horja2001derived}, and related to the B-brane supported on a sub-variety $E\subset X$ that collapses to another sub-variety $Z\subset X$. Actually, the type of spherical functor that will be relevant for us, will be the ones of the 
form\footnote{This corresponds to the special case 
of an EZ-spherical functor when $Z=\mathrm{pt}.$\cite{huybrechts2006fourier} and so the spherical object is $\mathcal{O}_{E}$. The mirror of these kind of functors where originally studied in \cite{seidel2000braid}.}
\begin{equation}\label{sphfunctorE}
T_E(\cB)=\Cone\left(\oplus_i \Hom(E,\cB[i])\otimes E[-i]\rightarrow\cB 
\right)\qquad \mathcal{B}\in\mathcal{C}_{I}=D^{b}Coh(X).
\end{equation}
where $E\in D^{b}Coh(X)$ denotes some spherical object\footnote{A spherical object is an object that satisfies $\mathrm{Hom}(E,E[i])=\mathbb{C}$ only if $i=0$ or $i=\mathrm{dim}X$ and $E\otimes K_{X}\cong E$, where $K_{X}$ is the canonical bundle.}. Let us review a few properties of the functors $T_{E}$ 
that will become useful later. At the cohomology level, the action of $T_{E}$ is given by a reflexive action \cite{huybrechts2006fourier} 
\begin{equation}
\Ch(\cT_E(\cB))=\Ch\cB-\langle v(E),\cB \rangle v(E)
\end{equation}
where $v(E):=\Ch(E)\hat\Gamma_X$ is called the Mukai vector of $E$, and $\langle-,-\rangle$ the Mukai pairing:
\begin{equation}
\langle \cB,\cB'\rangle:=\int_X \Ch(\cB)^\vee\Ch(\cB').
\end{equation}
Thus the action of $T_{E}$ on the hemisphere partition function is given by
\begin{equation}
T_{E}(Z_\cB)=Z_\cB-\chi(E,\cB)Z_E,\label{EZ}
\end{equation}
where $\chi(E,\cB)$ is the Euler characteristic:
\begin{equation}
\chi(E,\cB)=\sum_{k}(-1)^{k}\mathrm{dim}\mathrm{Ext}^{k}(E,\cB)
\end{equation}
We remark that the quantity $\chi(E,\cB)$ can be physically described by the open Witten index i.e. the partition function on a cylinder with boundary conditions given by the B-branes corresponding to $E$ and $\cB$. A localization 
formula for it (and so, valid in any phase) was derived in \cite{hori2013exact}:
\begin{equation}
\chi(E,\cB)=\frac{1}{|W_G|}\int_\Gamma\frac{d^{l_{G}}u}{
(2\pi 
i)^{l_{G}}}\frac{\prod_{\alpha>0}\left(2\sinh\frac{\alpha(u)}{2}
\right)^2 }
{\prod_i\left(2\sinh\frac{Q_i(u)}{2}\right)}f_E\left(\frac{u}{2\pi}
\right)f_{\cB}\left(\frac{-u}{2\pi}\right).\label{annu}
\end{equation}
The integration contour $\Gamma$ was not determined, in full generality, in \cite{hori2013exact}, but for the examples at hand, it can be taken as a torus encircling the origin (a general feature for models with weakly coupled geometric phases). The monodromy $\mathcal{T}$ we will encounter will not be in general (or ever, in the linear PAX examples we consider) of the form 
(\ref{sphfunctorE}), the reason being that the intersection between the discriminant components $\{e^{-t_{\alpha}}=0\}$ and $\Delta$ is not transversal. However, the monodromy actions we will found at at the phase boundaries $X$ and $Y$ are intimately related with the ones of the form (\ref{sphfunctorE}) as we will proceed to explain. The reason being mainly because when taking a path around a point that is not a transversal intersection point, we are going around 
'multiple components' of $\Delta$ at the same time. To be more precise, from a mirror perspective where the relevant moduli space is the complex structure moduli (or marginal deformations coming from the $(c,c)$ ring), in order to compute the monodromy of solutions around the singularities of the Picard-Fuchs equations will be necessary to resolve the non-transversal intersection by blowing up, until the complex moduli space becomes a smooth variety, as was done in \cite{Candelas:1993dm,Candelas:1994hw}. However, this smoothing procedure obscures the geometric meaning between monodromy, paths and discriminant and it is not the approach we will take here. We will rather use the approach of \cite{aspinwall2001some}: we will resolve the intersection $\{e^{-t_{\alpha}}=0\}\cap \Delta$ by taking a small 3-sphere around the 
intersection point. In general, suppose we have two divisors $D_f$ and $D_g$, defined by the polynomial equations $f(z_1,z_2)=0$ and $g(z_1,z_2)=0$ respectively, and WLOG assume they intersect at 
$(0,0)$. Then we define the 3-sphere $S^3_{\varepsilon}$, centered at their intersection point:
\begin{equation}
S^{3}_{\varepsilon}:=\{(z_{1},z_{2})\in\mathbb{C}^{2}
: |z_1|^2+|z_2|^2=\varepsilon^2\}
\end{equation}
for sufficiently small $\varepsilon$. Then $D_f\cap S^3_\varepsilon$ and $D_g\cap S^3_\varepsilon$ are real dimension one curves winding 
each other in $S^3_\varepsilon$ i.e. their union forms a link in 
$S^3_\varepsilon$ which we can visualize as a link $\mathcal{L}$ in 
$\mathbb R^3$. 

Under Wirtinger representation (see for example\cite{stillwell1993classical}), we can study the fundamental group of the link complement $\pi_{1}(S^3_\varepsilon\setminus\mathcal{L})$, using as generators small loops around the components of $\mathcal{L}$. In our context these links components are components of $\Delta$ or $\{e^{-t_{\alpha}}=0\}$ and the single loops that generates $\pi_{1}(S^3_\varepsilon\setminus\mathcal{L})$ can be interpreted as 
loops corresponding to spherical twists of the form $T_{E}$ as in 
(\ref{sphfunctorE}) when it corresponds to a loop around a component of $\Delta$, or of the form\footnote{This is also an example of a spherical twist under the general definition of 
\cite{rouquier2006categorification,anno2007spherical}.} $\otimes 
\mathcal(D_{\alpha})$ when it is a loop around $\{e^{-t_{\alpha}}=0\}$. The path defining $\mathcal{T}$ will correspond to a particular path in $S^3_\varepsilon\setminus\mathcal{L}$ so, its decomposition in terms of the generators of $\pi_{1}(S^3_\varepsilon\setminus\mathcal{L})$  will provide a very nontrivial relation between $\mathcal{T}$, and the spherical functors $T_{E}$, $\otimes 
\mathcal(D_{\alpha})$. This serves as a very strong test on the window categories/grade restriction rules found in section \ref{sec:section4}. Last but not least, we remark that in the nonabelian GN model, since we do not have the full machinery of CY complete intersections on toric varieties, we cannot determine the objects $E$, or more precisely the functors $T_{E}$ corresponding 
to loops around independent components of $\Delta$, from first principles. However, we make an educated guess that is consistent with all results that can be computed independently, namely window categories and localization computations. In the GN case, we can then frame this as a solid conjecture about the decomposition of $\mathcal{T}$ into spherical functors $T_{E}$ and twists.

\subsection{Empty branes and window equivalences}

In order to perform the monodromy computation, as pointed out above, a key ingredient is to determine the empty brane for each phase in PAX model. In abelian GLSMs, the classical solution of the D-terms equations, for a fixed value of the FI parameters $\zeta$ is a toric variety which, topologially, can be written as $(\mathbb{C}^{N}\setminus Z_{\zeta})/(\mathbb{C}^{*})^{k}$. The 
closed set $Z_{\zeta}$ corresponds to the unstable loci of the symplectic quotient implemented by the D-terms. Then a natural candidate for the empty branes exists for $\zeta$ in the interior of a phase, namely the B-brane supported on $Z_{\zeta}$. This B-brane can be lifted to a an object in $MF_{G}(W)$ \cite{herbst2008phases}. In the linear PAX models, the solution of the D-terms will take the form $\mathbb{P}^{n(n-k)-1}\times G(n-k,n)$, reflecting on the fact that the model is not abelian in general. Therefore, we cannot just use $Z_{\zeta}$ to construct empty branes, as we will see below. 
Let us first construct the empty branes relevant to the $Y$ phase boundary. In the $X_{A}$ phase, the ambient $\mathbb{P}^{n(n-k)-1}$ is spanned by the $\phi$ fields, hence clearly we have a distinguished unstable loci of the D-terms, given by $\phi_a=0$, for all $a=1,\ldots, n(n-k)$. A free resolution of this loci, in the $G$-equivariant derived category of the linear space $\mathcal{V}$ spanned by the chiral fields is straightforward to find, using a 
Koszul resolution:
\begin{equation}
\cK_\phi:\quad 0\rightarrow 
\cW(-n(n-k),0)\xrightarrow{\phi}\cW(-n(n-k)+1,0)^{\oplus 
n(n-k)}\xrightarrow{\phi}\cdots\xrightarrow{\phi}\cW(-1,0)^{\oplus 
n(n-k)}\xrightarrow{\phi}\cW(0,0)\rightarrow0.
\end{equation}
As a GLSM B-brane the complex $\cK_\phi$ corresponds to an object of the category $MF_{G}(W\equiv 0)$. However it is easily lifted to an object in $MF_{G}(W)$ by adding the appropriate terms to the tachyon profile corresponding to $\cK_\phi$, explicitly:
\begin{equation}
Q_{\cK_\phi}=\sum_a\left(p\cdot A^a\cdot x 
\right)\bar{\eta}_a+\phi_a\eta^a\in MF_{G}(W).
\end{equation}
with brane factor (upon taking a trivial Clifford vacuum)
\begin{equation}
f_{\cK_\phi}(\sigma)=\left( 1-e^{-2\pi\sigma_0} \right)^{n(n-k)}.
\end{equation}
Let us remark, that by the D-term equations (\ref{Dterm}), the boundary potential evaluated at $Q_{\cK_\phi}$ is given by
\begin{equation}
\begin{aligned}
 &\left\lbrace Q_{\cK_\phi},{Q^\dagger_{\cK_\phi}} \right\rbrace
 \\
=&\sum_a \left(|pA^ax|^2+|\phi_a|^2\right)
\\
=&\left\lbrace\begin{array}{l}
\sum_a |pA^ax|^2+\sum_{i,\alpha}|p_i^\alpha|^2+\zeta_0
\\
\text{or}
\\
\sum_a |pA^ax|^2+\sum_{i,\alpha}|x_{i\alpha}|^2+\zeta_0+(n-k) \zeta_1
\end{array}\right.
\end{aligned}
\end{equation}
Thus $Q_{\cK_\phi}$ boundary potential is non-vanishing in the $X_{A}$ and $X_{A^{T}}$ phases where $\zeta_0+(n-k) \zeta_1>0$ and $\zeta_0>0$, making this B-brane empty at the IR fixed point: it will RG-flow to an exact complex in $D^{b}Coh(X_{A})$ or  $D^{b}Coh(X_{A^{T}})$. In particular its central charge 
identically vanishes \cite{herbst2008phases}, which can be checked 
explicitly since the zeroes of $f_{\cK_\phi}(\sigma)$ cancels all the relevant poles coming from the gamma function contribution coming from the $\phi$ fields. The empty brane $Q_{\cK_\phi}$ (and its twists and shifts) can be used to restrict any B-brane in the $X_{A}$ phase to a window category $\mathbb W_Y(l)$, without changing its IR image. We would like to find now an empty brane that allows us to restrict any B-brane in the $Y_{A}$ phase to the window category $\mathbb W_Y(l)$ in the same fashion. Consider the matrix factorization of $Q_{\cO_X}$ studied in section 
\ref{sec:section3}. Its boundary potential is given by:
\begin{equation}
\begin{aligned}
 &\left \lbrace Q_{\cO_X},{Q^\dagger_{\cO_X}}  \right\rbrace
\\
=&\sum_{\alpha,i}\left(|p_1^\alpha|^2+|A_i(\phi)x_\alpha|^2\right)
\\
=&\left\lbrace\begin{array}{l}
\sum_a|\phi_a|^2+\sum_{i,\alpha}|A_i(\phi)x_\alpha|^2-\zeta_0  
\\
\text{or}
\\
\sum_{i,\alpha}\left(|x_{i\alpha}|^2+|A_i(\phi)x_\alpha|^2\right)-(n-k)\zeta_1
\end{array} \right.
\end{aligned}
\end{equation}
Thus similarly, $Q_{\cO_X}$ is empty in the $Y_A$ phase, where 
$\zeta_0,\zeta_1<0$. This brane can be used then to restrict general B-branes to the $\mathbb W_Y(l)$ in the $Y_A$ phase. In order to restrict B-branes to the window category $\mathbb W_{X}(l)$. This is when we need to make a distinction between abelian and nonabelian cases. In the abelian PAX model, the resolution of the loci $x\equiv 0$, given by:
\begin{equation}
\begin{gathered}
\cK_x:\quad 
\cW(0,5)\rightarrow\cW(0,4)^{\oplus5}\rightarrow\cdots\rightarrow\cW(0,1)^{
\oplus5}\rightarrow\cW(0,0),
\end{gathered}
\end{equation}
its lift to $Q_{\cK_x}$ is evident and this empty brane can be shown to be enough to restrict any B-brane to $\mathbb{W}_{X}(l)$. However, for the nonabelian model, even though $Q_{\cK_x}$ (and all its twists and shifts) is a perfectly well defined empty brane, it is easy to see it is not enough for restricting any B-brane to $\mathbb{W}_{X}(l)$. Essentially, because now we have 
more than one weight to restrict. Indeed, the $U(2)$ D-term equation implies that for $\zeta_1>0$, for example, the unstable loci corresponds to the points $x$ where the matrix $x_i^\alpha$ has rank one or zero. Thus the empty brane must have support in a variety that is not even a complete intersection, hence a Koszul type free resolution is not going to work. The free resolution of a determinantal variety is called an Eagon-Northcott complex and it was first studied in \cite{eagon1962ideals}. For instance, let $\cO_{\det}$ the sheaf of the determinantal variety $Z(A,k-1)$ 
where $A$ is a $k\times n$ matrix, or a section of $\Hom(S,V)$. Then the Eagon-Northcott resolution of $\cO_{\det}$ is given by the exact sequence:
\begin{equation}
\wedge^NV\otimes\wedge^kS\otimes\Sym^{N-k}S\rightarrow\cdots\rightarrow\wedge^{
k+1}V\otimes\wedge^kS\otimes S\rightarrow\wedge^kV\otimes\wedge^k S\rightarrow 
\C\rightarrow \cO_{\det}
\end{equation}
Thus empty branes used for grade restriction to $\mathbb{W}_{X}(l)$ will correpond to resolutions of the unstable loci inside $Gr(2,4)$, which will be given by the variety $Z(x,1)$. However, we will also need to consider the resolution of bundles over this loci. This problem was already addressed in \cite{donovan2013grassmannian}. For $G(2,4)$, there are three exact sequences 
that can be lifted to empty branes, namely
\begin{equation}
\begin{gathered}
\cE^{(1)}_x:\quad \cW_{\yng(3,1)}(0)\rightarrow 
\cW_{\yng(2,1)}(0)^{\oplus4}\rightarrow\cW_{\yng(1,1)}(0)^{\oplus6}
\rightarrow\cW_\cdot(0)
\\
\cE^{(2)}_x:\quad \cW_{\yng(3,2)}(0)\rightarrow 
\cW_{\yng(2,2)}(0)^{\oplus4}\rightarrow\cW_{\yng(1,1)}(0)^{\oplus4}
\rightarrow\cW_{\yng(1)}(0)
\\
\cE^{(3)}_x:\quad \cW_{\yng(3,3)}(0)\rightarrow 
\cW_{\yng(2,2)}(0)^{\oplus6}\rightarrow\cW_{\yng(2,1)}(0)^{\oplus4}
\rightarrow\cW_{\yng(2)}(0)
\end{gathered}\label{Ex}
\end{equation}
exchanging $x$ by $p$ is straightforward and the resulting complexes are given by:
\begin{equation}\label{Epcomplexes}
\begin{gathered}
\cE^{(1)}_p:\quad 
\cW_{\overline{\yng(3,1)}}(4)\rightarrow\cW_{\overline{\yng(2,1)}}(3)^{\oplus4}
\rightarrow\cW_{\overline{\yng(1,1)}}(2)^{\oplus6}\rightarrow\cW_\cdot(0)
\\
\cE^{(2)}_p:\quad 
\cW_{\overline{\yng(3,2)}}(5)\rightarrow\cW_{\overline{\yng(2,2)}}(4)^{\oplus4}
\rightarrow\cW_{\overline{\yng(1,1)}}(2)^{\oplus4}\rightarrow\cW_{\overline{
\yng(1)}}(1)
\\
\cE^{(3)}_p:\quad 
\cW_{\overline{\yng(3,3)}}(6)\rightarrow\cW_{\overline{\yng(2,2)}}(4)^{\oplus6}
\rightarrow\cW_{\overline{\yng(2,1)}}(3)^{\oplus4}\rightarrow\cW_{\overline{
\yng(2)}}(2)
\end{gathered}
\end{equation}
where the $\overline{( \ \ )}$ above the Young diagrams denotes the conjugate representations as it should be since the maps depend on the $p$-fields. In \cite{donovan2013grassmannian} a general formulation for building these complexes is given, for any Grassmannian. We present in appendix \ref{sec:proof} an argument for the existence of the lift of these complexes to $MF_{G}(W)$, adapted from one presented originally in \cite{EHKR}. For practical purposes, we only need the complexes (\ref{Ex}) and
(\ref{Epcomplexes}) in order to study the monodromy of the hemisphere partition function and find the autoequivalences associated to them.

We summarize the empty branes on each phase of the relevant examples:
\begin{itemize}
\item Determinantal quintic:
\begin{equation}
\begin{aligned}
X_A:&\ \cK_\phi,\cK_x
\\
X_{A^T}:&\ \cK_\phi,\cO_X
\\
Y_A:&\ \cK_x,\cO_X
\end{aligned}
\end{equation}

\item GN Calabi-Yau:
\begin{equation}
\begin{aligned}
X_A:&\ \cK_\phi,\cE_x
\\
X_{A^T}:&\ \cK_\phi,\cE_p
\\
Y_A:&\ \cE_x,\cO_X
\end{aligned}
\end{equation}
\end{itemize}
In the process of computing the monodromy around the discriminant, we must map objects from one window category to another. We refer to this process as 'window shifting'. We chose then the theta angles in a way that the window shifting we will be performing corresponds to map objects already belonging to $\mathbb{W}_{Y}$ into  
$\mathbb{W}_{Y}(-1)$ and from $\mathbb{W}_{X}$ to $\mathbb{W}_{X}(1)$ for the $Y$ and $X$ boundary monodromy, respectively. 
These two window shifts can be achieved by using the empty branes to replace certain factors of the complex by taking cones with empty branes. We then summarize, for reference, what are the factors of the complex that should be replaced in the following window shifts (denoted with an arrow $\Rightarrow$):
\begin{itemize}
\item Determinantal quintic:
\begin{equation}
\begin{gathered}
\mathbb W_Y\Rightarrow\mathbb W_Y(-1):\quad \cW(0,q)\mapsto
\\
\left(\cW(-5,q)\rightarrow\cW(-4,q)^{\oplus5}\rightarrow\cdots\rightarrow\cW(-1,
q)^{\oplus5}\right)
\end{gathered}
\end{equation}
\item GN Calabi-Yau:
\begin{equation}
\begin{gathered}
\mathbb W_X\Rightarrow\mathbb W_X(1):
\\
\cW_\cdot(q)\mapsto\left( 
\cW_{\yng(1,1)}(q-2)^{\oplus6}\rightarrow\cW_{\yng(2,1)}(q-3)^{\oplus{4}}
\rightarrow\cW_{\yng(3,1)}(q-4) \right)
\\
\cW_{\yng(1)}(q)\mapsto \left( 
\cW_{\yng(1,1)}(q-1)^{\oplus4}\rightarrow\cW_{\yng(2,2)}(q-3)^{\oplus{4}}
\rightarrow\cW_{\yng(3,1)}(q-4) \right)
\\
\cW_{\yng(2)}(q)\mapsto \left( 
\cW_{\yng(2,1)}(q-1)^{\oplus4}\rightarrow\cW_{\yng(2,2)}(q-2)^{\oplus{6}}
\rightarrow\cW_{\yng(3,3)}(q-4) \right)
\end{gathered}\label{windowshift}
\end{equation}
\\
\begin{equation}
\begin{gathered}
\mathbb W_Y\Rightarrow\mathbb W_Y(-1):\quad \cW_\lambda(0)\mapsto
\\
\left(\cW_\lambda(-8)\rightarrow\cW_\lambda(-7)^{\oplus8}
\rightarrow\cdots\rightarrow\cW_\lambda(-1)^{\oplus8}\right)
\end{gathered}
\end{equation}
\end{itemize}

\subsection{Monodromies from window shift}

We are now ready to compute the monodromies via brane transport through the window categories. We will consider monodromies around paths with base point 
at the $X_{A}$ phase. Schematically the process of transporting an arbitrary 
B-brane 
$\mathcal{B}\in MF_{G}(W)$ through the $X$ and $Y$ phase boundaries, can be 
written as:
\begin{equation}
\begin{gathered}
T_X:\quad \cB\xrightarrow{\cK_x/\cE_x}[\cB]_{\mathbb W_X}\xrightarrow{\cO_X/\cE_p}[\cB]_{\mathbb W_X(1)}
\\
T_Y:\quad \cB\xrightarrow{\cK_\phi}[\cB]_{\mathbb W_Y}\xrightarrow{\cO_X}[\cB]_{\mathbb W_Y(-1)}
\end{gathered}\label{transportsch}
\end{equation}
where $[\cB]_{\mathbb W}$ denotes the equivalent B-brane, restricted to the 
window category $\mathbb{W}$ upon taking the corresponding cones: the objects 
above the arrows in (\ref{transportsch}) denote the empty branes 
necessary to restrict, via the cone construction, the branes on the left to the 
category on the right. At this point, given any $\mathcal{B}$, computing the 
monodromy as in (\ref{transportsch}) is algorithmic. In order to illustrate the 
process we perform the transport along a loop around the $X$ phase 
boundary for the GN model, for the B-brane $Q_{\cO_X}$. Recall the brane factor 
of $Q_{\cO_X}$: 
\begin{equation}
f_{\cO_X}=\left( 1- e^{2\pi(\sigma_1-\sigma_0)} \right)^4\left( 1- e^{2\pi(\sigma_2-\sigma_0)} \right)^4.
\end{equation}
From it, we can read that such $Q_{\cO_X}$ is not in $\mathbb W_X$ (neither in 
any other window category). Using 
(\ref{Ex}) and their twists, the brane factor of $Q_{\cO_X}$, restricted to 
$\mathbb W_X$ is given by
\begin{equation}
\begin{aligned}
  &f_{[\cO_X]_{\mathbb W_X}}
  \\
=&\left( 1-16\mathrm{q}^{-3}+30\mathrm{q}^{-4}-16\mathrm{q}^{-5} +\mathrm{q}^{-8} \right)
\\
+&\left(-4\mathrm{q}^{-1}+24\mathrm{q}^{-3}-20\mathrm{q}^{-4} -20\mathrm{q}^{-5}+24\mathrm{q}^{-6}-4\mathrm{q}^{-8} \right)f_{\yng(1)}
\\
+&\left( 6\mathrm{q}^{-2}-16\mathrm{q}^{-3}+10\mathrm{q}^{-4}+10\mathrm{q}^{-6}-16\mathrm{q}^{-7}+6\mathrm{q}^{-8} \right)f_{\yng(2)}
\\
+&\left( 10\mathrm{q}^{-2}-90\mathrm{q}^{-4}+160\mathrm{q}^{-5}-90\mathrm{q}^{-6}+10\mathrm{q}^{-8} \right)f_{\yng(1,1)}
\\
+&\left( -20\mathrm{q}^{-3}+60\mathrm{q}^{-4}-40\mathrm{q}^{-5}-40\mathrm{q}^{-6}+60\mathrm{q}^{-7}-20\mathrm{q}^{-8} \right)f_{\yng(2,1)}
\\
+&\left( 20\mathrm{q}^{-4}-80\mathrm{q}^{-5}+120\mathrm{q}^{-6}-80\mathrm{q}^{-7}+20\mathrm{q}^{-8} \right)f_{\yng(2,2)}
\end{aligned}
\end{equation}
where $\mathrm{q}:=\exp(2\pi\sigma_0)$, $f_{\yng(1)}$ and $f_{\yng(1,1)}$ are 
given by (\ref{rep1}) and (\ref{rep11}), respectively, and
\begin{equation}
\begin{gathered}
f_{\yng(2)}=f^2_{\yng(1)}- f_{\yng(1,1)},\quad f_{\yng(2,1)}=f_{\yng(1,1)}f_{\yng(1)},\quad f_{\yng(2,2)}=f^2_{\yng(1,1)}.
\end{gathered}
\end{equation}
It can be cheked directly that $Z_{[\cO_X]_{\mathbb W_X}}=Z_{\cO_X}$, in the 
$X_{A}$ phase. 
Then one can directly apply the transport rule (\ref{windowshift}) to obtain
\begin{equation}
\begin{aligned}
  &f_{T_X(\cO_X)}
  \\
=&\left( 1-16\mathrm{q}^{-3}+30\mathrm{q}^{-4}-16\mathrm{q}^{-5} +\mathrm{q}^{-8} \right)
\\
 &\times\left( 6\mathrm{q}^{-2}f_{\yng(1,1)}-4\mathrm{q}^{-3}f_{\yng(2,1)}+\mathrm{q}^{-4}f_{\yng(3,1)} \right)
\\
+&\left(-4\mathrm{q}^{-1}+24\mathrm{q}^{-3}-20\mathrm{q}^{-4} -20\mathrm{q}^{-5}+24\mathrm{q}^{-6}-4\mathrm{q}^{-8} \right)
\\
 &\times\left( 4\mathrm{q}^{-1}f_{\yng(1,1)}-4\mathrm{q}^{-3}f_{\yng(2,2)}+\mathrm{q}^{-4}f_{\yng(3,2)} \right)
\\
+&\left( 6\mathrm{q}^{-2}-16\mathrm{q}^{-3}+10\mathrm{q}^{-4}+10\mathrm{q}^{-6}-16\mathrm{q}^{-7}+6\mathrm{q}^{-8} \right)
\\
 &\times\left( 4\mathrm{q}^{-1}f_{\yng(2,1)}-6\mathrm{q}^{-2}f_{\yng(2,2)}+\mathrm{q}^{-4}f_{\yng(3,3)} \right)
\\
+&\left( 10\mathrm{q}^{-2}-90\mathrm{q}^{-4}+160\mathrm{q}^{-5}-90\mathrm{q}^{-6}+10\mathrm{q}^{-8} \right)f_{\yng(1,1)}
\\
+&\left( -20\mathrm{q}^{-3}+60\mathrm{q}^{-4}-40\mathrm{q}^{-5}-40\mathrm{q}^{-6}+60\mathrm{q}^{-7}-20\mathrm{q}^{-8} \right)f_{\yng(2,1)}
\\
+&\left( 20\mathrm{q}^{-4}-80\mathrm{q}^{-5}+120\mathrm{q}^{-6}-80\mathrm{q}^{-7}+20\mathrm{q}^{-8} \right)f_{\yng(2,2)},
\end{aligned}
\end{equation}
where
\begin{equation}
f_{\yng(3,1)}=f_{\yng(1,1)}f_{\yng(2)},\quad f_{\yng(3,2)}=f_{\yng(2,2)}f_{\yng(1)},\quad f_{\yng(3,3)}=f^3_{\yng(1,1)}.
\end{equation}
By evaluating $Z_{T_X(\cO_X)}$, the monodromy result on $K(X)$ is given by
\begin{equation}
Z_{T_X(\cO_X)}=Z_{\cO_X}-4Z_{\cO_{D_\phi}}+120Z_{0}+120Z_{1}-200Z_{\cO_P}
\end{equation}
We can repeat this procedure for the rest of the B-branes in the GN model and 
the determinantal quintic likewise. Finally we obtain the matrices associated 
with the monodromies obtained by the window shift operations 
(\ref{transportsch}):
\begin{itemize}
\item Determinantal quintic:
\begin{equation}
M_{T_X}=\begin{pmatrix}
1 & &&&& 
\\
-5 & 1 & &&&
\\
0&0&1&&&
\\
50 & -25 & -50 & 1 & &
\\
100 & -50 & -100 & 0 & 1 &
\\
-100 & 75 & 150 & -5 & 0 & 1
\end{pmatrix},
\end{equation}
\begin{equation}
M_{T_Y}=\begin{pmatrix}
1 & &&&&
\\
0 & 1 &&&&
\\
-5 & 0 & 1 &&&
\\
100 & -100 & -50 & 1 & &
\\
50 & -50 & -25 & 0 & 1 &
\\
-100 & 150 & 75 & 0 & -5 & 1
\end{pmatrix}.
\end{equation}

\item GN Calabi-Yau:
\begin{equation}
M_{T_X}=\begin{pmatrix}
1 & & & & & 
\\
-4 & 1 & & & &
\\
0 & 0 & 1 & & &
\\
120 & -80 & -80 & 1 & &
\\
120 & -80 & -64 & 0 & 1 &
\\
-200 & 200 & 192 & -4 & 0 & 1
\end{pmatrix},
\end{equation}
\begin{equation}
M_{T_Y}=\begin{pmatrix}
1 & & & & & 
\\
0 & 1 & & & &
\\
-4 & 0 & 1 & & &
\\
96 & -136 & -64 & 1 & &
\\
48 & -64 & -32 & 0 & 1 &
\\
-80 & 168 & 80 & 0 & -4 & 1
\end{pmatrix}.\label{TY}
\end{equation}
\end{itemize}
The monodromy matrices are computed on the basis (\ref{chargeofBB}) and it 
action on the charge vector is given by:
\begin{equation}
T\circ \vec{a}=\vec{a}\cdot M_{T}\qquad 
\vec{a}:=(a^{0},a^{\alpha},a_{\alpha},a_{0})
\end{equation}
Last but not least, we remark that the procedures (\ref{transportsch}) allow to 
determine $T_{X}$ and $T_{Y}$ exactly as functors, i.e. their exact action on 
the objects of $D^{b}Coh(X_{A})$. Here we just present their action on 
$K(X_{A})$ illustrated in the monodromy matrices. We will develop more details 
on the functors  $T_{X}$ and $T_{Y}$ in the next subsection.

\subsection{Decomposition of monodromies}

In this subsection we explain how the previously computed monodromy around the 
$X$ and $Y$ boundaries decomposes in simpler ones, as in the cases\footnote{Further applications of this apporach to the resolution of nontransversal intersection in $\Delta\subset \mathcal{M}_{K}(X)$ have been used in \cite{Cota:2019cjx}, in other abelian models.} analyzed in
\cite{aspinwall2001some}. For this purpose we have to intersect the 
discriminant with a 3-sphere and analyze the resulting link, as we outlined in 
\ref{subsec:monwin}. We begin with 
the abelian model (determinantal quintic) where the massless objects 
associated to each component of the discriminant are easier to identify. For 
the nonabelian example (GN model), there are extra components in the 
discriminant whose associated functors are not straightforward to identify, but 
we have a proposal for them. Due to the complexity of the link of the $X$ 
boundary in the nonabelian case, we only have a proposal for the $Y$ boundary.

\subsubsection{Determinantal quintic}

We start by computing the monodromy matrix for the loop around the divisors $v=0$ and $w=0$ (using the notation of section \ref{sec:section2}). The 
functors are Serre twists
\begin{equation}
T_{(v)}=(-)\otimes \mathcal{O}_{X_{A}}(1,0),\qquad T_{(w)}=(-)\otimes 
\mathcal{O}_{X_{A}}(0,-1),
\end{equation}
respectively. For objects in 
$MF_{G}(W)$, this corresponds to shift the gauge charges of the factors 
$\mathcal{W}(q^{0},q^{1})$ by $q^0\mapsto q^0+1$ and $q^1\mapsto 
q^1-1$ respectively. The corresponding matrices acting on the basis of charges 
(\ref{chargeofBB}) are
\begin{equation}
v\mapsto v e^{2\pi i}:\quad M_{T_{(v)}}:=\begin{pmatrix}
1 & & & & &
\\
1 & 1 & & & &
\\
0 & 0 & 1 & & & &
\\
5 & 5 & 10 & 1 & &
\\
10 & 10 & 10 & 0 & 1 &
\\
0 & 0 & 0 & 1 & 0 & 1
\end{pmatrix}
\end{equation}
\begin{equation}
w\mapsto w e^{2\pi i}:\quad M_{T_{(w)}}=\begin{pmatrix}
1 & & & & &
\\
0 & 1 & & & &
\\
1 & 0 & 1 & & & &
\\
10 & 10 & 10 & 1 & &
\\
5 & 10 & 5 & 0 & 1 &
\\
0 & 0 & 0 & 0 & 1 & 1
\end{pmatrix}
\end{equation}

\begin{figure}[h]
\centering
\includegraphics[scale=0.4]{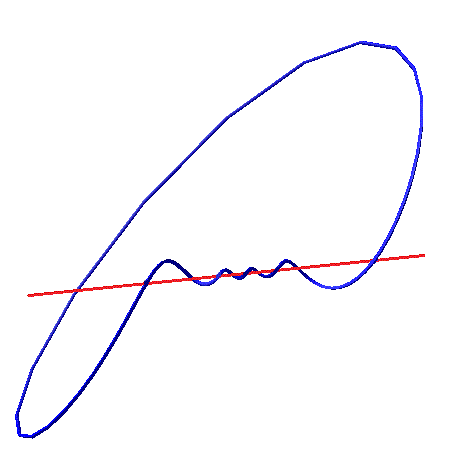}
\caption{The 5-link between discriminant divisor $\Delta$ (Blue) and toric 
divisor $D=\{v=0\}$ (Red) for the determinantal quintic model.}
\label{linkAb}
\end{figure}

\begin{figure}[h]
\centering
\includegraphics[scale=0.3]{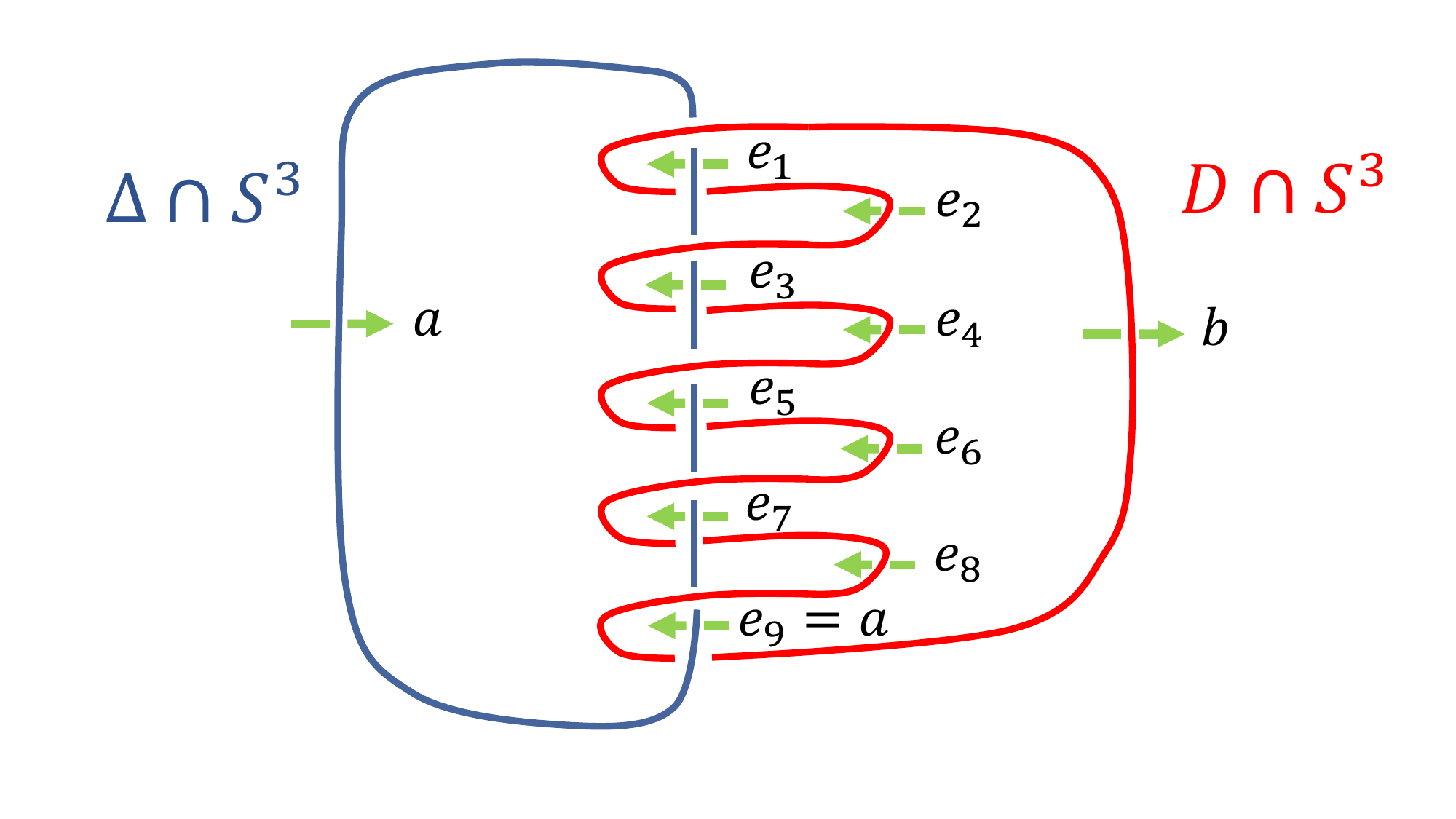}
\caption{A projection of the 5-link in Fig. \ref{linkAb}}
\label{bblinkAb}
\end{figure}

Now we compute the action of the spherical twist $T_{\mathcal{O}_{X_{A}}}$, 
since it also is going to become handy. We can compute the action on the 
basis of brane charges, by the use of the annulus partition function 
(\ref{annu}). For the determinantal quintic is given by
\begin{equation}
\begin{aligned}
\chi(\cB,\cB')=\oint_{u=0}\frac{du_0du_1}{(2\pi i)^2}&\left( 
2\sinh\frac{u_0}{2} \right)^{-5}\left( 2\sinh\frac{u_1-u_0}{2} 
\right)^{-5}\left( 2\sinh\frac{-u_1}{2} \right)^{-5}
\\
&\times f_\cB\left(\frac{u}{2\pi}\right)f_{\cB'}\left(\frac{-u}{2\pi}
\right)
\end{aligned}
\end{equation}
which is evaluated by taking the residue at the origin. By abuse of notation, 
if we label the B-branes by their charge vector component in $\vec{a}$, as in 
(\ref{chargeofBB}), we obtain the following matrix:
\begin{equation}
\chi(a^\alpha,a'^\beta)=\begin{pmatrix}
0 & 5 & 5 & 0 & 0 & 1
\\
-5 & 0 & 0 & -1 & 0 & 0
\\
-5 & 0 & 0 & 0 & -1 & 0
\\
0 & 1 & 0 & 0 & 0 & 0
\\
0 & 0 & 1 & 0 & 0 & 0
\\
-1 & 0 & 0 & 0 & 0 & 0
\end{pmatrix}.
\end{equation}
Then, we can compute the action of $T_{\mathcal{O}_{X_{A}}}$ in the B-brane 
central charges using the formula (\ref{EZ}):
\begin{equation}
\begin{aligned}
\cB\mapsto & T_{\cO_{X_{A}}}(\cB):
\\
M_{T_{\cO_X}}=&\begin{pmatrix}
1&-5&-5&0&0&-1\\
&1&&&&\\
&&1&&&\\
&&&1&&\\
&&&&1&\\
&&&&&1
\end{pmatrix}
\\
=&\Id-\begin{pmatrix}
1\\0\\0\\0\\0\\0
\end{pmatrix}
\circ
\begin{array}{c}
\begin{pmatrix}
1 & 0 & 0 & 0 & 0 & 0
\end{pmatrix}
\\\\\\\\\\\\
\end{array}
\circ
\begin{pmatrix}
0 & 5 & 5 & 0 & 0 & 1
\\
-5 & 0 & 0 & -1 & 0 & 0
\\
-5 & 0 & 0 & 0 & -1 & 0
\\
0 & 1 & 0 & 0 & 0 & 0
\\
0 & 0 & 1 & 0 & 0 & 0
\\
-1 & 0 & 0 & 0 & 0 & 0
\end{pmatrix}
\end{aligned}.
\end{equation}
where $(1,0,0,0,0,0)$ is identified with the charge vector of the 
B-brane $Q_{\mathcal{O}_{X}}$.
Now, we are ready to plot the link diagram around intersection points of 
$\Delta$ and $\{v=0\}$ or $\{w=0\}$. The situation is symmetric, and the link 
inside the $S^{3}$ centered either at $\Delta\cap\{v=0\}$ or 
$\Delta\cap\{w=0\}$ looks identical. They form a toric 
link of degree 5, whose embedding on $\mathbb{R}^{3}$ we draw 
in fig. \ref{linkAb}, thanks to the \verb=plot_knot= command of Maple. In 
addition, we show in fig. \ref{bblinkAb} the projection of the link to the 
plane along 
with all loop generators that we denote $e_{i}$, $i=1,\ldots,9$, $a$ and $b$ 
(all the green arrows in fig. \ref{bblinkAb} must be understood as loops with a 
base point outside the plane)

The generators are subject to relations. Indeed, the fundamental group of the 
link in fig. \ref{linkAb} is generated by $a,b$ and $e_{i}$'s can be expressed 
in terms of $a.b$ as
\begin{equation}
\begin{gathered}
e_1=b^{-1}\circ a\circ b\\
e_3=b^{-1}\circ a^{-1}\circ b^{-1}\circ a\circ b\circ a\circ b\\
\vdots\\
e_9=(babababab)^{-1}\circ a\circ (babababab)=a
\end{gathered}
\end{equation}
Thus the link relation of generators from $e_9$ is
\begin{equation}
(ab)^5=(ba)^5.\label{linkrel}
\end{equation}
The path going from the $X_{A}$ to the $X_{A^{T}}$ or $Y_{A}$ phase 
and back to $X_{A}$ (i.e. a loop around the $X$ or $Y$ boundary) is 
homotopically equivalent to $\{v=0\}\cap S^{3}$ and $\{w=0\}\cap S^{3}$, 
respectively. Therefore, we can express these paths in terms of generators of 
the fundamental group of the link, as:
\begin{equation}
\begin{aligned}
 &e_1\circ e_3\circ e_5\circ e_7\circ e_9
\\
=&b^{-5}(ab)^5.
\end{aligned}\label{monodromyrel}
\end{equation}

Finally, the test of the relations (\ref{linkrel}) and (\ref{monodromyrel}) 
using matrices is straightforward. Since $a$ is the generator of toric divisor 
and $b$ is that of conifold divisor, it can be checked effortlessly that the 
corresponding monodromy matrices indeed satisfy the following highly 
non-trivial relation of degree 5:
\begin{equation}
\begin{gathered}
\left(T_{(v)}T_{\cO_X}\right)^5=\left(T_{\cO_X}T_{(v)}\right)^5
\\
\left(T_{(w)}T_{\cO_X}\right)^5=\left(T_{\cO_X}T_{(w)}\right)^5
\end{gathered}. 
\end{equation}
And as expected, the window shift monodromy matrices are decomposed as
\begin{equation}
\begin{gathered}
T_X=T_{(v)}^{-5}(T^{}_{(v)}T^{}_{\cO_X})^5,
\\
T_Y=T_{(w)}^{-5}(T^{}_{(w)}T^{}_{\cO_X})^5.
\end{gathered}\label{TxTyDet}
\end{equation}
Then, the relations (\ref{TxTyDet}) can be checked directly, at the level of 
monodromy matrices, by replacing the matrices for $T_{(v)}$, $T_{(w)}$ and 
$T_{\cO_X}$ just computed above. Finally, we remark that the functor $T_{X}$ 
and $T_{Y}$ are very similar to the twists $(-)\otimes 
\mathcal{O}_{X_{A}}(-5,0)$ and $(-)\otimes 
\mathcal{O}_{X_{A}}(0,5)$, respectively. Indeed they act by a twist on every 
object for our choice of generators except for
\begin{equation}
\begin{gathered}
T_X(\cO_{D_{x}})=\cO_{X_{A}}(0,1)\oplus \cO_{X_{A}}^{\oplus 
4}(-5,0)\rightarrow \cO_{X_{A}}^{\oplus 5}(-4,0),
\\
T_Y(\cO_{D_{\phi}})=\cO_{X_{A}}(-1,0)\oplus \cO_{X_{A}}^{\oplus 
4}(0,5)\rightarrow \cO_{X_{A}}^{\oplus 5}(0,4)
\end{gathered}
\end{equation}

\subsubsection{GN Calabi-Yau}

By an analogous computation to the abelian case, we can associate the Serre 
twists 
\begin{equation}
T_{(z)}=(-)\otimes{\cO_{X}}(1), \qquad T_{(w)}=(-)\otimes{\det}^{-1}{S}_{X}
\end{equation} 
to the monodromy around the  
toric divisors $\{z=0\}$ and $\{w=0\}$, respectively. Here, $S_{X}$ denotes the 
pullback to $X_{A}$ of the tautological sub-bundle (of rank $2$) $S\rightarrow 
G(2,4)$. Therefore, at the level of the gauge charges, 
$T_{(z)}$ and 
$T_{(w)}$ simply act as
$q^0\mapsto q^0+1$ and $q^{1,2}\mapsto q^{1,2}-1$, respectively. The 
corresponding monodromy matrices, acting on $Z_{\cB}$, are given by
\begin{equation}
z\mapsto ze^{2\pi i}:\quad M_{T_{(z)}}=\begin{pmatrix}
1 & & & & &
\\
1 & 1 & & & &
\\
0 & 0 & 1 & & & &
\\
20 & 20 & 20 & 1 & &
\\
20 & 20 & 16 & 0 & 1 &
\\
0 & 0 & 2 & 1 & 0 & 1
\end{pmatrix},
\end{equation}
\begin{equation}
w\mapsto we^{2\pi i}:\quad M_{T_{(w)}}=\begin{pmatrix}
1 & & & & &
\\
0 & 1 & & & &
\\
1 & 0 & 1 & & & &
\\
16 & 20 & 16 & 1 & &
\\
8 & 16 & 8 & 0 & 1 &
\\
0 & -2 & 0 & 0 & 1 & 1
\end{pmatrix}.\label{matrixTw}
\end{equation}
The annulus partition function for GN model is given by
\begin{equation}
\begin{gathered}
\chi(\cB,\cB')=\frac{1}{2}\oint_{u=0} \frac{du_0du_1du_2}{(2\pi 
i)^3}Z_{U(2)}Z_{\text{matter}}f_\cB\left(\frac{u}{2\pi}\right)f_{\cB'}
\left(\frac{-u}{2\pi}\right),
\\
Z_{\text{matter}}:=\left( 2\sinh\frac{u_0}{2} 
\right)^{-8}\prod_{\alpha=1}^2\left( 2\sinh\frac{u_\alpha-u_0}{2} 
\right)^{-4}\left( 2\sinh\frac{-u_\alpha}{2} \right)^{-4},
\\
Z_{U(2)}:=\left(2\sinh\frac{u_1-u_2}{2}\right)^2,
\end{gathered}
\end{equation}
which gives the following open Witten index matrix:
\begin{equation}
\chi(a^\alpha,a'^\beta)=\begin{pmatrix}
0 & 8 & 6 & 0 & 0 & 1
\\
-8 & 0 & -2 & -1 & 0 & 0
\\
-6 & 2 & 0 & 0 & -1 & 0
\\
0 & 1 & 0 & 0 & 0 & 0
\\
0 & 0 & 1 & 0 & 0 & 0
\\
-1 & 0 & 0 & 0 & 0 & 0
\end{pmatrix}.
\end{equation}
Then, for later use, we compute the matrices corresponding to the  
spherical twists $T_{\cO_X}$ and $T_{S_X}$. The charge vector 
$\vec{a}$ of the object $\cO_{X_A}$ is given by 
$\vec{a}(\mathcal{O}_{X_A})=(1,0,0,0,0,0)$ and the charge vector of $S_{X}$ by 
$\vec{a}(S_{X})=(2,0,-1,-6,-4,2)$ \footnote{Note that $\vec{a}(S_{X})$ or the 
charge vector of the pullback to $X_{A}$ of any other vector bundle over 
$G(2,4)$ can be straightforwardly obtained by replacing the trivial 
representation Clifford vacuum of $Q_{\cO_X}$ by a Clifford vacuum in the 
corresponding representation. In the case at hand, this will be the 
representation $\mathbf{2}$ of $U(2)$.}. The corresponding matrices then, are 
given by
\begin{equation}
Z_\cB\mapsto Z_{T_{\cO_X}(\cB)}:\quad M_{T_{\cO_X}}=\begin{pmatrix}
1&-8&-6&0&0&-1\\
&1&&&&\\
&&1&&&\\
&&&1&&\\
&&&&1&\\
&&&&&1
\end{pmatrix},\label{matrixTX}
\end{equation}
\begin{equation}
Z_\cB\mapsto Z_{T_{S_X}(\cB)}:\quad M_{T_{S_X}}=
\begin{pmatrix}
-7&-16&-16&0&-2&-4
\\
0 & 1 & 0 & 0 & 0 & 0
\\
4 & 8 & 9 & 0 & 1 & 2
\\
24 & 48 & 48 & 1 & 6 & 12
\\
16 & 32 & 32 & 0 & 5 & 8
\\
-8 & -16 & -16 & 0 & -2 & -3
\end{pmatrix}.\label{matrixSX}
\end{equation}

\begin{figure}[h]
\centering
\includegraphics[scale=0.25]{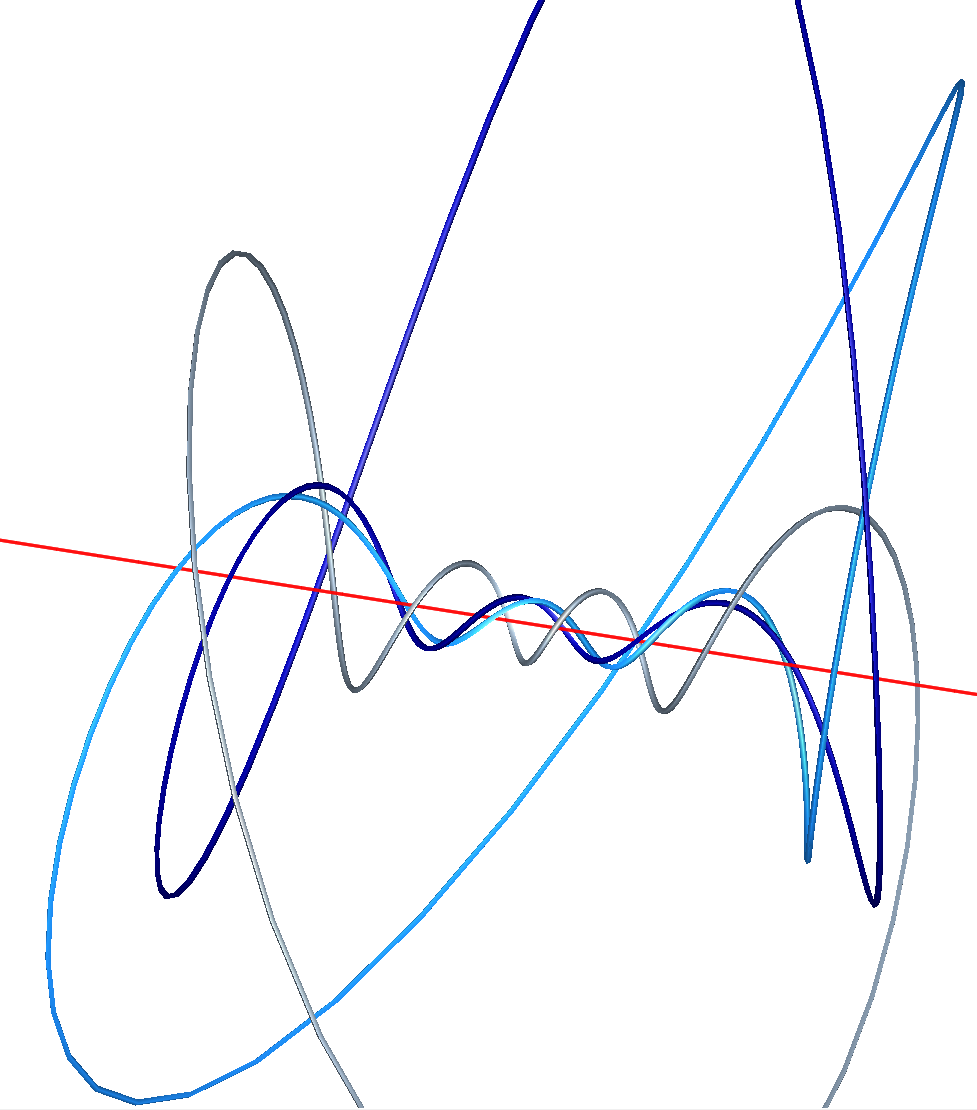}
\hspace{1in}
\includegraphics[scale=0.15]{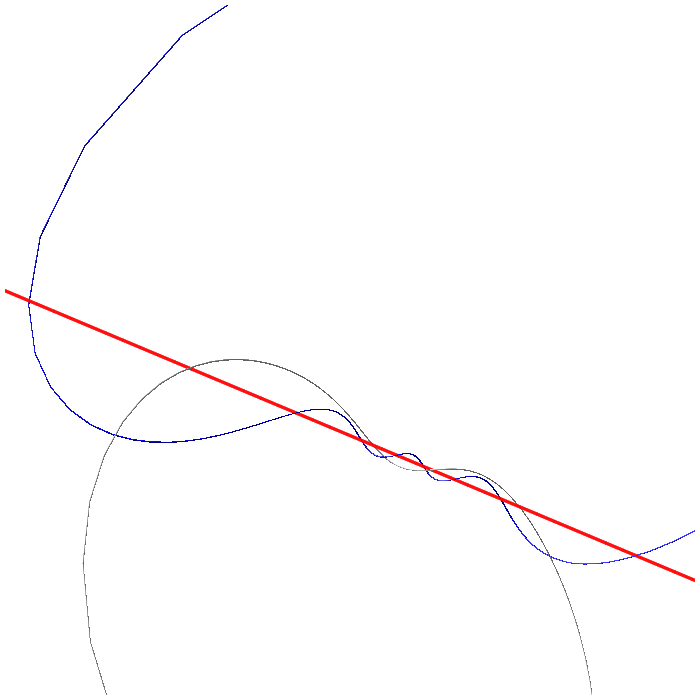}
\caption{Illustration of $\Delta_1$ (Gray) and $\Delta_2$ (Blue and Darkblue), 
linking toric divisor (Red) of $\{z=0\}$ (left) and $\{w=0\}$ (right).}
\label{linkNAb}
\end{figure}

\begin{figure}[h]
\centering
\includegraphics[scale=0.23]{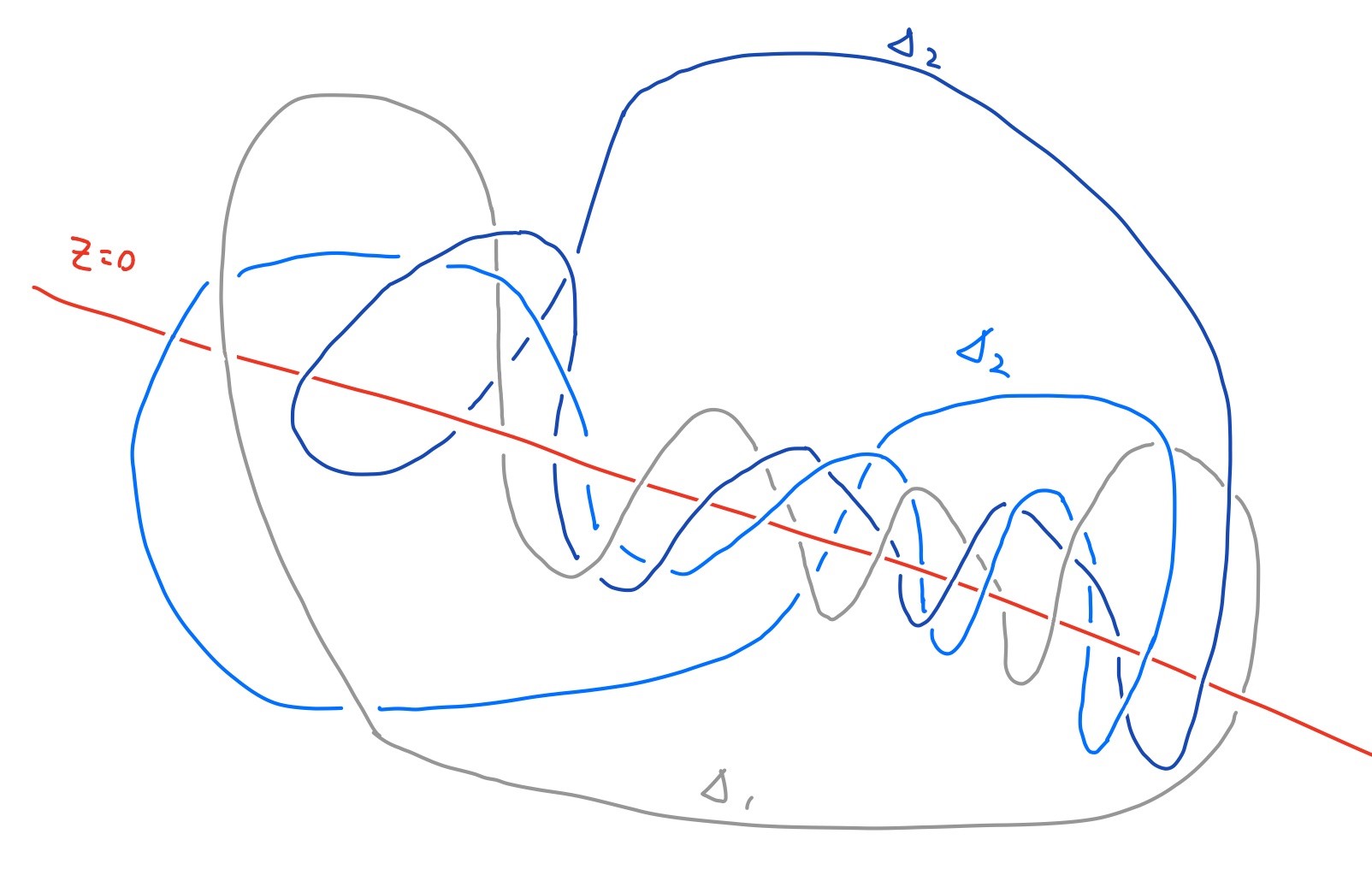}
\hspace{1in}
\includegraphics[scale=0.25]{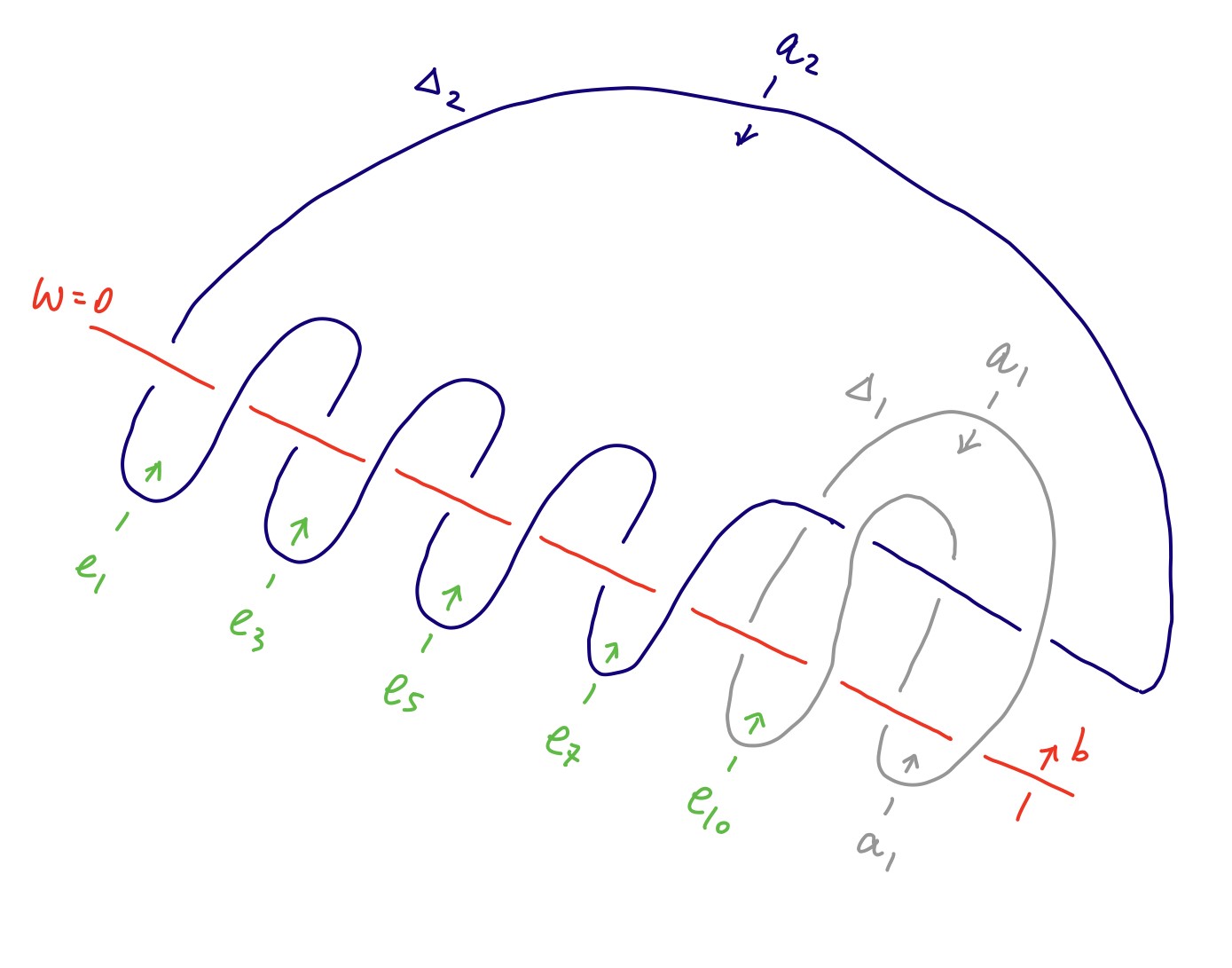}
\caption{Projection of links around $\{z=0\}$ (left) and $\{w=0\}$ (right)}
\label{linkNAbres}
\end{figure}

The link of divisors are more entangled in the GN model, we plotted them in 
$\mathbb{R}^{3}$, in fig. \ref{linkNAb} and their projection to $\mathbb{R}^{2}$ 
in fig. \ref{linkNAbres}. We find immediately differences with the 
abelian case. For instance, the primary divisor $\Delta_2$ splits into two 
components around the $X$ boundary, with one of the components presenting 
self-linking. In such case, the fundamental group of the link 
should be generated by loops around each component, hence we will have two 
generators associated to the same discriminant component, namely $\Delta_{2}$. At 
present we do not have a proposal to which functors these generators associated 
to the components on which $\Delta_{2}$ splits, corresponds to. We can remark 
however, that $T_{X}$ can be identified with the twist:
\begin{equation}
T_X=T_{(z)}^{-4}
\end{equation}

On the other hand, the link around the $Y$ boundary is a nested toric link that 
can be analyzed in a similar manner as in the abelian case. The generators of 
the fundamental group are given in fig. \ref{linkNAbres} and their relations can be straightforwardly obtained:
\begin{equation}
\begin{aligned}
(a_1a_2b)^2(a_2b)^2b=&b(a_1a_2b)^2(a_2b)^2,
\\
(a_1a_2b)^2(a_2b)^2a_2=&a_2(a_1a_2b)^2(a_2b)^2,
\\
(a_1a_2b)^2a_1=&a_1(a_1a_2b)^2.
\end{aligned}.\label{linkrel2}
\end{equation}
The loop around the $Y$ boundary (with base point at the $X_{A}$ phase) is 
homotopically equivalent to $\{w=0\}\cap S^{3}$, hence expressing it in terms of fundamental group generators is straightforward:
\begin{equation}
e_1e_3e_5e_7e_{10}a_1=b^{-4}(a_1a_2b)^2(a_2b)^2.\label{monodromyrel2}
\end{equation}
We then propose the following association of functors to generators:
\begin{equation}\label{assigGN}
b=T_{(w)},\quad a_1=T_{S_X},\quad a_2=T_{\cO_X}.
\end{equation}
It is direct to check (\ref{linkrel2}) and (\ref{monodromyrel2}) by assigning 
the corresponding matrices to the functors (\ref{assigGN}). At the level of functors, we propose then the following non-trivial relation:
\begin{equation}
T_{Y}=T_{(w)}^{-4}\left( T_{\mathcal{S}_X}T_{\cO_X}T_{(w)} \right)^2\left( T_{\cO_X}T_{(w)}\right)^2,
\end{equation}
which can be immediately checked by assigning the matrices (\ref{matrixTw}), (\ref{matrixTX}) and (\ref{matrixSX}) according to (\ref{assigGN}).

\section{\label{sec:section6} Future directions}
In this section we list some challenges and possible future directions to 
explore.

\begin{itemize}
\item Even though we can find the monodromies around discriminant components 
with relative ease in linear PAX models, we do not have a systematic way to 
assign functors to each discriminant component, particularly in the nonabelian 
case, where the techniques of \cite{gelfanddiscriminants} no longer applies. 
The examples analyzed here and in \cite{EHKR}, as well as several unpublished 
results, seems to suggest that these monodromies remain relatively simple in 
the nonabelian models, namely they still have the form of a spherical 
functor $T_{E}$ where the object $E$ is torsion sheaf or locally free sheaf 
associated with that component of the discriminant. It should be then possible 
to find some correspondence between GLSM data and number of discriminant components similar to the one that exist in the abelian models \cite{gelfanddiscriminants}.

\item In our analysis, all the links coming from discriminant in 
PAX models are found to be of the type of 'nested torus links', whose fundamental groups were systematically studied in 
\cite{argyres2019fundamental}. It would be interesting to further explore how these fundamental group relations impose conditions on the autoequivalences of linear PAX models. In addition, a valid question is how general the phenomenon of these discriminants being nested torus links is.

\item A completely new direction would be to study the interaction between the 
$T_{Y}$, $T_{X}$ functors and their decomposition in the context of 
Seiberg-like dualities \cite{Hori:2011pd}. For instance, the monodromy matrices 
are expected to remain invariant under duality, but their interpretation can 
change as the base point can be mapped to a phase whose IR dynamics have a 
different description.

\item In the two-parameter models analyzed here, our derivation of the window categories suggests that we can consider an effective one -parameter model, close to the phase boundaries but keeping $|\zeta_{0}|$ or $|\zeta_{1}|$ large but finite. Then the corresponding FI-theta parameter that is frozen to a large value should act as an effective coupling, splitting the Coulomb branch singularities of the corresponding one-parameter model. This can provide a systematic and unambiguous way to determine the assignment between the generators of the fundamental group of the link and functors.

\item The assignment of $T_{\cO_{X}}$ to the monodromy around a conifold point as a well established mirror interpretation as Dehn twists around a lagrangian spheres \cite{seidel2000braid}. It would be interesting to study the mirror of the functor $T_{S_{X}}$ we found, for instance, in the GN example. Similar functors are also found in \cite{EHKR}.

\item More complicated examples of determinantal varieties 
are presented in \cite{2012determinantal} and their link groups are much more 
complex. These models can give relations between functors 
that are mathematically novel.

\end{itemize}

\appendix
\section{\label{sec:proof} Proof of lift $D^{b}Coh(X_{A})\rightarrow MF_{G}(W)$}

In this section we will present an argument to show that any B-brane in the IR can be lifted to a GLSM B-brane i.e. to an object in $MF_{G}(W)$. This derivation originally appears in \cite{EHKR} and we adapted it to our examples. We will focus on the $X_{A}$ phase, since other phases of linear PAX models will work analogously. Let us start by considering an object $\mathcal{E}\in D^{b}Coh(X_{A})$. For the linear PAX models considered in the present work, we have an embedding $\iota: X_{A}\hookrightarrow Z:=\mathbb{P}^{n(n-k)-1}\times Gr(n-k,n)$. Therefore, we can pushforward $\mathcal{E}$ to $\iota_{*}\mathcal{E}\in D^{b}Coh(Z)$. Given a full exceptional collection for $\mathbb{P}^{n(n-k)-1}$ and for $Gr(n-k,n)$, generated by the objects $\{L_{q}\}$ and $\{E_{\lambda}\}$ respectively, we can construct a full exceptional collection for $D^{b}Coh(Z)$, generated by the objects $\mathcal{W}_{\lambda}(q):=\{L_{q}\boxtimes E_{\lambda}\}$ \cite{kuznetsov2011base}. Indeed, we can consider $L_{q}:=\mathcal{O}(q)$, $q=p,\ldots, n(n-k)-1+p$, for some $p\in \mathbb{Z}$ (i.e. a twist of Beilinson's exceptional collection) and $E_{\lambda}:=\Sigma_{\lambda}S$, corrresponding to Kapranov's expectional collection, as reviewed in section \ref{sec:section4} or a twist of it. Then, we can always find an object $\mathcal{F}$, quasi-isomorphic to $\iota_{*}\mathcal{E}$ of the form
\begin{equation}
\mathcal{F}:\begin{tikzcd}
\cdots \arrow{r}{d^{i-1}_{\mathcal{F}}}
    & \mathcal{F}_{i}\arrow{r}{d^{i}_{\mathcal{F}}}
        & \mathcal{F}_{i+1} \arrow{r}{d^{i+1}_{\mathcal{F}}}
       & \cdots,
\end{tikzcd}
\end{equation}
where each factor takes the form
\begin{equation}
\mathcal{F}_{i}=\bigoplus_{\lambda_{i},q_{i}}\mathcal{W}_{\lambda_{i}}(q_{i}).
\end{equation}
Then, we define the Chan-Paton vector space (see section \ref{sec:section3}) of our lift by
\begin{equation}
M_{0}:=\bigoplus_{\lambda_{i},q_{i},i \text{ even\ }}\mathcal{W}_{\lambda_{i}}(q_{i})\qquad M_{1}:=\bigoplus_{\lambda_{i},q_{i},i \text{ odd\ }}\mathcal{W}_{\lambda_{i}}(q_{i})
\end{equation}
where each $\mathcal{W}_{\lambda}(q)$ can be straightforwardly understood as a $U(1)\times U(2)$ representation, hence defining $\rho_{M}$ straightforwadly. Moreover, the differential $d_{\mathcal{F}}$ is $U(1)\times U(2)$-equivariant and its components $d^{i}_{\mathcal{F}}$ can be written as polynomial functions on the fields $\phi$ and $x$. Next, we have to find $\mathbf{T}$, a matrix factorization of the GLSM superpotential $W$. For this, we propose the following ansatz:
\begin{equation}
\mathbf{T}=\mathbf{T}_{0}+\sum_{l>0}p_{[l]}\mathbf{T}^{[l]}_{l},\qquad p_{[l]}\mathbf{T}^{[l]}_{l}:=p_{i_{1}}\cdots p_{i_{l}}\mathbf{T}^{i_{1}\cdots i_{l}}_{l}
\end{equation}
and
\begin{equation}
\mathbf{T}_{0}:=d_{\mathcal{F}},\qquad \mathbf{T}^{[l]}_{l}(x,\phi)\in \mathrm{End}^{1}_{\mathrm{Sym}(V^{\vee})}(M),\quad l>0.
\end{equation}
Then, by construction
\begin{equation}
\rho_{M}^{-1}(g)\mathbf{T}_{0}(\rho_{m}(g)\cdot(\phi,x))\rho_{M}(g)=\mathbf{T}_{0}(\phi,x).
\end{equation}
Note that the endomorphisms $\mathbf{T}^{[l]}_{l}$ are only functions of $(x,\phi)$ and belong to the representation $(l,\mathrm{Sym}^{l}\overline{\mathbf{n}-\mathbf{k}})$. We will solve inductively in $l$ the equation
\begin{equation}\label{eqind}
\mathbf{T}^2=W\cdot \mathrm{id}_{M}
\end{equation}
more precisely, (\ref{eqind}) gives the following set of equations for $\mathbf{T}^{[l]}_{l}$:
\begin{eqnarray}\label{eqindd}
\{\mathbf{T}_{0},\mathbf{T}^{i}_{1}\}&=& A^{ij}(\phi)x_{j}\nonumber\\
\{p_{[l]}\mathbf{T}^{[l]}_{l},\mathbf{T}_{0}\}&=&-\frac{1}{2}\sum_{\substack{s+r=l \\ s,r\geq 1}}p_{[s]}p_{[r]}\{\mathbf{T}^{[s]}_{s},\mathbf{T}^{[r]}_r\}
\end{eqnarray}
for $l=1$, $\mathbf{T}^{i}_{1}$ must be in the representation $(1,\overline{\mathbf{2}})$ of $U(1)\times U(2)$ and had weight $-1$ under $U(1)_{R}$. Then when restricted to $X_{A}$,
\begin{eqnarray}
\{\mathbf{T}_{0},\mathbf{T}^{i}_{1}\}|_{X_{A}}=\{d_\mathcal{F},\mathbf{T}^{i}_{1}\}|_{X_{A}}=0
\end{eqnarray}
this means that $\mathbf{T}^{i}_{1}$ is a cochain map of homological degree $-1$. This is true, because the equivariant map $A^{ij}(\phi)x_{j}$ is nullhomotopic in $D^{b}Coh(X_{A})$. Assume we have then solutions $\mathbf{T}_{2}^{[2]},\ldots,\mathbf{T}^{[l-1]}_{l-1}$, $l>1$ to the eqs. (\ref{eqindd}), then is straightfoward to show
\begin{eqnarray}
-\frac{1}{2}[\mathbf{T}_{0},\mathbf{T}']=[\mathbf{T}_{0},\{\mathbf{T}^{[l]}_{l},\mathbf{T}_{0}\}]=0\qquad \mathbf{T}':=\sum_{\substack{s+r=l \\ s,r\geq 1}}\{\mathbf{T}^{[s]}_{s},\mathbf{T}^{[r]}_r\}
\end{eqnarray}
since $(\mathbf{T}_{0})^{2}=0$. Then $\mathbf{T}'$ is a cochain map of $U(1)_{R}$ charge i.e. homological degree $2-2l<0$. Moreover
\begin{eqnarray}
\mathrm{Hom}(\mathcal{F},\mathcal{F}[2-2l])=\mathrm{Ext}^{0}(\mathcal{F},\mathcal{F}[2-2l])=\mathrm{Ext}^{2-2l}(\mathcal{F},\mathcal{F})=0
\end{eqnarray}
therefore $\mathbf{T}'$ must be nullhomotopic and so, a solution $\mathbf{T}^{[l]}_{l}$ to (\ref{eqindd}) must exist. So, we have shown the existence of an object $\mathcal{B}=(M,\rho_{M},R_{M},\mathbf{T})$. As we showed in section \ref{sec:section3}, it is always possible to construct an admissible contour $\gamma$ in this chamber, therefore we finished the construction of an object $(\mathcal{B},\gamma)\in MF_{G}(W)$.

\section*{Declarations}

\textbf{Conflict of interest}: The authors declared that they have no conflicts 
of interest to this work.

\bibliographystyle{fullsort}
\bibliography{refGN}

\end{document}